\documentclass{aa}  

\usepackage{graphicx}
\usepackage{txfonts}
\def\figref#1{Fig.\,\ref{#1}}

\newcommand{\rSii}{[S~{\sc ii}] $\lambda$6731/6716}
\newcommand{\rOiii}{[O~{\sc iii}] $\lambda$4363/5007}
\newcommand{\rNii}{[N~{\sc ii}] $\lambda$5755/6583}

\newcommand{\Opp}{O$^{++}$}
\newcommand{\Op}{O$^{+}$}

\newcommand{\Te}{$T_{e}$}
\newcommand{\Ne}{$n_{e}$}

\newcommand{\lam}{$\lambda$}

\begin{document}

   \title{Spectroscopy of southern Galactic disk planetary nebulae
   \thanks{Based on observations made at the South African Astronomical
           Observatory}
   }

   \subtitle{Notes on chemical composition and emission-line stars}

   \author{S.K. G\'orny}

   \institute{N. Copernicus Astronomical Center, 
              Rabia\'nska 8, 87-100 Toru\'n, Poland\\
              \email{skg@ncac.torun.pl}
             }

   \date{Received June 2013; accepted June 2014}

 
  \abstract
   {}
   {We present low resolution spectroscopic observations for a sample of
    53 planetary nebulae (PNe) located in the southern sky between
    Vela and Norma constellations and pertaining to the Galactic disk
    with expected Galactocentric distance range of 5 to 10\,kpc.
   }
   {We derive nebular chemical composition and plasma parameters with the
    classical empirical method. For most of the observed objects this has
    been done for the first time.  We compare our results to published data
    for PNe of the Galactic bulge and PNe in the inner-disk region with
    expected typical Galactocentric distance of about 3\,kpc.  We use the
    spectra to search for emission-line central stars in the observed
    sample.
   }
   {The distributions of the chemical abundances of the observed disk sample
    are generally indistinguishable from Galactic bulge and inner-disk PNe
    populations.  The exceptions are possible differences in the He/H
    distribution, as compared to bulge PNe and Ne/Ar, as compared to
    the inner-disk PNe sample.  The derived O/H ratios for the observed disk
    PNe fit to the concept of flattening of the chemical gradient in the
    inner parts of the Milky Way.
      Investigating the spectra, we found six new emission-line central
    stars comprising examples of all known types: WEL, VL, and [WR].  We
    confirm that these types represent three evolutionary unconnected forms
    of enhanced mass-loss in the central stars of PNe.  We note on the
    problem of high ionisation PNe with nebular C\,IV emission that can
    mimic the presence of WEL central stars in 1D spectra.
   }
   {}


   \maketitle
%

\section{Introduction}

Planetary nebulae (PNe) are a very short evolutionary phase in the life of
low- and intermediate-mass stars (0.8 -- 8\,M$_{\odot}$) that occur after
they leave the asymptotic giant branch (AGB) and before they end their lives
as white dwarfs \citep{Iben1995}.  During the AGB and in particular, at the
tip of this phase of evolution, stars experience strong winds that
efficiently enrich the surrounding interstellar medium with huge amounts of
gas, dust, and molecules from the outer layers of the star
\citep[e.g.,][]{Herwig2005}.  When mass loss stops, these stars leave the
AGB and rapidly increase their effective temperatures preserving a roughly
constant luminosity.  When the ionisation of the ejected gas by the central
star becomes possible, a new PN emerges composed of the matter formerly
expelled from the progenitor.

A better knowledge of the PNe phase is necessary for understanding the final
fate of stars like the Sun but also to study the formation and the chemical
evolution of the Milky Way and other galaxies.  The PNe have several
properties that make them very good tools in this respect.  They are good
indicators of the chemical composition of the interstellar matter from which
the progenitor stars were born \citep[see, e.g.,][]{Chiappini2009}.  Due to
the wide mass range of their progenitors, the PNe that we now observe have
been created by stars that formed at very different epochs.  In addition,
their narrow emission-line nature makes them observable at very large
distances.

Thanks to a considerable effort, hundreds of new PNe have been discovered in
the Milky Way in the last two decades \citep[for a review
see][]{Parker2011}.  However, published analysis based on the observed PNe
properties still frequently use limited and biased samples.  The first
observing bias is the natural condition that the most easily accessible
objects (brightest or simply near-by) are first to be observed and analysed. 
Even if a substantial number of such data is collected, it does not
necessarily mean they properly represent the PNe population across the
Galaxy.

The problem that observational selection effects have to be properly taken
into account has been underlined many times in the past
\citep[e.g.,][]{Stasinska1994}.  Nevertheless, they can remain one of the
major concerns, while investigating the abundance gradients in the Milky
Way.  By analysing the samples collected by different authors, it can be
noted that they are often dominated by the relatively close objects.

In some cases, selection effects may appear unexpectedly. For example,
\cite{Gorny2009} reported that Wolf-Rayet ([WR]) type central stars are
probably related to the intrinsically brightest PNe in a given population. 
The reason for this is unknown.  It is important here that this 
property remained unnoticed until a dedicated, sufficiently deep search of
new [WR] objects was undertaken that also incorporated the fainter Galactic
bulge PNe.

For this reason, we think it is still necessary to collect new observational
data to further minimise the presence and influence of selection effects in
PNe related studies.  In this paper, we present low resolution optical
spectroscopic observations of 53 PNe that are located in the southern sky
between Vela and Norma constellations with an expected Galactocentric
distance range of 5 to 10\,kpc.  The sample and observations are described
in Sect.~2.  The results of the search for new emission-line central stars
is presented in Sect.~3.  In Sect.~4, we derive the chemical composition of
the observed PNe and compare them briefly with abundance distributions for
Galactic bulge and inner-disk PNe populations.  The main results of this
work are summarised in Sect.~5.

\begin{table*}[t]
\caption{ Log of observations presenting PN\,G numbers, usual names, 
          year of the observation, exposure time in seconds; mean 
          airmass during observation, and nebular diameter in arcsec. 
}
\begin{tabular}{ l l @{\hspace{0.10cm}} l r r r c l l @{\hspace{0.10cm}} l r r r }
\cline{1-6} \cline{8-13} 
 PN\,G      & name          & date & exp  & Z    & diam    & ~~~~~~ &   PN\,G      & name          & year & exp. & Z    & diam   \\
\cline{1-6} \cline{8-13} 
 268.4+02.4 & PB 5          & 2004 & 3600 & 1.10 &    1.6  &  &   308.2+07.7 & MeWe 1- 3     & 2004 & 5400 & 1.11 &   18.0 \\ 
 269.7-03.6 & PB 3          & 2003 &  900 & 1.06 &    7.0  &  &   309.0+00.8 & He 2- 96      & 2006 & 1050 & 1.15 &    2.8 \\ 
 279.6-03.1 & He 2- 36      & 2004 & 1800 & 1.11 &   19.5  &  &   309.5-02.9 & MaC 1- 2      & 2011 & 3600 & 1.19 &     st.\\ 
 283.8+02.2 & My 60         & 2006 & 1800 & 1.17 &    8.0  &  &   310.4+01.3 & Vo 4          & 2003 & 1800 & 1.14 &    8.0 \\ 
 283.8-04.2 & He 2- 39      & 2006 & 1800 & 1.19 &   12.3  &  &   310.7-02.9 & He 2-103      & 2003 & 1680 & 1.19 &   21.5 \\ 
 285.4+01.5 & Pe 1- 1       & 2004 & 2400 & 1.14 &    3.0  &  &   311.4+02.8 & He 2-102      & 2004 & 2400 & 1.12 &   11.5 \\ 
 286.0-06.5 & He 2- 41      & 2011 & 2400 & 1.24 &     st. &  &   312.6-01.8 & He 2-107      & 2004 & 2400 & 1.17 &    9.4 \\ 
 289.8+07.7 & He 2- 63      & 2011 & 2400 & 1.09 &    2.9  &  &   314.4+02.2 & PM 1- 81      & 2004 & 1800 & 1.12 &    4.7 \\ 
 291.4+19.2 & ESO 320-28    & 2004 & 5400 & 1.02 &   28.8  &  &   315.1-13.0 & He 2-131      & 2004 & 1200 & 1.30 &    9.8 \\ 
 292.8+01.1 & He 2- 67      & 2003 & 1800 & 1.20 &    3.6  &  &   315.4+05.2 & He 2-109      & 2004 & 2400 & 1.09 &    9.1 \\ 

 293.1-00.0 & BMPJ1128-61   & 2005 & 2700 & 1.19 &    0.1  &  &   315.7+05.5 & LoTr 8        & 2005 & 1200 & 1.09 &   26.7 \\ 
 293.6+01.2 & He 2- 70      & 2003 & 1500 & 1.13 &   21.7  &  &   316.1+08.4 & He 2-108      & 2004 & 2400 & 1.07 &   12.9 \\ 
 294.6+04.7 & NGC 3918      & 2004 & 2400 & 1.11 &   17.9  &  &   318.3-02.5 & He 2-116      & 2004 & 2400 & 1.15 &   47.3 \\ 
 294.9-04.3 & He 2- 68      & 2004 & 2400 & 1.25 &    2.5  &  &   319.2+06.8 & He 2-112      & 2004 & 3600 & 1.07 &    6.6 \\ 
 295.3-09.3 & He 2- 62      & 2006 & 1200 & 1.29 &     st. &  &   320.9+02.0 & He 2-117      & 2004 & 2400 & 1.10 &    4.9 \\ 
 296.3-03.0 & He 2- 73      & 2005 & 2400 & 1.21 &    2.9  &  &   321.3+02.8 & He 2-115      & 2005 & 2400 & 1.09 &    2.9 \\ 
 297.4+03.7 & He 2- 78      & 2003 &  900 & 1.12 &    3.5  &  &   321.8+01.9 & He 2-120      & 2006 & 3600 & 1.10 &   30.9 \\ 
 299.0+18.4 & K 1- 23       & 2004 & 3600 & 1.03 &   60.2  &  &   322.5-05.2 & NGC 5979      & 2004 & 2400 & 1.14 &   19.6 \\ 
 299.5+02.4 & He 2- 82      & 2003 & 1200 & 1.13 &   28.4  &  &   323.9+02.4 & He 2-123      & 2005 & 2400 & 1.11 &    6.7 \\ 
 300.2+00.6 & He 2- 83      & 2004 & 2400 & 1.15 &    4.6  &  &   324.2+02.5 & He 2-125      & 2006 & 2700 & 1.14 &    3.3 \\ 

 300.4-00.9 & He 2- 84      & 2004 & 1800 & 1.18 &   29.1  &  &   324.8-01.1 & He 2-133      & 2004 & 1800 & 1.10 &    1.8 \\ 
 300.5-01.1 & He 2- 85      & 2003 & 2700 & 1.18 &    8.5  &  &   327.1-01.8 & He 2-140      & 2005 & 2100 & 1.14 &    4.1 \\ 
 300.7-02.0 & He 2- 86      & 2004 & 2400 & 1.19 &    3.2  &  &   327.8-01.6 & He 2-143      & 2004 & 2400 & 1.10 &    3.7 \\ 
 300.8-03.4 & ESO 095-12    & 2006 & 3600 & 1.21 &   18.0  &  &   330.9+04.3 & Wray 16-189   & 2005 & 1800 & 1.04 &   15.0 \\ 
 302.2+02.5 & Wray 16-120   & 2005 & 1200 & 1.14 &   12.0  &  &   336.9+08.3 & StWr 4-10     & 2006 & 2100 & 1.07 &     st.\\ 
 305.1+01.4 & He 2- 90      & 2003 & 1620 & 1.14 &    3.1  &  &   338.1-08.3 & NGC 6326      & 2004 & 2400 & 1.06 &   16.8 \\ 
 307.3+05.0 & Wray 16-128   & 2004 & 3600 & 1.11 &   18.0  &  &              &               &      &      &      &        \\ 
\cline{1-6} \cline{8-13} \\
\end{tabular}\\
\end{table*}

\begin{figure} \resizebox{\hsize}{!}{\includegraphics{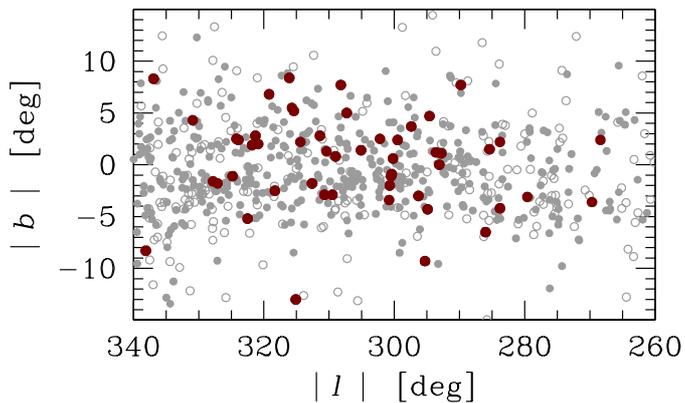}}
\caption[]{
 The distribution of the observed PNe on the sky in galactic coordinates,
 dark circles.  Open grey circles mark other PNe catalogued by 
 \cite{Acker1992} and filled grey circles PNe from \cite{Parker2006} and
 \cite{Miszalski2008}.  }
\label{lb} \end{figure}

\begin{figure}
\resizebox{\hsize}{!}{\includegraphics{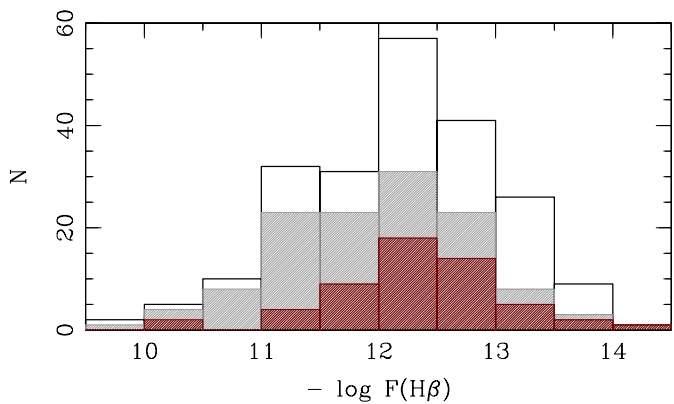}} 
\caption[]{
 The distribution of the nebular F(H$\beta$) fluxes for PNe observed in this
 work, dark hatched bars; PNe included in other large spectroscopic surveys
 (see text for the references) - grey hatched bars; and other PNe from
 \cite{Acker1992}, as open bars.  }
\label{fhb}
\end{figure}

\begin{figure*}[t]
\resizebox{\hsize}{!}{\includegraphics{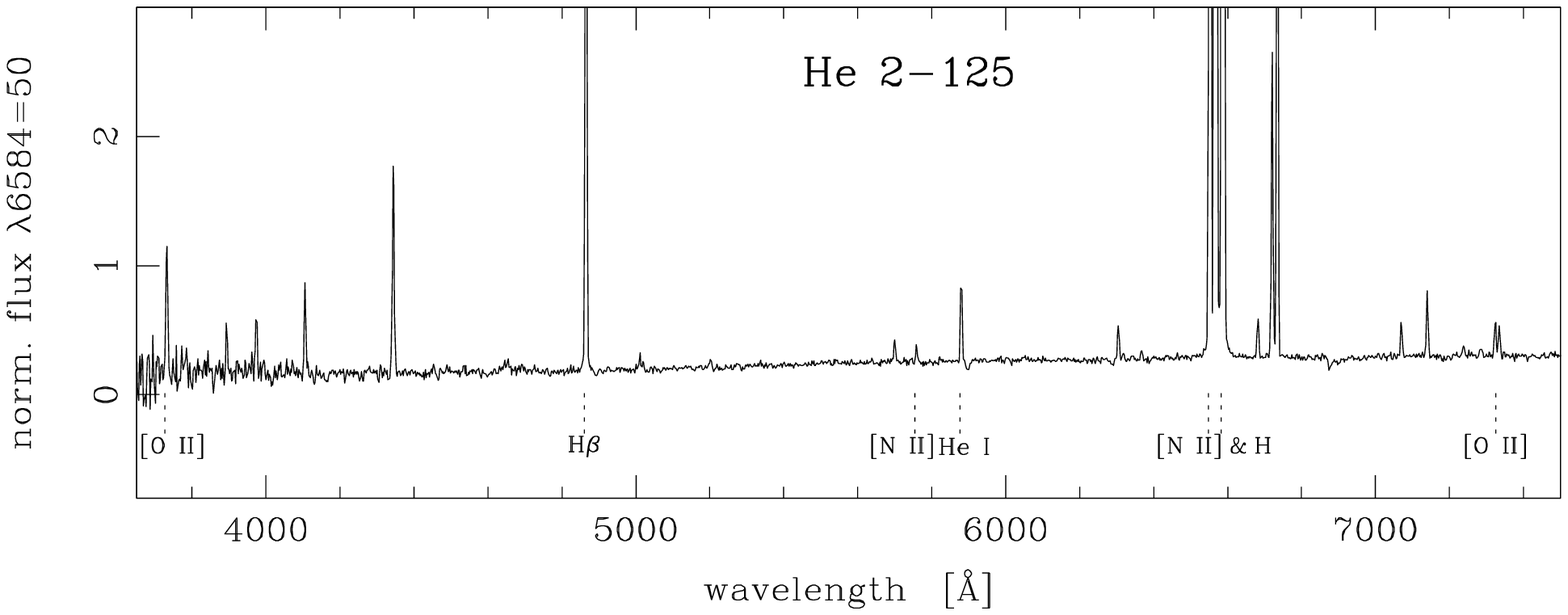}}
\resizebox{\hsize}{!}{\includegraphics{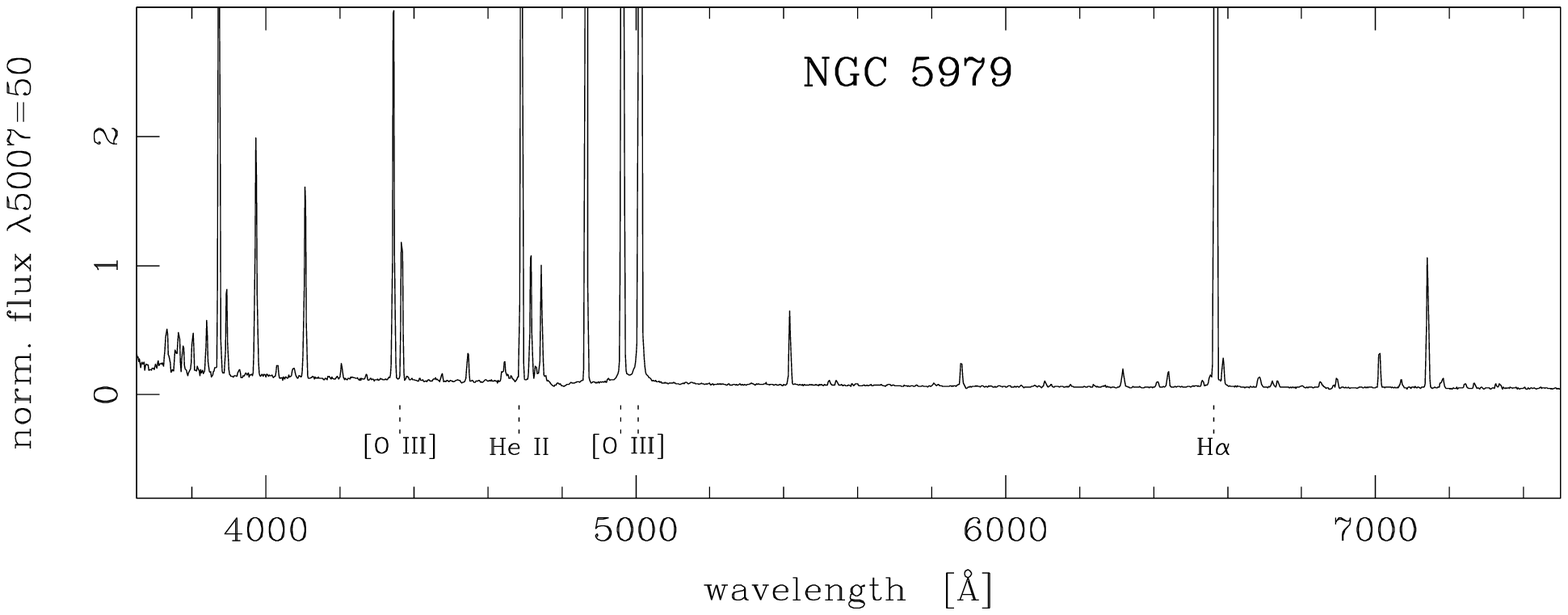}}
\caption[]{
 Illustrative examples of the acquired spectra of low ionisation planetary
 nebula (top panel) and a high ionisation nebula (bottom panel) in our
 observing programme.
}
\label{spectra_all}
\end{figure*}

\section{Low resolution optical spectra}

\subsection{Observations and data reduction}

We present observations of 53 PNe that are located near the plane of the
Milky Way disk with Galactic longitude from {\it l}=260$\degr$ to
340$\degr$.  Their on-sky distribution is presented in \figref{lb}.  Spectra
of these objects were secured with the 1.9-meter telescope of the South
African Astronomical Observatory in April 2003, May 2004, June 2005, and May
2006 as an auxiliary program during observing runs dedicated to the search
of new small PNe towards the Galactic bulge \citep{Gorny2006} when that part
of the sky was not accessible.  No strict selection criteria were employed
to choose the targets; however, objects not listed in large surveys like
\cite{FMCB,FPMC,KB94,Cu96,Mi02,Kw03} were preferred and, therefore, we have
only 13 sources in common with these authors.  Some additional preference
was given to PNe with an unknown type of the central star and without
information on measured stellar flux.  Three additional PNe with Spitzer
spectra were observed in April 2011.

In \figref{fhb} we present the distribution of total apparent nebular
H$\beta$ fluxes for our observed sample, as seen in the dark hatched bars. 
We compare them to the distribution for the PNe located in the same
direction of the sky and presented in above mentioned papers or
included in more recent papers, such as \cite{Ca10,He10,Mi10}.  For a
reference, the remaining PNe of this region from Strasbourg-ESO Catalogue of
\cite{Acker1992} are shown.  As a result of our selection approach, the
sample we observed is composed of fainter PNe.  Our targets were also
smaller.  The median diameter of PNe we observed was about 8\,arcsec,
whereas the PNe included in other papers had a median diameter of
10\,arcsec, as compared to almost 16\,arcsec for a general PNe population
in that region \citep[238 PNe from ][]{Acker1992}.

In Table~1, the detailed log of observations is presented. The PNe are
identified by their PN\,G identifier (Galactic coordinates) and usual names. 
The typical spectral coverage was 3500-7400\AA\ with an average resolution
of 1000 which allows for detection of all important nebular lines without
serious blending.  The slit width was usually set to 1.8\,arcsec, which is
slightly larger than the typical seeing conditions at the site.  The slit
orientation was always E-W, and it was positioned directly on the central
star if it could be identified or roughly at the geometrical center of the
nebula.  The CCD was read with a 1x2 pre-binning in the spatial direction,
which is along the slit.  The slit width and its orientation could in
principle lead to some loss of the stellar light due to seeing variations
and differential atmospheric refraction.  The latter effect could have
influence on relative line ratios for the smallest observed PNe.  To avoid
this, we tried to observe all objects at their minimum air-masses.  Their
values are given in column 5 of Table~1.

The integration time of the observations varied from 20 to 90 minutes and
was divided into two or three sub-exposures.  In many cases, additional
shorter exposure lasting from 3~minutes to only 15 seconds were executed to
ensure unsaturated detection of the strongest lines: H$\alpha$,
[O\,III] \lam5007, and/or [N\,II] \lam6583.

Each night, at least two different spectroscopic standard stars were
observed (LTT\,3218, LTT3864, LTT4364, LTT\,7987, LTT\,9239 CD\,-32 or
Eg\,274).  Since no order blocking filters were used, some contamination to
the red light of the standard stars is possible from the second order blue
continuum.  The spectrograph was, however, characterised by a general low
sensitivity to blue light.  We checked during tests in 2011 and 2012 with
the same instrument and configuration for stars, LTT\,3218 and CD\,-32, that
contamination was adding only up to 4\% of additional light at 7600\AA.  The
effect was linear in wavelength (from about 6000\AA\ upward) but would
sometimes disappear depending on the sky conditions of a given night.  We
have therefore not corrected the measured lines for this effect.  In
practise, its influences for our PNe would be measurable only for [O\,II]
\lam7325 doublet adding some additional uncertainty to its
measurements\footnote{We anyway correct \lam7320/30 lines to measurements of
their blue counterpart [O\,II] \lam3725 - see in Sect.\,4}.

We used the long-slit package procedures of ESO-MIDAS\footnote{ESO-MIDAS is
developed and maintained by the European Southern Observatory {\tt
http://www.eso.org/sci/software/esomidas/}} to reduce and calibrate the
spectra.  These included bias subtraction, flat-field correction,
atmospheric extinction correction, wavelength, and flux calibration and
extraction of the 1-dimensional spectra.  For the standard stars and most of
the observed PNe, this was done in the usual manner by summing the
appropriate rows of the sky subtracted frames.  In the case of some PNe, in
particular, those located close to the Galactic plane in crowded fields a
multi-step method similar to the one described in \cite{Gorny2009} had to be
applied.  In this method before extracting 1D spectra, all the background
sources have to be removed.  This includes not only sky continuum emission
and telluric and interstellar lines but also stellar continua, including
that of the PN central star.  Intensities of the nebular lines were finally
measured from the 1-dimensional spectra by employing the REWIA
package\footnote{REWIA is a data processing program developed by
J.\,Borkowski at Copernicus Astronomical Center {\tt
http://www.ncac.torun.pl/$\sim$jubork/rewia/}}, assuming Gaussian profiles
and performing multi-Gaussian fits when necessary.

The line intensities have been corrected for extinction by adopting the
extinction law of \cite{Seaton1979} to reproduce the theoretical case B of
Balmer line ratios at the electron temperature and density derived for the
object.  The same iterative procedure, as the one described in
\cite{Gorny2009}, was applied using the measured H$\alpha$/H$\beta$ ratio. 
As a result the procedure was usually not giving the theoretically expected
ratios of H$\gamma$/H$\beta$ and/or H$\delta$/H$\beta$.  This can be due to
deviations from the adopted extinction law for individual PNe or higher
uncertainty of flux calibration in the blue part of the spectra.  If the
deviations in H$\delta$ and/or H$\gamma$ were systematical in one direction
and larger than expected from purely observational uncertainties, we applied
an additional correction procedure \citep{Gorny2009} to bring the
H$\gamma$/H$\beta$ and H$\delta$/H$\beta$ ratios to their theoretically
expected values.  The correction factors typically amounted to 7\% and 11\%
for H$\gamma$ and H$\delta$ lines, respectively.  A proportional wavelength
dependent correction was than applied to all the nearby lines, including the
important [O~{\sc iii}] $\lambda$4363 and [O~{\sc ii}] $\lambda$3727 lines,
to secure that they are properly reddening corrected.

In Table~B.1 in Appendix B\footnote{All appendices are available
online}, we present the measured and dereddened intensities of all important
nebular lines on the scale of H$\beta$ = 100 for the observed PNe.  In cases
when the additional correction described above was necessary, the lines are
marked with "c" in Table~B.1.

\begin{figure}[t]
\resizebox{\hsize}{!}{\includegraphics{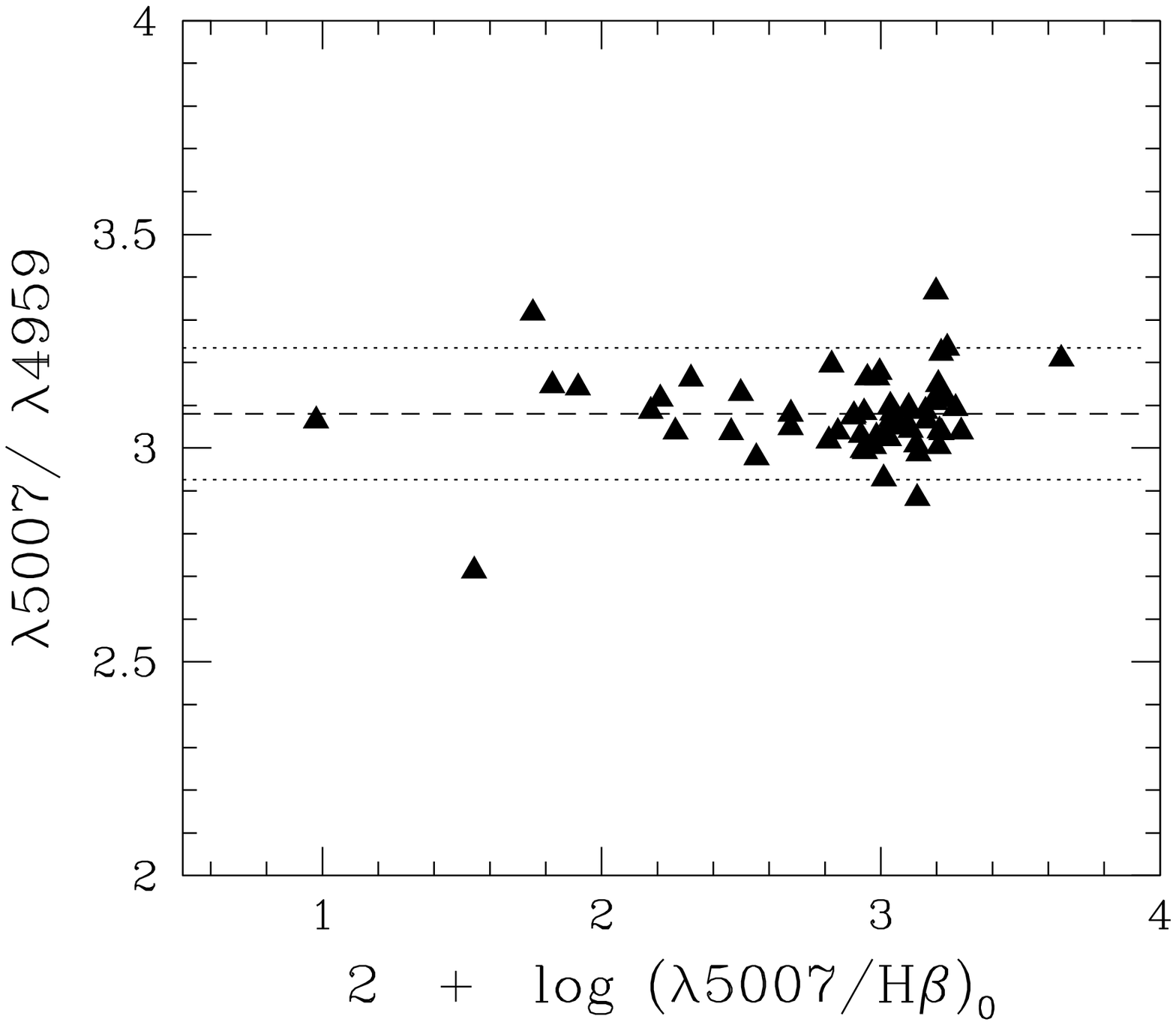}}
\caption[]{ 
  Intensity ratio of the [O~{\sc iii}] $\lambda$5007 to $\lambda$4959 lines
  as the function of the flux of [O~{\sc iii}] $\lambda$5007 (in units of
  H$\beta$).  The dashed line represents mean value for the presented
  observations, and the dotted lines are 5\% deviations from it.
}
\label{o3}
\end{figure}

\begin{figure}[t]
\resizebox{\hsize}{!}{\includegraphics{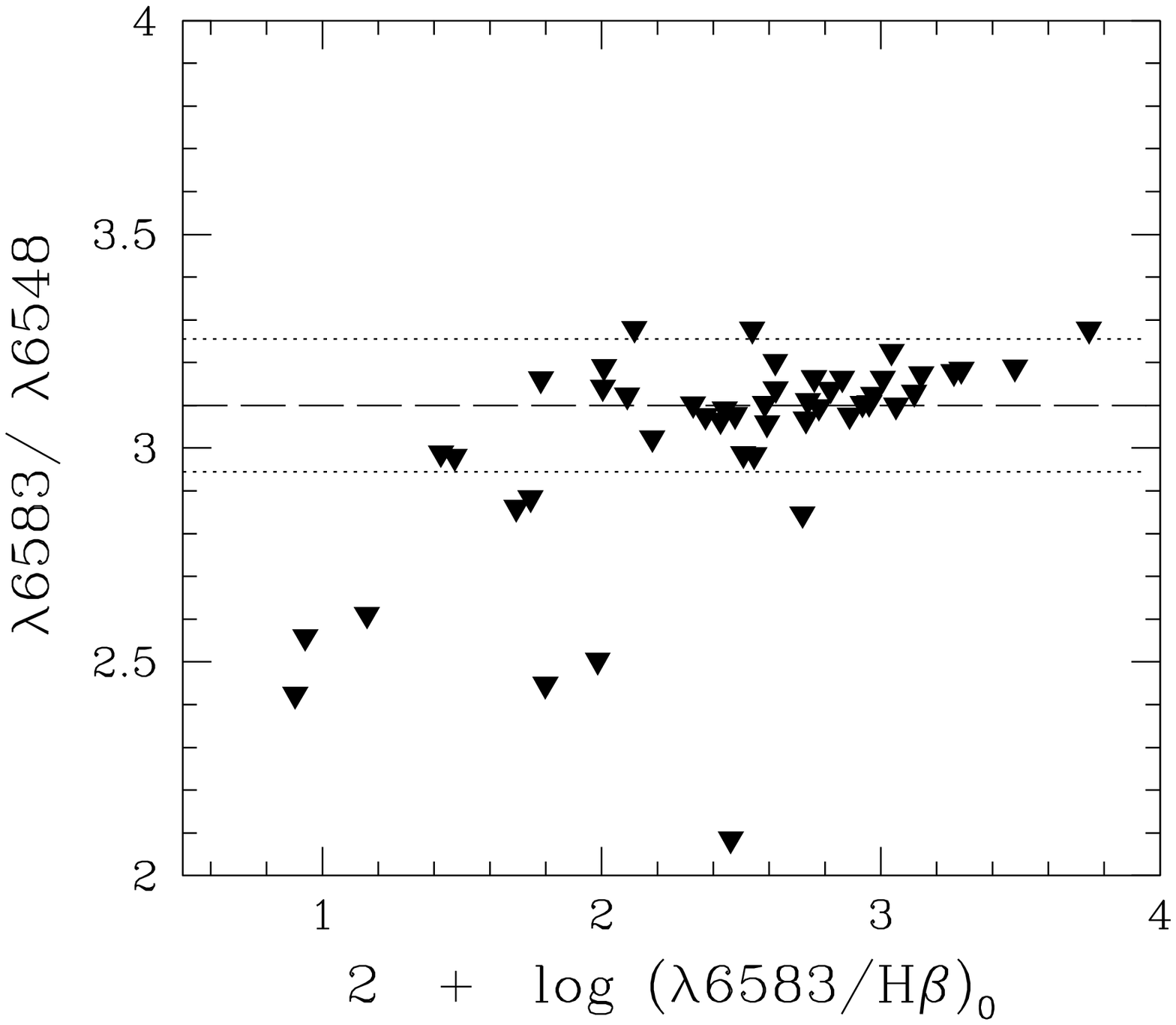}}
\caption[]{ 
  Intensity ratio of the [N~{\sc ii}] $\lambda$6583 to $\lambda$6548 lines
  as the function of the flux of [N~{\sc ii}] $\lambda$6548 (in units of
  H$\beta$). The long dashed line represents median value for the presented
  observations, and the dotted lines are 5\% deviations from it.
}
\label{n2}
\end{figure}

\subsection{Quality of the data}

Figure \ref{o3} presents intensity ratios of the [O\,III] doublet lines
\lam5007 and \lam4959 for the observed objects.  The mean value for this
sample is marked with a dashed line in \figref{o3}, and the 5\% deviations
from it are marked with dotted lines.  The \lam5007 to \lam4959 ratio should
be practically constant for all PNe, therefore, the observed deviations can
be regarded as the evidence of the quality of the registered spectra.  Since
the [O\,III] \lam5007 line is usually one of the brightest in the PN
spectrum, it often saturates during long exposures.  Therefore its measured
value comes usually from an additional short exposure.  It is of some
difficulty to precisely relate the measured value to the other much fainter
lines.  In the ratio shown in \figref{o3}, the larger source of error is
indeed measurements of the \lam4959 line.  This line is three times
fainter than the \lam5007, but both lines had to be measured from the
snapshot spectra.  From the deviations of the intensity ratio plotted in
\figref{o3}, the accuracy of the \lam4959 line can be estimated to be
typically better than 5\% for most of cases.

Similar exercises can be performed for the [N\,II] doublet of \lam6583 and
\lam6548 lines at wavelengths considerably further to the red.  Their
measured ratios are displayed in \figref{n2}.  Analogous conditions to the
described above apply to this doublet but an additional disturbing factor
can be the strong H$\alpha$ line separating them.  It can be seen in
\figref{n2} that the majority of the observed ratios are well confined
between the 5\% deviations.  There are, however, some outliers, in
particular, for cases when [N\,II] lines are fainter, and it can be noted
that the intensity of the \lam6548 line is apparently overestimated. 
Indeed, this is possible when a stronger blending with nearby prominent
H$\alpha$ line takes place.  However, this has to be a very minor effect on
our results, since [N\,II] \lam6548 line is not used for calculations of
nitrogen abundances.

An independent test of the quality of observations can be reached through a
comparison with other observers.  However, since selecting the targets, we
were trying to avoid objects already present in large spectroscopic surveys
of the Galactic disk PNe, the existing duplicates with other authors are not
as numerous and distributed through several different references.  We found
instead that our list of targets has 11 objects in common with
\cite{Dopita}, who measured absolute fluxes of selected lines for a large
sample of southern sky PNe.  In Appendix\,A, we present and discuss the
comparison of our measurements with those of \cite{Dopita}.

As a result of all above mentioned evaluations, we conclude that the basic
accuracy of all the line measurements from our spectra can be adopted as
roughly 5\%.  In cases marked with a colon in Table~B.1, the uncertainty is
estimated around 20\%, and in the rare cases of extremely weak lines or lines
contaminated with sky features or field stars (marked with semicolon), this
can be as high as 40\%.

\section{Search for PNe with emisssion-line stars}

\subsection{New and reclassified objects}

A large fraction of PNe exhibit emission-line central stars (CSs). As can be
inferred from the work of \cite{Weidmann2011a}, they compose about one third
of PNe with classified CSs.  Two main types of emission-line central stars
are widely recognised in the literature \citep{TAS93}: the Wolf-Rayet type
central stars ([WR]) and weak emission-line CSs (WELs).  The spectra of the
first group closely resemble those of the genuine massive population~I
Wolf-Rayet stars of the WC spectral subclass with strong and wide emissions
of C, O, and He.  The WELs seem to be a less homogeneous group with usually
only narrow and much fainter C\,IV doublet $\lambda$5801/12.  Recently, the
existence of a separate third group of PNe with emission-line central stars
has been proposed by \cite{Gorny2009}, who investigated a large sample of
PNe in the Galactic bulge and inner-disk region.  This new class of CSs was
named VL or {\it very late} type and was found in the low ionisation PNe
with a stellar spectra represented by C\,III and C\,II emission lines, but,
again narrow and less strong.  In the classical scheme, they could be
assigned to [WC\,11] subtype.  The sample of PNe analysed in the present
work gave us the possibility to check if VL type of objects are also present
among Galactic disk PNe.

We used our acquired data to search for new emission-line CSs. We checked
our reduced 2D and integrated 1D spectra for the presence of characteristic
emission-lines utilised for the spectral classification as listed in
\cite{Gorny2004} and the compound of C and N lines around 4650\AA,
which is usually associated with the WELs type objects \citep{TAS93}.

We have found three new WELs, two new [WR]-type, and one new VL PNe. The
spectra of the new WELs PNe are presented in \figref{spectra_wel} and of the
new [WR] and VL type objects in \figref{spectra_WC} in Appendix C.  In both
figures, the expected locations of the characteristic emission lines are
indicated with dotted lines, and if positively identified, they are marked
with the ion name of their origin.

In addition, we also registered stellar emission lines in almost all cases
when the previous emission-line classification already existed
\citep{Weidmann2011a}.  The only exception was He\,2-63, which is probably
due to the short exposure time of our observation and the NGC\,5979 that
turned out not to be a true but a mimic of a WEL object (see below).  We
hereby confirmed the [WR]-type nature of two CSs but reclassified three
other objects to be a VL-type PNe.  In \figref{spectra_reob}, we present
spectra of the reobserved and confirmed PNe with previously known
emission-line CSs.

In Table~B.2 in Appendix~B, we present a complete list of new and previously
known emission-line CSs that are present among PNe from our observing list. 
In the first column we give their usual names.  In column 2, the full width
at half maximum of the main stellar emission line is presented but only if
it was exceeding the instrumental width.  In practise, these measurements in
Table~B.2 are always referring to C\,IV $\lambda$5805, since in the cases
that the C\,III $\lambda$5695 line was present, it was of the same width as
the nebular lines.  In column 3, we give our classifications of the central
star.  We tried to use the standard classification framework, as in
\cite{Gorny2004} and \cite{Gorny2009}, which are based on ratios of
intensities of the main emission features of the C and O ions.  In practise,
however, only single lines of C\,IV or C\,III were present in analysed
spectra.  In column 5, previous classification from the compilation of
\cite{Weidmann2011a} is given.

We were able to positively or tentatively recognise the presence of the
stellar continuum of the central star in about 83\% of our spectra. 
Although this is not a necessary condition for the prominent features of
[WR]-type objects to be discovered nevertheless, in particular for the WELs,
the stellar emission-lines should be detectable if the stellar continuum is
observed.  We, therefore, assume that the list of PNe with emission-line CSs
presented in Table~B.2 is probably complete in the same proportion.

\subsection{Notes on individual PNe}

Objects PB\,5, My\,60, and He\,2-115, whose spectra are shown in
\figref{spectra_wel} present good examples of WELs PNe, as defined by
\cite{TAS93}.  The C\,IV lines at $\lambda$5801/12 are visible as separate
doublet components and are much fainter and narrower than in the case of
[WR] PNe.  Other lines often associated with WELs, such as
$\lambda$7235 and some elements of the $\lambda$4650 aggregate, are also
present.  The same situation can be observed in the case of He 2-123, whose
spectra is shown in \figref{spectra_reob} and was already classified as
WEL by \cite{Lee2007}

He\,2-67 and He\,2-117 are two PNe discovered in our spectra to
posses classical broad emission structures of C\,IV at \lam5805.  No
other emission lines of C\,III or oxygen ions used in the classical
classification scheme can be seen.  For this reason, we use the FWHM of the
C\,IV line as an additional classification parameter and following
suggestions of \cite{AN2003}, we tentatively classify He\,2-67 as a
[WC\,4] and He\,2-117 as a [WC\,5-6] objects.  In the latter case, we reject
the possibility that the star is of an even earlier subclass, since no
trace of C\,III \lam5695 line is seen in our spectra.

The object Pe\,1-1 has been classified by \cite{AN2003} in their modified
scheme as [WO\,4].  Their spectrum was apparently of better quality than
ours since it was acquired with the 3.6m telescope, and they were able to
measure both C\,IV \lam5805 feature and the O\,V \lam5590 line.  The
intensity of the latter is at about 1/10 of \lam5805, according to
\cite{AN2003} but could not be identified in our spectra.  Using their
measurements, we would classify that object in the pure classical scheme of
[WC] classes as [WC\,4-5].  It can be added that the measured FWHM of C\,IV
line would also place it as a rather late [WO\,4] class in the \cite{AN2003}
scheme.  The nebula of Pe\,1-1 was recently analysed with an actually deeper
high resolution spectra by \cite{GarciaRojas2012}.

The object He\,2-86 has been classified by \cite{AN2003} as [WC\,4]. They
registered broad stellar C\,IV 5805 and He\,II 4686 features and C\,III at
\lam4649.  While the former two are clearly visible in our spectra (see in
\figref{spectra_reob}), the latter seems to be a compound of much narrower,
and, therefore, probably nebular recombination lines.  Taking the width of
the C\,IV line of 22\AA\ \citep[][derived a similar value]{AN2003} into
account, this object should be classified as [WC5-6].

The spectrum of He 2-125 presented in \figref{spectra_WC} clearly shows two
stellar emission lines of carbon: C\,III at \lam5695 and C\,II at \lam7235. 
The intensity ratio of the two lines is about 1.6, which is half of what is
predicted for [WC\,11] objects, according to the classification criteria. 
However, it fits well the definition of VL type PNe, since the central star
is surrounded by a very low ionisation nebula, with only a trace of [O\,III]
\lam5007 line detectable in the spectra.  This can be seen in
\figref{spectra_all}, presenting it in its whole wavelength range.

\cite{Weidmann2011b} who use their own spectra, have classified He\,2-107 as
[WC\,10-11] object.  In our spectra, the ratio of C\,III at \lam5695 to C\,II
at \lam7235 is 0.31, but it cannot be precisely derived by how much the
nebular emission contributes to the latter line.  This object is also of
relatively low ionisation, and although the [O\,III] lines are present, the
intensity of \lam5007 is at only 76\% of H$\beta$.  We, therefore, rate this
object to the VL-type PNe with emission-line CSs.

The objects He\,2-131 and He \,2-108 are well studied emission-line objects
with CSs models available in literature
\citep{Hultzsch2007,Pauldrach2004,Kudritzki1997}.  The object He\,2-108 was
recently analysed by \cite{Pottasch2011} with apparently lacking good
optical spectra.  We feel both PNe are worth to be included in the sample
for at least illustrative purposes how they would be processed and classified
with the kind of spectra we collected.

He\,2-131 and He \,2-108 were first classified as WELs objects in
\cite{TAS93}.  However, both objects display not narrow C\,IV \lam5801/12
emissions but a C\,III \lam5695, and C\,II \lam7235, and He\,II \lam4685
line that is also identified as stellar in our low resolution spectra.  The
additional interesting property of He\,2-131 is that there are traces of
C\,IV \lam5801/12 in absorption.  The central star of He\,2-131 is
surrounded by a low ionisation nebula with [O\,III] \lam5007 at the level of
only 10\% of H$\beta$, whereas the star is presumably hotter and, therefore,
the intensity of [O\,III] 5007 is almost two times larger than H$\beta$ in
the case of He\,2-108.  We would classify them both, however, as VL type.

\begin{figure}
\resizebox{0.5\hsize}{!}{\includegraphics{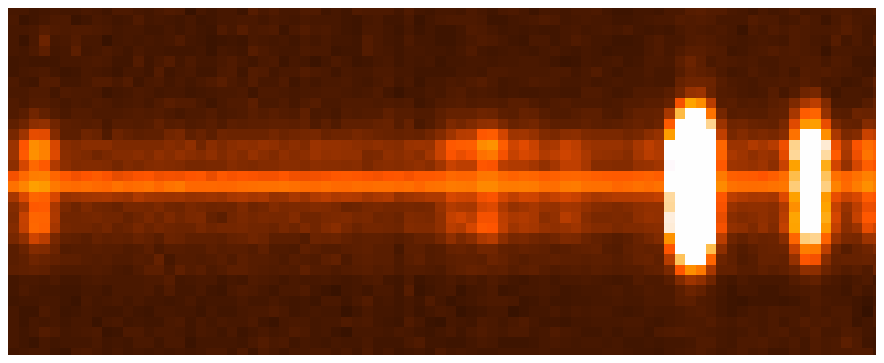}} 
~~~~~
\resizebox{0.40\hsize}{!}{\includegraphics{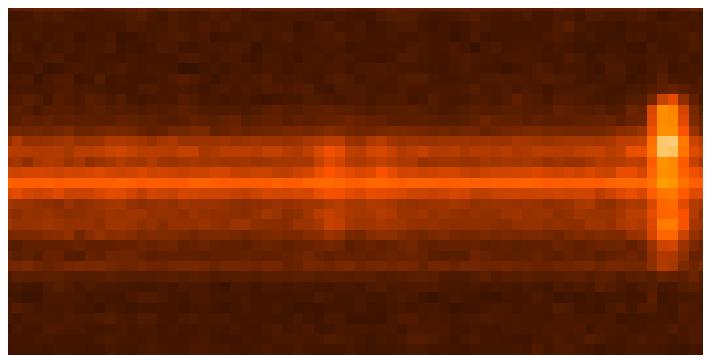}}
\caption[]{
 Spectrum of NGC 5979 around 4640\AA\ (left) and 5800\AA\ (right).
}
\label{spectra_NGC5979}
\end{figure}

\begin{figure}
\resizebox{0.5\hsize}{!}{\includegraphics{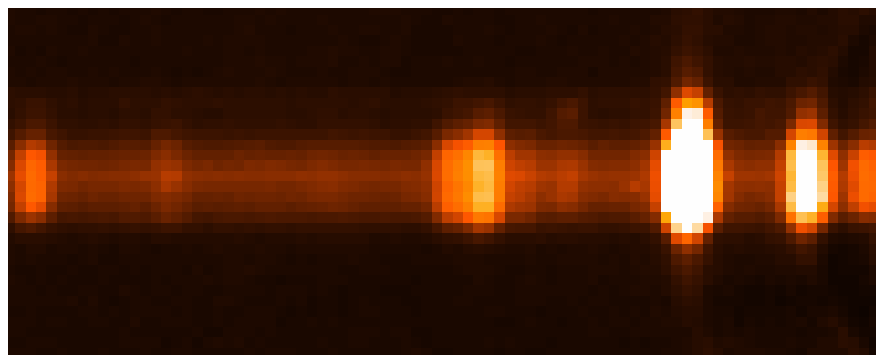}} 
~~~~~
\resizebox{0.40\hsize}{!}{\includegraphics{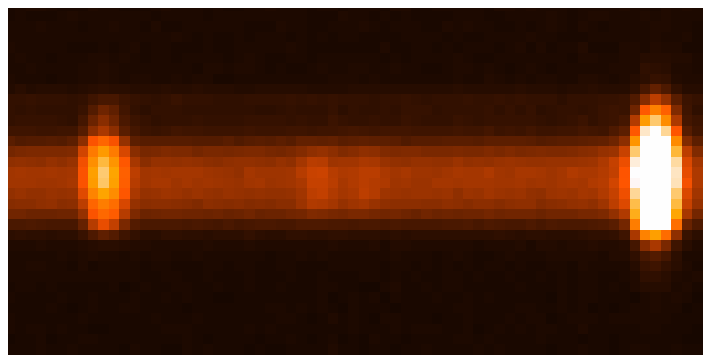}}
\caption[]{
 Spectrum of NGC 3918 around 4640\AA\ (left) and 5800\AA\ (right).
}
\label{spectra_NGC3918}
\end{figure}

\subsection{Mimics of emission-line stars}

The WELs type PNe are probably a heterogeneous group of objects that has
already been suggested by \cite{TAS93}.  Recently, it has been proposed by
\cite{Corradi2011} that the stellar emission lines of C and N attributed in
WELs to the enhanced mass loss may originate from the irradiated zone in
close binary systems.  \cite{Miszalski2011} argue this is actually the case
of NGC\,6326, which they observed both photometrically and spectroscopically. 
In the bottom panel of \figref{spectra_reob}, we present our low resolution
spectra of NGC\,6326.  Faint but clearly visible emission lines of C\,IV at
\lam5801/12 would normally be identified as stellar and clearly would allow
us to classify this object as a typical WEL.  The C\,II \lam7325 and, in
particular, the complex of C and N lines near 4650\AA\ are clearly of nebular
origin, since they appear as emitted from a spatially extended region in our
2D spectra.  Could the C\,IV \lam5801/12 lines also have the nebular
origin in some PNe?

The central star of NGC\,5979 has been classified by \cite{Weidmann2011a}
using the low resolution spectra they acquired as a WEL object.  In
\figref{spectra_NGC5979}, we present our fully reduced 2D spectrum of this
PN.  In the right panel, we show the part of the spectrum around 5805\AA,
and in the left panel, the region centred on 4640\AA.  The lines of N\,III,
C\,III, and C\,IV ions are clearly detectable at the expected wavelengths. 
They all are, however, emitted from the spatially extended region and are,
therefore, not of the stellar but nebular origin.  The stellar continuum
emission can also be recognised in the centre of the images.  It is
broadened in the spatial direction by the seeing but obviously much narrower
than the emission lines.  If analysed using only the final summed 1D
spectrum, this object would perfectly mimic WEL type spectra.

In \figref{spectra_NGC3918}, we show the same regions of spectrum for the
NGC\,3918.  In the case of this object, the stellar continuum is detectable
only in the blue part.  The emission lines of C\,IV at 5801/12 and the
combination of lines around 4650\AA\ are again clearly not of stellar
origin.  In the case of the presumed C\,IV line identified at 4658\AA, it
can be noticed that the emission has two separate spatial components.

The objects NGC3918 and NGC5979 are high ionisation PNe.  From our line
measurements, we derived the ratio of selective to total He$^{++}$/He
abundance to be 0.38 and 0.81, respectively.  In \figref{spectra_all}, we
presented the full spectrum of NGC\,5979, where lines of highly ionised ions
can be identified.  The effective temperature of NGC\,5979 calculated with
Zanstra He method is T$_{\star}$=100k\,K, and the luminosity is
log\,L$_{\star}$=3.68 L$_{\odot}$.  In the case of NGC\,3918, those
parameters can be derived as 148k\,K and 3.52\,L$_{\odot}$.  Apparently, in
both these PNe some carbon is ionised up to C$^{4+}$ in the large
part of the nebula and can give rise to lines that are otherwise expected to
be observed from stellar atmospheres.

\begin{figure}
\resizebox{\hsize}{!}{\includegraphics{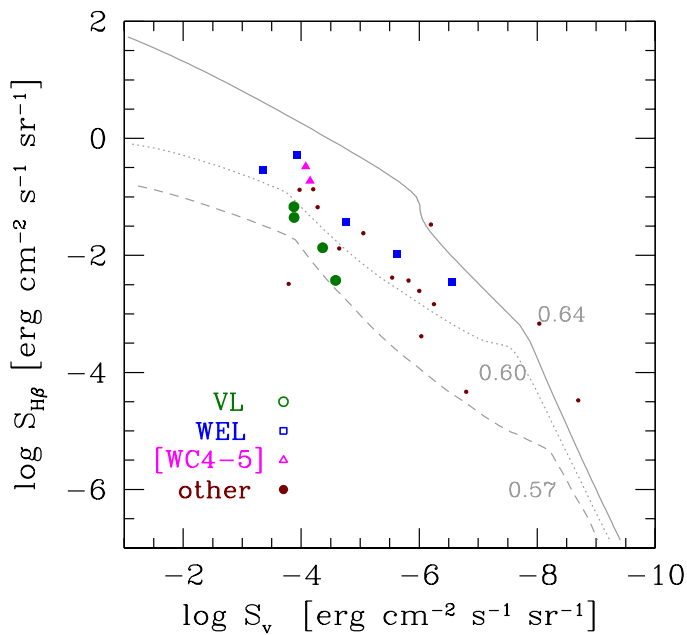}} 
\caption[]{
  Nebular surface brightness S$_{H\beta}$ versus parameter S$_V$ as defined
  in an analogous way using stellar V flux.  Data presented for different
  types of disk PNe from the observed sample: magenta triangles --
  [WR]\,PNe, blue squares -- WEL\,PNe, and green circles -- VL\,PNe
  ([WC\,11]-like spectra).  Small symbols mark normal PNe without
  emission-line CSs detected.  The lines present model calculations for
  central stars of 0.57, 0.60, and 0.64M$_{\sun}$, adopting a simple nebular
  model (see text).
}
\label{ss}
\end{figure}

\subsection{Occurrence rate and evolutionary status}

It was noticed by \cite{Gorny2001} that there is an important difference in
the distribution of spectral classes among [WR]\,PNe located in the bulge as
compared to the PNe in the Galactic disk.  An apparent underpopulation of
[WR]\,CSs in the [WC5]-[WC7] range for disk objects was observed.  In the
bulge, on the contrary, [WR]\,CSs were found to be mostly of the
intermediate spectral types suggesting that the Wolf-Rayet phenomenon
depends on the characteristics of the investigated stellar population giving
rise to the observed PNe.  The new observations of bulge PNe by
\cite{Gorny2004} and \cite{Gorny2009} did not solve the problem since the
comparable search surveys for the Galactic disk were not available. 
Instead, a potential new class of emission-line objects (VL) had to be
introduce \citep{Gorny2009}.

Recently,  \cite{Depew2011} published a list of newly discovered [WR]\,PNe
that contains some possible intermediate [WC] class objects in the Galactic
disk\footnote{Originally, \cite{Depew2011} used different classification
schemes of [WC] and [WO] subclasses, whereas we continue to use the
classical [WC] scheme that range from [WC11] to [WC2]}.  Additionally, we found
or reclassified three intermediate [WC] spectral-class objects in our disk
sample (see Table~B.2).  The first important result from the present work is
therefore that the difference in spectral class distribution, although still
present, may be not that pronounced.  Possibly more such objects could be
found in a deeper dedicated search.  The second piece of information is that
the VL objects are also present in the Galactic disk and not confined only
to the inner parts of the Milky Way and its bulge in particular.

In \figref{ss}, we present the locations of PNe from the observed sample
in the S$_{H\beta}$ versus S$_V$ plot, where S$_{H\beta}$ is the nebular
surface brightness in H$\beta$ and S$_V$ is defined in an analogous way but
using the stellar flux in the visual V band \citep[see in][]{Gorny1997}. 
The evolutionary tracks shown in \figref{ss} have been calculated with a
simple model of PNe by assuming total nebular mass M$_{neb}$=0.20M$_{\sun}$,
a filling factor $\epsilon$=0.75 and an expansion velocity V$_{exp}$=20km/s
for central stars of 0.57, 0.60, and 0.64M$_{\sun}$ evolving in agreement
with \cite{Bloecker1995} models.

It can be seen in \figref{ss} that the locations of VL\,PNe clearly differ
from those of [WR] and WEL objects.  In particular, VL PNe cannot be
evolutionary predecessors of either of the two other groups, as first noted
by \cite{Gorny2009}.  Because S$_{H\beta}$ and S$_V$ are distance
independent, one can directly compare locations of PNe analysed in this work
with Fig.11 of \cite{Gorny2009}, which present bulge and inner-disk
emission-line CSs that they discovered.  Analogous locations are occupied in
the case of any of the three groups in both plots.  We therefore can confirm
the conclusion of \cite{Gorny2009} that [WR], VL, and WELs form three
independent types of emission-line CSs phenomenon and that it is also true
for the PNe in the Galactic disk.

The rate of occurrence of the different types of PNe with emission-line CSs
is, of course, of a large importance if one wants to discuss the possible
evolutionary paths of their creation.  At the same time, it is very difficult
to be established, as the number of objects that are discovered in a
given sample is unavoidably subject to serious bias.  The bias
originates from the observational details, selection effects when choosing
targets, and obvious physical constraining factors, such as the distance to
the given population.

A natural sample to compare with the present one would be the 44 Galactic
bulge PNe observed by \cite{Gorny2004}.  Both groups have been actually
observed with the same instrument and almost identical setup, but there are
also differences.  The Galactic bulge PNe are further away and smaller,
which should work against efficient discovery of emission-line CSs in the
bulge sample due to the stronger contamination effect from the nebulae.  On
the other hand, the \cite{Gorny2004} sample was preselected by choosing
objects with infrared IRAS colours typical for [WR] PNe \citep{Gorny2001}. 
This could have an important consequence since no such preselection was
active in the present sample.  Nevertheless, taking the numbers of objects
into account discovered from original new observations in both studies
\citep[i.e., comparing only with sample G of][see column 3 in their
Table~4]{Gorny2004} one findss that seven WELs and two [WR] were discovered
in the bulge by \cite{Gorny2004}, as compared to three WEL, two [WR] and one
new VL in the present sample.  Clearly, no final conclusions about the rate
of occurrence of emission-line phenomenon in both environments can be
reached in this way.

The precise answer to the question of what exactly is the rate of occurrence
of each type of emission-line phenomenon in different environments requires
constricting a more complete samples in both regions of the Milky Way. 
Possible selection effects need to be carefully considered but is outside
the scope of the present data paper.

\section{Chemical abundances and plasma parameters}

We start this section with the description of the method we used. Then we
analyse the calculated plasma parameters and present the derived chemical
abundances.  We finalise the result by comparing them to the abundances of
the Galactic bulge and inner-disk PNe populations that were analysed
recently.

\subsection{Method}

We use the code ABELION that was developed by G.\,Stasi\'nska, which is
based on the classical empirical method to derive the plasma parameters and
nebular chemical abundances.  The version of the code is identical with that
used by \cite{Gorny2009} and \cite{Chiappini2009}.  We first derive the
electron densities from the \rSii\ ratio and electron temperatures
\Te(O~{\sc iii}) from the \rOiii\ and/or \Te(N~{\sc ii}) from \rNii\ ratios. 
We used them to refine the inferred reddening corrections, as described
above.  Later, we use \Te(N~{\sc ii}) to derive abundances of ions with low
ionisation potential and \Te(O~{\sc iii}) for hydrogen and lines from other
ions with intermediate and high ionisation potentials.  If \Te(N~{\sc ii})
was very uncertain (three cases) we used the \Te(O~{\sc iii}) for all the
ions instead.  If the observational data did not allow for the stimation of
the electron temperature, the object was rejected from further
considerations (five cases).

As discussed in detail in \cite{Gorny2009}, the O$^+$ ionic abundances
derived from $\lambda$3727 line as compared to $\lambda$7325 differ
frequently.  The latter can be affected by recombination from \Opp\ ions,
but this does not solve the problem, as checked by \cite{Gorny2009}.  In
\figref{opop}, we compare the \Op\ ionic abundances derived from
$\lambda$7325 to \Op\ from $\lambda$3727 by showing their ratio as a
function of electron density.  It can be noted that the median difference
for the presented observations is 0.17\,dex, and we use this value to
correct $\lambda$7325 measurements, as proposed by \cite{Gorny2009}.  In
\figref{opop2}, we plot the same ratios as a function of electron
temperature \Te(N~{\sc ii}).  A weak correlation can be seen in the sense
that the discrepancy between $\lambda$7325 and $\lambda$3727 also becomes
larger for PNe with larger \Te(N~{\sc ii}).  We, however, attempted no
fitting to this correlation when correcting the $\lambda$7325 line.  The
final \Op\ ionic abundance is a mean of the two values weighted by their
reversed uncertainties.

After the ionic abundances are computed, the elemental abundances are
obtained using the ionisation correction factors (ICFs), as in
\cite{Gorny2009} and based mainly on the scheme of ICFs from \cite{KB94}. 
The uncertainties in abundance ratios and other derived parameters were
obtained by propagating uncertainties in the observed emission line
intensities using Monte-Carlo simulations by assuming that the principle
line intensity errors are of at least 5\%, as explained above.

\begin{figure}[t]
\resizebox{\hsize}{!}{\includegraphics{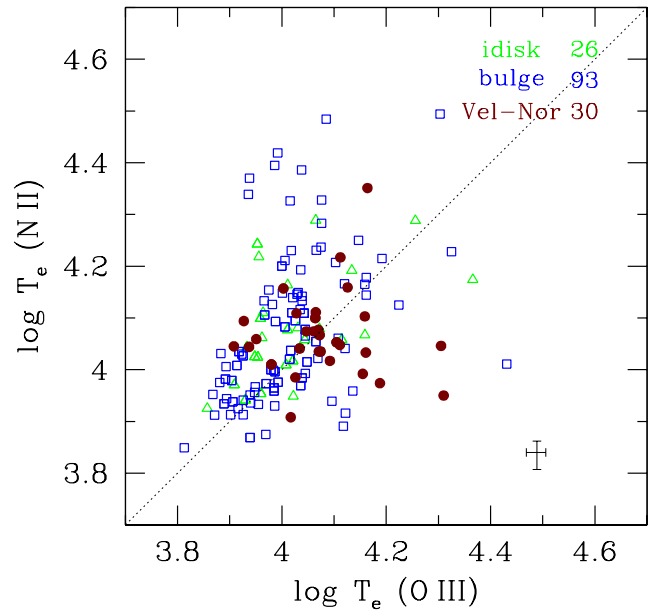}} 
\caption[]{
 The electron temperature derived from \rOiii\ versus electron temperature
 from \rNii. Dark dots mark the Galactic disk PNe observed in this work
 (Vela-Norma); blue open squares - Galactic bulge PNe from
 \cite{Chiappini2009}; green open triangles - Galactic inner-disk PNe from
 the same reference. The numbers in the top right give the numbers of 
 objects presented in the plot for each sample.
}
\label{to_tn}
\end{figure}

In Table~B.3 in Appendix B, we give the derived plasma diagnostics and ionic
and elemental abundances for 48 objects ordered by PN\,G numbers.  For each
object, there are three rows of data.  The first row gives the values of
parameters computed from the nominal values of the observed lines and their
ratios.  The second and third row give the upper and lower limits of these
parameters, respectively.  Column (1) of Table~B.3 gives the PN\,G number and
Column (2) gives the usual name of the object; Column (3) gives the electron
density deduced from \rSii, Columns (4) and (5) give the electron
temperature deduced from \rOiii\ and \rNii, respectively.  If the value of
\Te(N~{\sc ii}) is in parenthesis, \Te(O ~{\sc iii}) was chosen for all ions. 
Column (6) gives the He/H ratio and Columns (7) to (12) the N/H, O/H, Ne/H,
S/H, Ar/H, Cl/H ratios, respectively.  To avoid dealing with values that are
too uncertain, we removed from further consideration any plasma parameter or
abundance ratio for which the error (at two-sigma level) is larger than 0.3
dex.  We marked these values with a colon in Table~B.3.

\subsection{Plasma parameters}

In this section, we present the plasma parameters derived for our observed
disk PNe from Vela-Norma direction and compare it to the Galactic bulge and
inner-disk samples of \cite{Chiappini2009} that were computed with exactly
the same code and assumptions.  The inner-disk sample is less well defined
and includes PNe that are observed in the direction of the Galactic centre
but do not pertain to the bulge, according to the standard criteria.  They
can be located at Galactocentric distances estimated\footnote{These
estimates are based on distances from \cite{Cahn1992}.} as less than 2\,kpc
up to almost 7\,kpc but the median Galactocentric distance is about 3\,kpc.
   In \figref{to_tn}, we plot \Te(O~{\sc iii}) versus \Te(N~{\sc ii}). 
First, it can be noted that all disk PNe that we show here seem to have \Te\
larger than 10$^4$\,K.  Among the bulge and inner-disk samples, there is a
considerable number of objects with temperatures below that value.  From
\figref{to_tn}, the correlation between \Te(O~{\sc iii}) and \Te(N~{\sc ii})
seems to be at the same level for disk PNe as for the other two groups but
there are two PNe in this sample that have \Te(O~{\sc iii})\,$\gg$\Te(N~{\sc
ii}).  In the bulge PNe sample, on the contrary, there is a group with
\Te(N~{\sc ii})\,$\gg$\Te(O~{\sc iii}).  Calculating a median value of
log\,\Te(O~{\sc iii}) for the disk PNe (39 objects with useful data) we
checked that it is 4.09 and 0.06\,dex larger than for the bulge PNe.  The
tests indicate the difference is statistically significant\footnote{We used
Kolmogorov-Smirnov and Wilcoxson tests for all the distributions analysed in
this work.} at more than 99\% confidence level.

Analysing the distributions of the electron density derived from \rSii\
ratio for the three samples of PNe, we found no statistically significant
differences, although our disk sample from Vela-Norma direction seems to
have more members with smaller densities.  The median log\,N$_e$=3.26 for
the objects presented here are 0.07 and 0.20 dex smaller than for the
Galactic bulge and inner-disk PNe, respectively \citep[][]{Chiappini2009}. 
This qualitatively agrees with the ionisation level of these nebulae that we
checked measuring the relative ionic He$^{++}$ to total He abundance.  The
median He$^{++}$/He for the Galactic disk PNe investigated here is 0.11.  A
similar value have been derived for the Galactic inner-disk sample, whereas
very few observed PNe have any He$^{++}$ ions at all for the bulge PNe.

Summarising, we would tentatively describe the disk PNe from the sample
analysed here as more evolved, of higher electron temperature, and also of
presumably higher ionisation level than the PNe in the other two samples. 
This is what could be expected, since it is generally more difficult to
observe evolved PNe at larger distances like the bulge or inner-disk
regions.

\begin{figure}
\resizebox{\hsize}{!}{\includegraphics{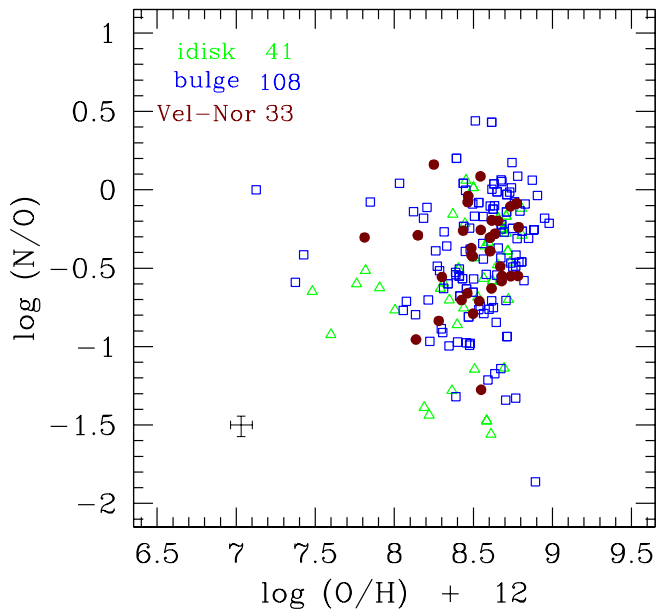}} 
\caption[]{
 The nebular abundance ratios N/O versus O/H for the three analysed Galactic
 PNe samples. The notation of symbols as in \figref{to_tn}.
}
\label{oh_no}
\end{figure}

\begin{figure}
\resizebox{\hsize}{!}{\includegraphics{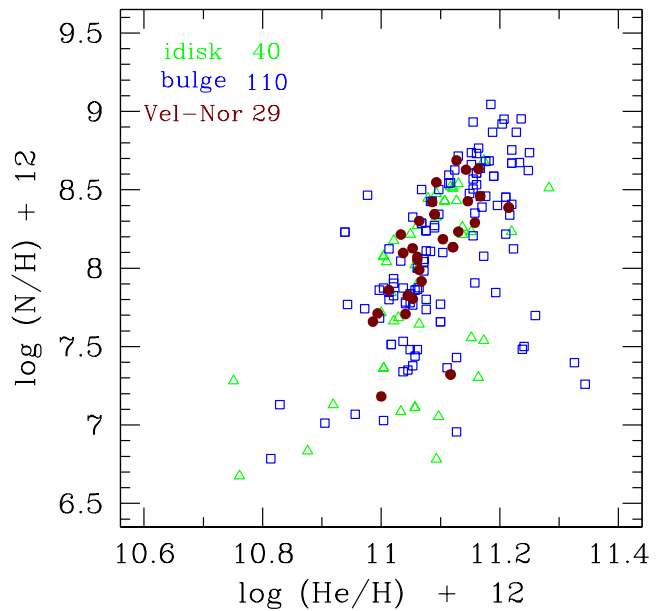}} 
\caption[]{
 The nebular abundance ratios He/H versus N/H for the three analysed Galactic
 PNe samples. The notation of symbols as in \figref{to_tn}.
}
\label{heh_nh}
\end{figure}

\subsection{Chemical abundances}

One of the advantages of studding chemical abundances of PNe is that some of
them bring information on the nucleosynthesis processes occurring in the
interiors of their progenitor stars (e.g.  nitrogen), whereas others are
assumed to remain unchanged during the life of the star and inform about the
primordial chemical composition of the matter that the parent star was born
from.

In \figref{oh_no} we show the relation of O/H versus N/O for the PNe
included in the sample investigated here as compared to Galactic bulge and
inner-disk samples of \cite{Chiappini2009}.  Similar behaviour can be seen
in this plot for the three groups of PNe.  There are also very few outliers. 
The object with the lowest O/H abundance in the disk PNe group in this plot
is He\,2-131 that we discussed already in Sect.~3.2\footnote{ 
  The low O abundance we calculate is the result of the high \Te(O~{\sc
  iii})$\approx$20200\,K that we derive from a very faint [O\,III] \lam4363 line,
  as compared to \Te(N~{\sc ii})$\approx$11100\,K.  If the true \Te\ is
  close to the latter value and the preference would be given to O$^+$ from
  \lam7325, then O/H=1.86\,10$^{-4}$ would be derived.}.
We derive median log\,O/H=-3.46 and median log\,N/O=-0.37 for the disk PNe
sample analysed here. 

In \figref{heh_nh}, we plot He/H versus N/H for the analysed PNe of the
three Galactic locations.  The median value log\,He/H=11.01 for our disk PNe
sample seems very similar to the other two samples \citep[see Table\,2
of][]{Chiappini2009}.  Nevertheless, analysing them with the statistical
tests, there is only 1--2\% chance that our disk sample and bulge sample of
\cite{Chiappini2009} originate in the same parent distribution.  This is
probably due to the group of objects with log He/H$>$11.2 that can be noted
in the bulge PNe sample.  Surprisingly, according to the standard view on
the chemical evolution of the AGB stars, this could mean there is a group of
PNe with higher mass central stars that is observed in the bulge but missing
in our disk sample.  This is, however, not confirmed by statistical test for
neither N/H nor N/O distributions, although a very convincing correlation of
He/H with N/H can be seen in \figref{heh_nh}.

\begin{figure}
\resizebox{\hsize}{!}{\includegraphics{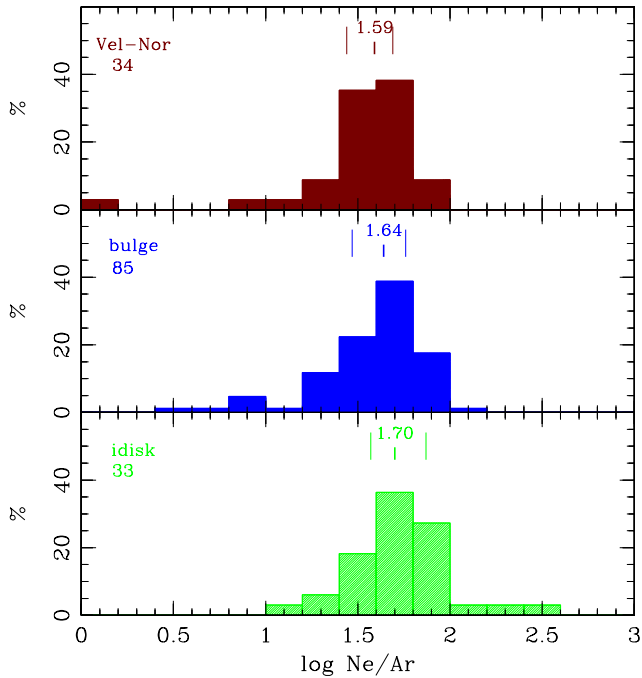}} 
\caption[]{
 Distributions of the nebular Ne/Ar abundance ratio for the three analysed
 Galactic samples: disk PNe from Vela-Norma direction (top), bulge PNe
 (middle), and inner-disk PNe (bottom).  The median values, the 25 and 75
 percentiles are marked with three short vertical lines above each
 histogram.  Numbers of objects used are shown in the left-hand parts of the
 panels below sample names.
}
\label{near}
\end{figure}

Figure \ref{near} presents the comparison of Ne/Ar abundance ratios for the
three samples of PNe.  In this case, a gradual increase of Ne/Ar can be
observed from disk sample with the smallest ratio through bulge PNe and
towards the inner-disk sample.  The difference between our disk sample and
the inner-disk PNe is confirmed at about 99\% significance level. We
noticed no statistically significant differences in distributions of
any other abundance ratios comparing the three PNe samples.

\begin{figure}
\resizebox{\hsize}{!}{\includegraphics{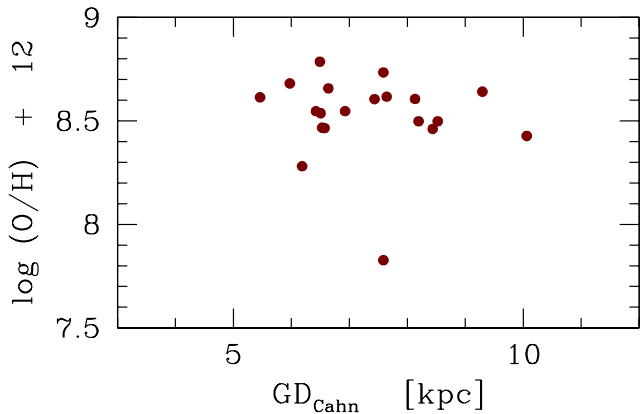}} 
\caption[]{
 The nebular abundance ratio O/H versus Galactocentric distance 
 for PNe observed in this work. 
}
\label{dc}
\end{figure}

Although the group of PNe we observed is not numerous enough and, as we
mentioned above, a large and properly selected samples should be used to
study gradients of chemical abundances in the Milky Way, we present the
derived O/H abundances versus the Galactocentric distances of these objects
in the \figref{dc}.  This is mostly done to demonstrate the usability of the
data we collected for this and other purposes.  We used the distances of
\cite{Cahn1992} and found them for 19 of our disk PNe.  As in previous
cases, we analysed only objects with an O/H ratio errors estimated to
be smaller than 0.3\,dex.  As seen in \figref{dc}, the observed PNe are
located from about 5 to 10 kpc from the Galactic center.  Even though the
presented sample of disk PNe is small, their locations in \figref{dc}
clearly support the idea of flattening the chemical gradient in the internal
parts of the Milky Way.

\section{Summary}

We presented low resolution optical spectroscopic observations for a
sample of 53 PNe located in the southern sky between Vela and Norma
constellations and pertaining to the Galactic disk.  We used the spectra
to analyse the chemical composition of PNe and search for new emission-line
central stars.  Our main results are as follows:

\begin{enumerate}
      \item 
We derived chemical abundances of 48 observed PNe, for the first time for
most of them or allowed abundances of additional, 
previously unobserved elements to be measured.
      \item We compared the nebular chemical abundances in our
sample with the results of \cite{Chiappini2009} for Galactic bulge and
inner-disk PNe populations.  We found no statistically meaningful
differences in abundance distributions, except for He/H, that seems more
abundant in the bulge PNe as compared to our disk
sample and Ne/Ar with larger ratios found in the inner-disk PNe population.
      \item 
The oxygen abundances derived for the
analysed sample favour flattening of the O/H gradient within the inner parts of
the Galactic disk.
      \item
We performed extensive search for central stars with emission-lines and 
found three new WEL, two [WR] and
one VL type object. Re-analysing the spectra of previously classified objects
additional three members of the VL group have been proposed.
      \item 
We identify NGC\,3918 and NGC\,5979 as
the possible mimics of WEL objects with C\,IV 5801/12 emitted from highly
ionised nebulae.
      \item
We confirmed for the first time that the VL type PNe are also located
outside the inner parts of the Milky Way. We identified three examples of 
[WR]-type central stars with intermediate spectral class [WC5]-[WC6]
pertaining to the Galactic disk.
   \item
We argue WELs, [WR], and VL objects are three evolutionary unrelated types of
emission-line phenomenon in central stars of PNe.
   \end{enumerate}

\begin{acknowledgements}
      Part of this work was supported from grant N203\,511838 of the Science
      and High Education Ministry of   
      Poland. I wish to thank M.\,Hajduk who assisted during observations in 
      June 2005 and A.D.\,Garc\'{\i}a-Hern\'andez for his help in selecting
      targets in 2011.
\end{acknowledgements}

\bibliographystyle{aa}
\bibliography{skg_paper}

\clearpage

\begin{appendix}

\section{Comparison with emission-line flux standards}

\begin{figure*} 
\resizebox{0.33\hsize}{!}{\includegraphics{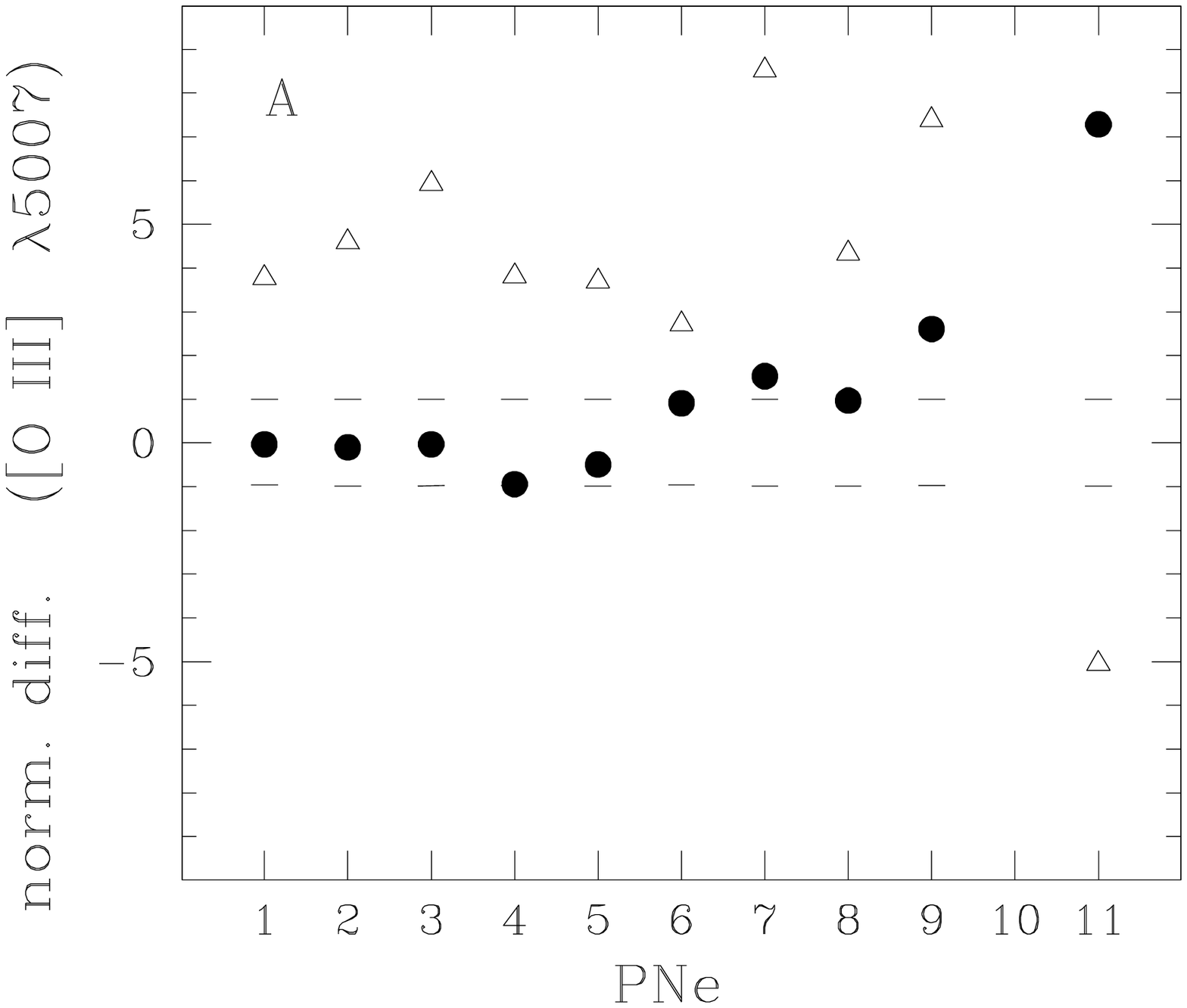}}
\resizebox{0.33\hsize}{!}{\includegraphics{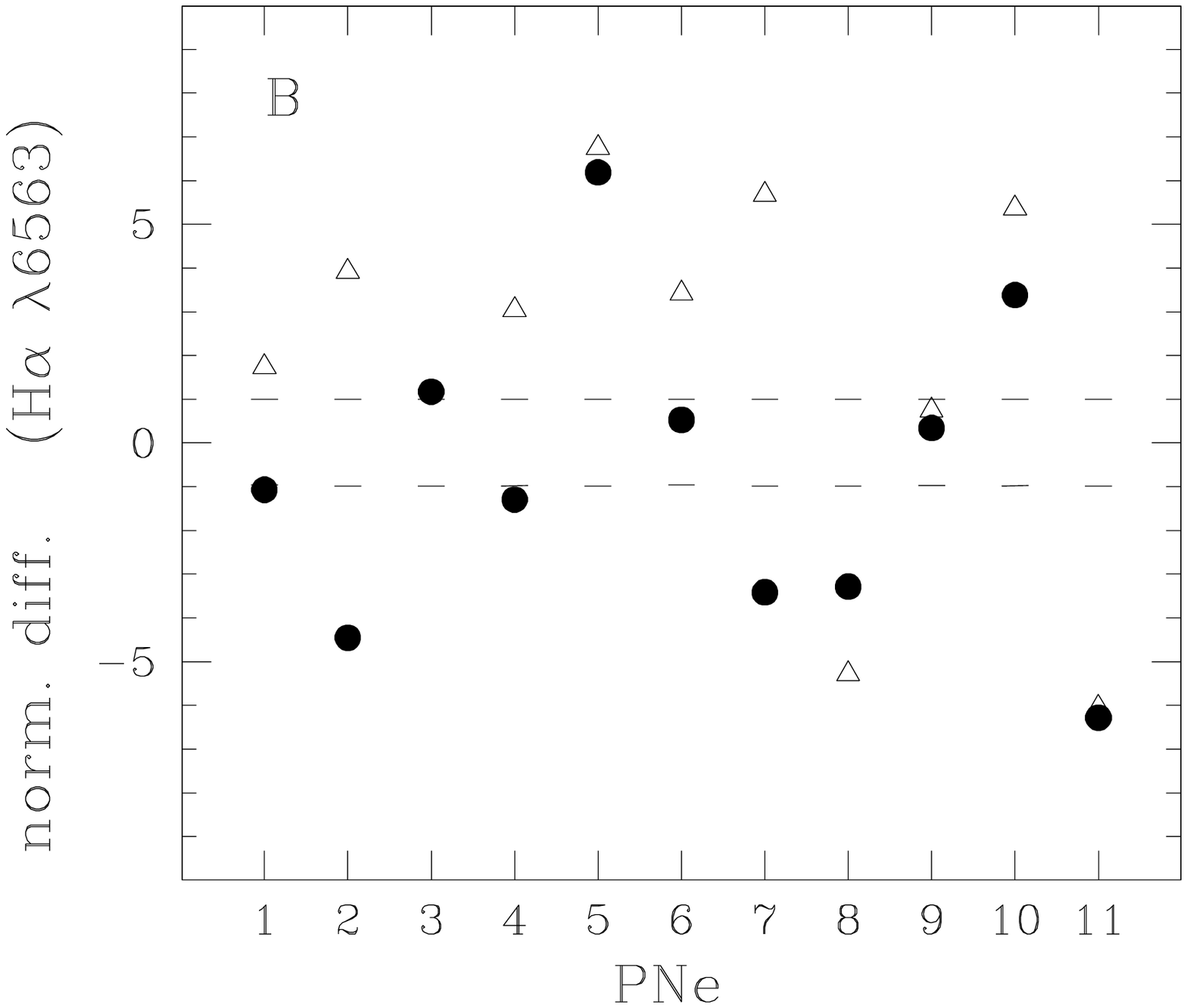}}
\resizebox{0.33\hsize}{!}{\includegraphics{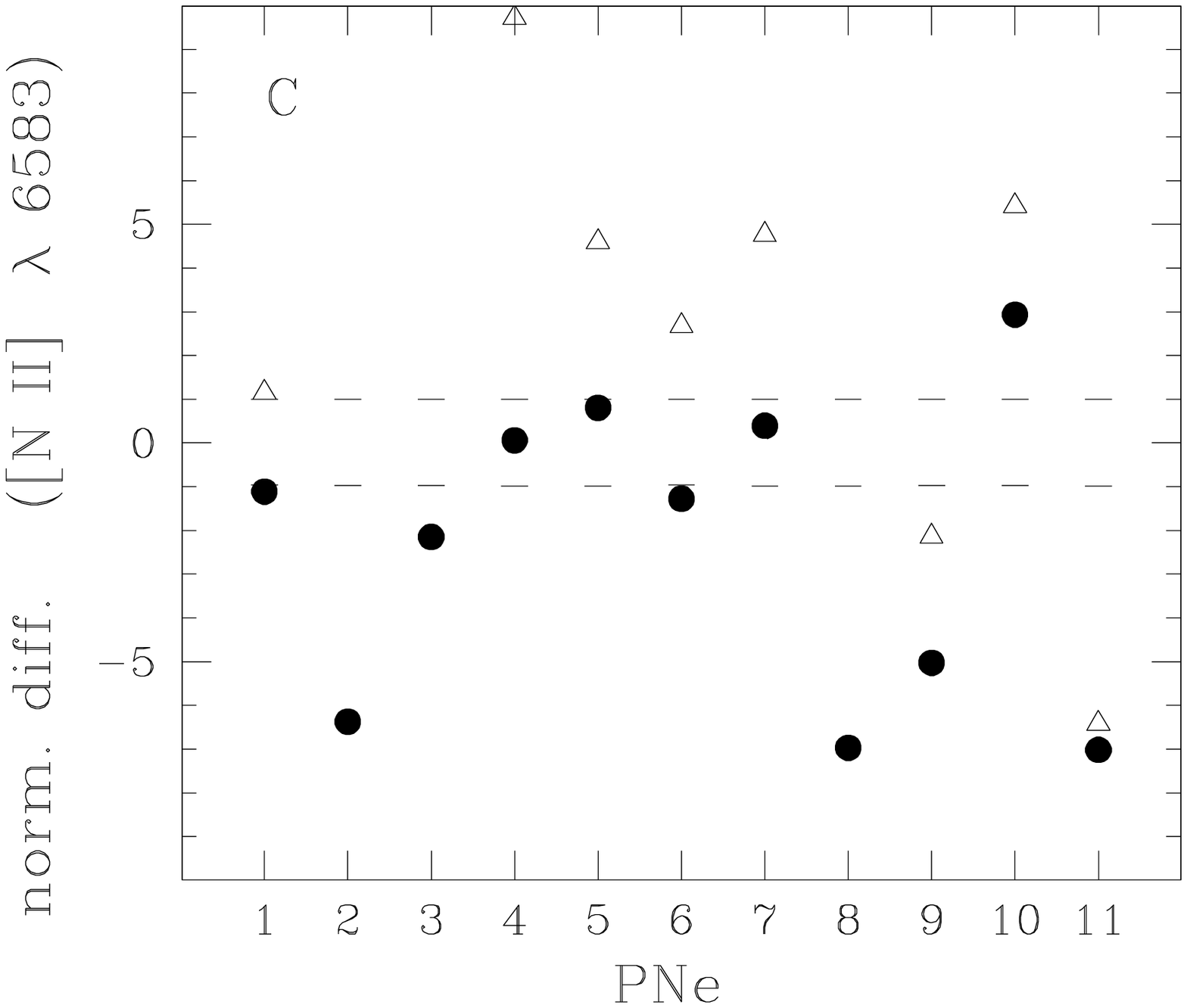}}

\resizebox{0.33\hsize}{!}{\includegraphics{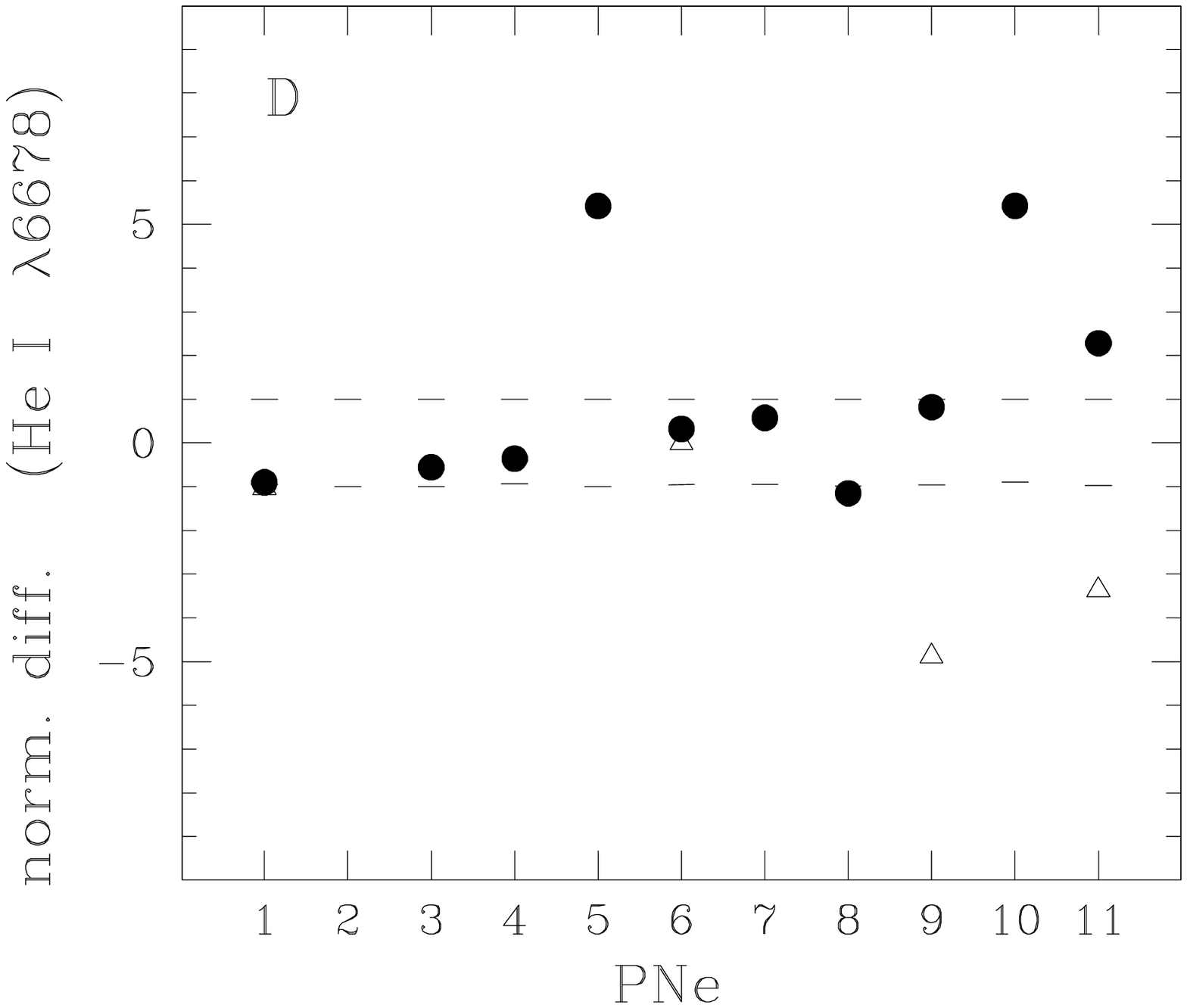}} 
\resizebox{0.33\hsize}{!}{\includegraphics{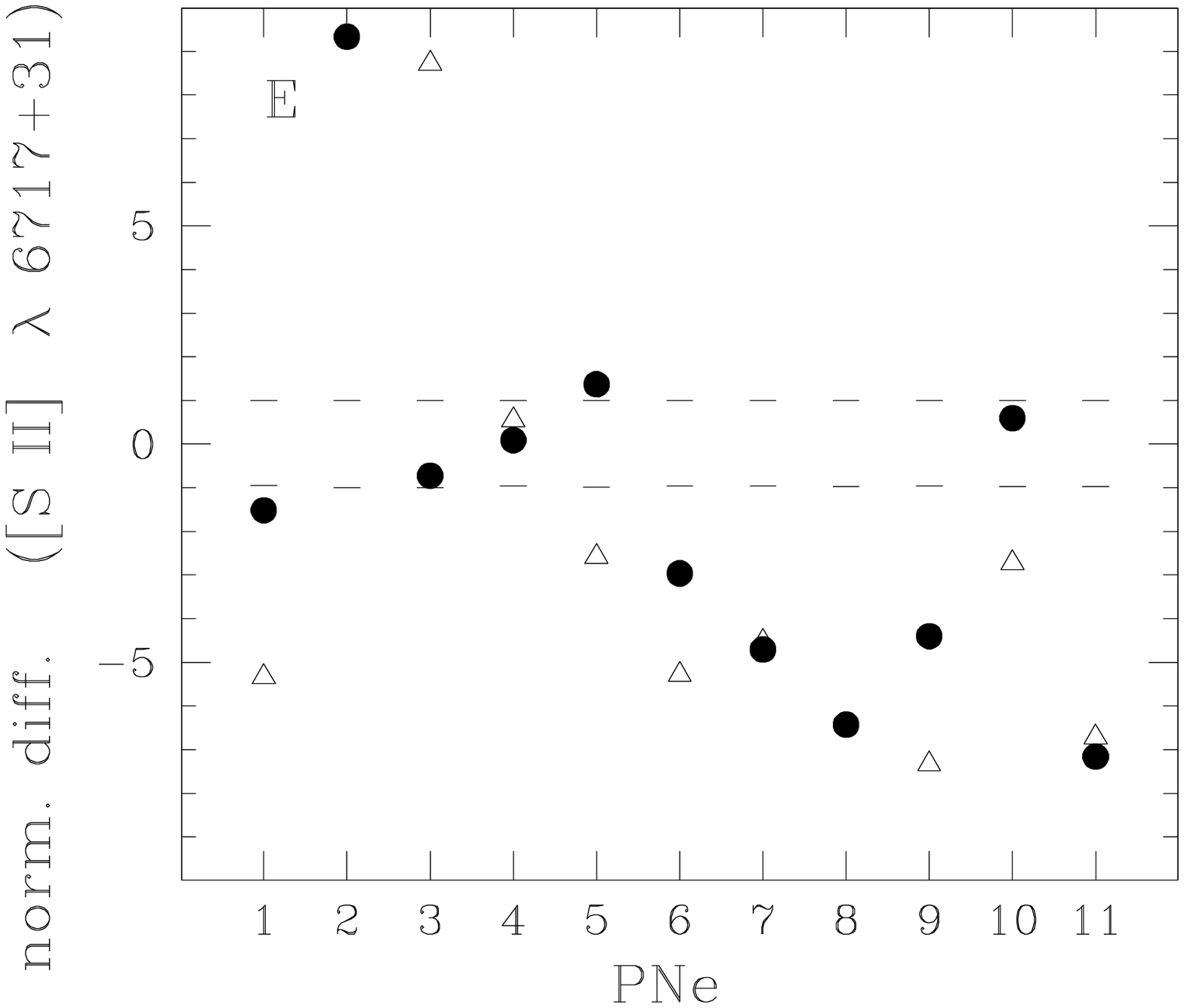}} 
\resizebox{0.33\hsize}{!}{\includegraphics{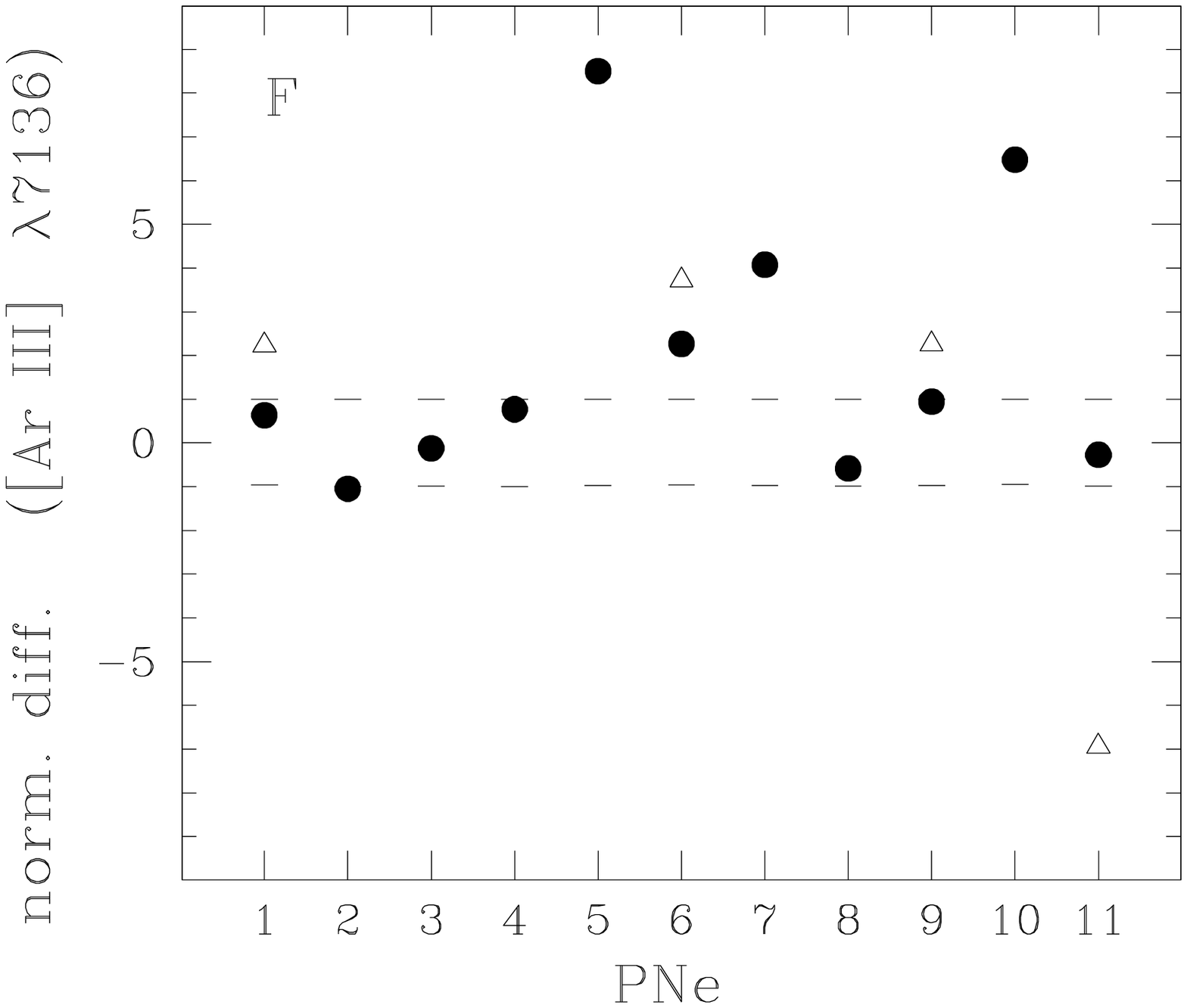}} 
\caption[]{
 Comparison of our line intensities with measurements of \cite{Dopita} for
 11 PNe in common that are listed in Table\,A.1.  The difference for each
 PNe (identified on horizontal axis) between our measurement and of
 \cite{Dopita} is marked with a black dot using a normalised scale (vertical
 axis).  The normalisation was done with the original errors from
 \cite{Dopita}.  If the normalised absolute difference is less than 1
 (dashed lines) our measurements agree within the uncertainty given by the
 latter authors.  Open triangles present an analogous comparison for data from
 \cite{Acker1992} compared to \cite{Dopita} for the same PNe and are given
 for reference.
}
\label{dop1} 
\end{figure*}

In this section, we compare our line measurements with data published by
\cite{Dopita}.  We have 11 PNe in common with their list of southern
emission-line flux standards. They are listed in Table~A.1.  In
\figref{dop1}, we present the comparison for the lines of [O\,III] \lam5007,
H$\alpha$, [N\,II] \lam6583, He\,I \lam6678, [S\,II] \lam6717+31, and
[Ar\,III] \lam7136.  The differences have been normalised separately for
each PNe and each line to the error of the line intensity as given by
\cite{Dopita}.

Because our observations are not absolutely flux calibrated, we first had to
recalculate values of \cite{Dopita} to the same scale (H$\beta$=100) as in
our Table~B.1.  Therefore, the deviations shown in \figref{dop1} are
influenced not only by the uncertainty of the given line but also to some
degree by the errors of H$\beta$ in both our and \cite{Dopita} measurements. 
For the 11 PNe in common with our observations and presented in
\figref{dop1}, the error of the H$\beta$ flux\footnote{In practise, the
errors in H$\beta$ have in practise minor influence on our further
calculations since they cancel when ratios of lines are computed, except
through the H$\alpha$/H$\beta$ ratio used to deredden the observations} in
\cite{Dopita} varies between 1\%\ and 2\%.

The dashed lines in \figref{dop1} represent the normalised error limits of
\cite{Dopita} measurements.  Our results are shown with black-filled
circles.  The triangles in \figref{dop1} represent line measurements from
the Strasbourg-ESO catalogue of \cite{Acker1992} and are shown for
reference.  The latter have been derived from the Catalogue's spectroscopic
survey and are of considerably lower sensitivity than our spectra.

The panel\,A of \figref{dop1} presents the comparison of our measurement and
that of \cite{Dopita} of the [O\,III] \lam5007 line.  In almost perfect
agreement, it can be seen that most of our points are confined within the
error bars given by \cite{Dopita}.

The situation is different in panel\,B that presents the comparison of
H$\alpha$ measurements.  Although there is no systematic deviation between
our data and \cite{Dopita}, the scatter is larger than in the previous plot. 
It cannot be blamed on the measurement of H$\beta$ line, which is
sometimes 10 times fainter then the H$\alpha$ line, since it would then
have comparable impact in the first panel.  We checked that the internal
errors of Dopita measurements for H$\alpha$/H$\beta$ line ratio are largest
for objects No.  1, 6, and 9 (see Table A.1) and reach 4.0, 4.5, and 3.2\%,
respectively.  It can be noted that our measurements actually fall within
the normalised errors in these cases.  One possibility is therefore that the
errors of \cite{Dopita} are underestimated for other PNe.  The discrepancy
of H$\alpha$/H$\beta$ ratios between different authors are not uncommon.  In
this case, we can see that they are mostly influenced by H$\alpha$
measurements itself.

In panel\,C of \figref{dop1}, we show a comparison for the close [N\,II]
\lam6583 line.  The same objects No.  1, 6, and five others show
satisfactory agreement with \cite{Dopita}.  The objects that deviate most
are No.  2, 8, and 11.  Object 2 is a high excitation nebula with [N\,II]
lines very weak, and the error of this measurement could be larger than
calculated by \cite{Dopita}.  In the case of No.  8 and 11, another
explanation can be invoked.  Those two PNe are low ionisation objects with
[N\,II] dominating the spectra.  \cite{Dopita} used slitless spectroscopic
observations and measured the cumulative flux from the nebula.  Since we
used a narrow slit positioned on the central star, we could have missed some
of the [N\,II] radiation from the external parts of the
nebula.\footnote{This underestimation have, however, a minor effect on our
results, since N/O abundance ratio is derived by comparison to [O\,II] lines
that origin from a similar region in the nebula.} This difference in
observing technique will have smaller influence on ions of higher ionisation
potentials, like [O\,III], since they are naturally concentrated toward the
center of the nebula.

In panel\,D of \figref{dop1}, we compare the He\,I \lam6678 line.  Despite
that it is usually much fainter, we find a better agreement with the data of
\cite{Dopita}.  This again advocates in favour of the errors of some
H$\alpha$ and [N\,II] measurements being underestimated by these authors. 
However, some influence in which we frequently had to use short snapshot
spectra to measure such lines cannot be excluded.

In panels E and F of \figref{dop1}, we compare our results with
\cite{Dopita} for the lines of another two ions.  In the case of [S\,II]
doublet \lam6717+31, a similar picture to [N\,II] is seen strengthening our
arguments presented above.  For [Ar\,III] \lam7136, we found a generally
satisfactory agreement.  It is worth to note that no measurable influence by
second order contamination effects is noticeable in our data.  It would
systematically underestimate measurements of red lines, like [Ar\,III]
\lam7136, which is not observed.

In summary, the analysis presented above show that there are no
systematic deviations in our line measurements compared to \cite{Dopita}. 
It also demonstrates that the discrepancies in measured line intensities
between different authors can be a complex function of many factors
(wavelength, blending, location of emission zone within the nebula, etc.),
and the values do not always converge with increasing line strength.

\begin{table}[t]
\caption{ List of PNe in common with \cite{Dopita}.
        }
\begin{tabular}{ r l l }
\hline
 No.&  PN\,G     & name          \\
\hline
 1  & 285.4+01.5 & Pe 1- 1  \\
 2  & 289.8+07.7 & He 2- 63 \\
 3  & 286.0-06.5 & He 2- 41 \\
 4  & 295.3-09.3 & He 2- 62 \\
 5  & 296.3-03.0 & He 2- 73 \\
 6  & 300.7-02.0 & He 2- 86 \\
 7  & 309.0+00.8 & He 2- 96 \\
 8  & 320.9+02.0 & He 2-117 \\
 9  & 323.9+02.4 & He 2-123 \\
 10 & 324.2+02.5 & He 2-125 \\
 11 & 327.1-01.8 & He 2-140 \\
\hline
\end{tabular}
\end{table}

\clearpage

\section{Supplementary tables}

\setcounter{table}{0}
\begin{table*}[b]
\caption{     
  Main nebular lines on the scale H$\beta$=100.  The two columns for each
  object give measured (left) and dereddened (right) intensities,
  respectively.  Meaning of the symbols: (:) refers to line error at 20\%;
  (;) is a line error at 40\%; (p) is a line present but not measurable; (?)
  defines uncertain identyfication; (c) is a additional reddening correction
  applied.
}
\begin{tabular}{     l @{\hspace{0.60cm}}
                     l @{\hspace{0.80cm}}
                     r r r r r r r r r}
\hline
 PN\,G      &      &\multicolumn{2}{c}{268.4+02.4} &\multicolumn{2}{c}{269.7-03.6} &\multicolumn{2}{c}{279.6-03.1} &\multicolumn{2}{c}{283.8+02.2} \\
 name       &      &\multicolumn{2}{c}{PB 5}       &\multicolumn{2}{c}{PB 3}       &\multicolumn{2}{c}{He 2- 36}   &\multicolumn{2}{c}{My 60}      \\
            &      &               &               &               &               &               &               &               &               \\
 C(H$\beta$)&      &               &   1.82~~~~    &               &   1.41~~~~    &               &   0.87~~~~    &               &   0.84~~~~    \\
            &      &               &               &               &               &               &               &               &               \\
 3727 & {[O\ II]}  &    12.1 :     &    32.52 :c   &   158.4 ~     &   252.20 ~c   &    54.8 ~     &   137.91 ~    &    11.2 ;     &    18.37 ;c   \\
 3869 & {[Ne III]} &    49.9 ~     &   118.48 ~c   &    73.0 ~     &   110.57 ~c   &    66.2 ~     &   145.11 ~    &    58.2 ~     &    89.73 ~c   \\
 4068 & {[S II]}   &     6.5 :     &    12.96 :c   &     8.7 :     &    12.23 :c   &               &               &         p     &          p ~  \\
\hline
\end{tabular}
\end{table*} 

\setcounter{table}{0}
\begin{table*}
\caption{     
  Continued.
}
\begin{tabular}{     l @{\hspace{0.60cm}}
                     l @{\hspace{0.80cm}}
                     r r r r r r r r r}
\hline
 PN\,G      &      &\multicolumn{2}{c}{283.8-04.2} &\multicolumn{2}{c}{285.4+01.5} &\multicolumn{2}{c}{286.0-06.5} &\multicolumn{2}{c}{289.8+07.7} \\
 name       &      &\multicolumn{2}{c}{ He 2- 39 } &\multicolumn{2}{c}{ Pe 1- 1  } &\multicolumn{2}{c}{ He 2- 41 } &\multicolumn{2}{c}{ He 2- 63 } \\
            &      &               &               &               &               &               &               &               &               \\
 C(H$\beta$)&      &               &   0.44~~~~    &               &   1.72~~~~    &               &   0.47~~~~    &               &   0.13~~~~    \\
            &      &               &               &               &               &               &               &               &               \\
 3727 & {[O\ II]}  &   157.5 ~     &   261.58 ~c   &    26.1 ~     &    71.42 ~c   &    36.63 ~    &    48.17 ~    &    14.80 :    &    18.37 :c   \\
 3869 & {[Ne III]} &    97.9 ~     &   151.98 ~c   &    31.8 ~     &    76.76 ~c   &    51.12 ~    &    64.97 ~    &    56.85 ~    &    69.40 ~c   \\
 4068 & {[S II]}   &               &               &     2.5 ;     &     5.06 ;c   &     1.11 ;    &     1.34 ;    &               &               \\
\hline
\end{tabular}
\end{table*}

\setcounter{table}{0}
\begin{table*}
\caption{     
  Continued.
}
\begin{tabular}{     l @{\hspace{0.60cm}}
                     l @{\hspace{0.80cm}}
                     r r r r r r r r r}
\hline
 PN\,G      &      &\multicolumn{2}{c}{291.4+19.2} &\multicolumn{2}{c}{292.8+01.1} &\multicolumn{2}{c}{293.1-00.0} &\multicolumn{2}{c}{293.6+01.2} \\
 name       &      &\multicolumn{2}{c}{ ESO 320-2} &\multicolumn{2}{c}{ He 2- 67 } &\multicolumn{2}{c}{BMPJ1128-6} &\multicolumn{2}{c}{ He 2- 70 } \\
            &      &               &               &               &               &               &               &               &               \\
 C(H$\beta$)&      &               &   0.33~~~~    &               &   0.81~~~~    &               &   2.16~~~~    &               &   0.92~~~~    \\
            &      &               &               &               &               &               &               &               &               \\
 3727 & {[O\ II]}  &               &               &    44.6       &    61.45 ~c   &               &               &   244.7 :     &   419.46 :c   \\
 3869 & {[Ne III]} &    20.9 :     &    13.10 :c   &    85.2       &   114.90 ~c   &    77.1 ;     &    98.64 ;c   &   146.3 :     &   234.49 :c   \\
 4068 & {[S II]}   &               &               &     5.4       &     7.05 ~c   &               &               &               &               \\
\hline
\end{tabular}
\end{table*}

\setcounter{table}{0}
\begin{table*}
\caption{     
  Continued.
}
\begin{tabular}{     l @{\hspace{0.60cm}}
                     l @{\hspace{0.80cm}}
                     r r r r r r r r r}
\hline
 PN\,G      &      &\multicolumn{2}{c}{294.6+04.7} &\multicolumn{2}{c}{294.9-04.3} &\multicolumn{2}{c}{295.3-09.3} &\multicolumn{2}{c}{296.3-03.0} \\
 name       &      &\multicolumn{2}{c}{ NGC 3918 } &\multicolumn{2}{c}{ He 2- 68 } &\multicolumn{2}{c}{ He 2- 62 } &\multicolumn{2}{c}{ He 2- 73 } \\
            &      &               &               &               &               &               &               &               &               \\
 C(H$\beta$)&      &               &   0.27~~~~    &               &   0.33~~~~    &               &   0.33~~~~    &               &   1.4~~~~     \\
            &      &               &               &               &               &               &               &               &               \\
 3727 & {[O\ II]}  &    58.7 :     &    61.84 :c   &   126.7 ~     &   153.71 ~c   &    26.7 ~     &    38.90 ~c   &    20.8 ~     &    58.69 ~c   \\
 3869 & {[Ne III]} &   100.7 ~     &   106.75 ~c   &     2.8 ~     &     3.32 ~c   &    61.7 ~     &    85.03 ~c   &    56.4 ~     &   139.07 ~c   \\
 4068 & {[S II]}   &     2.1 :     &     2.24 :c   &     3.0 :     &     3.43 :c   &     2.5 :     &     3.18 :c   &     3.0 ~     &     6.10 ~c   \\
\hline
\end{tabular}
\end{table*}

\setcounter{table}{0}
\begin{table*}
\caption{     
  Continued.
}
\begin{tabular}{     l @{\hspace{0.60cm}}
                     l @{\hspace{0.80cm}}
                     r r r r r r r r r}
\hline
 PN\,G      &      &\multicolumn{2}{c}{297.4+03.7} &\multicolumn{2}{c}{299.0+18.4} &\multicolumn{2}{c}{299.5+02.4} &\multicolumn{2}{c}{300.2+00.6} \\
 name       &      &\multicolumn{2}{c}{ He 2- 78 } &\multicolumn{2}{c}{ K 1- 23  } &\multicolumn{2}{c}{ He 2- 82 } &\multicolumn{2}{c}{ He 2- 83 } \\
            &      &               &               &               &               &               &               &               &               \\
 C(H$\beta$)&      &               &   0.89~~~~    &               &   0.13~~~~    &               &   0.92~~~~    &               &   2.42~~~~    \\
            &      &               &               &               &               &               &               &               &               \\
 3727 & {[O\ II]}  &   138.9 ~     &   234.70 ~c   &   170.7 ~     &   221.26 ~c   &   293.7 ~     &   457.07 ~c   &    29.2 :     &    56.49 :c   \\
 3869 & {[Ne III]} &               &               &    75.1 ~     &    94.48 ~c   &    55.5 ~     &    81.64 ~c   &               &               \\
 4068 & {[S II]}   &               &               &               &               &               &               &               &               \\
\hline
\end{tabular}
\end{table*}

\setcounter{table}{0}
\begin{table*}
\caption{     
  Continued.
}
\begin{tabular}{     l @{\hspace{0.60cm}}
                     l @{\hspace{0.80cm}}
                     r r r r r r r r r}
\hline
 PN\,G      &      &\multicolumn{2}{c}{300.4-00.9} &\multicolumn{2}{c}{300.5-01.1} &\multicolumn{2}{c}{300.7-02.0} &\multicolumn{2}{c}{300.8-03.4} \\
 name       &      &\multicolumn{2}{c}{ He 2- 84 } &\multicolumn{2}{c}{ He 2- 85 } &\multicolumn{2}{c}{ He 2- 86 } &\multicolumn{2}{c}{ ESO 095-1} \\
            &      &               &               &               &               &               &               &               &               \\
 C(H$\beta$)&      &               &   1.27~~~~    &               &   1.76~~~~    &               &   2.00~~~~    &               &   0.67~~~~    \\
            &      &               &               &               &               &               &               &               &               \\
 3727 & {[O\ II]}  &    91.8 ~     &   161.37 ~c   &    18.5 :     &    49.45 :c   &     7.8 ~     &    37.40 ~c   &   134.1 :     &   161.27 :c   \\
 3869 & {[Ne III]} &    76.2 ~     &   125.10 ~c   &    46.4 ~     &   108.72 ~c   &    21.7 ~     &    83.58 ~c   &    91.2 :     &   107.46 :c   \\
 4068 & {[S II]}   &               &               &               &               &     1.6 :     &     4.48 :c   &               &               \\
\hline
\end{tabular}
\end{table*}

\setcounter{table}{0}
\begin{table*}
\caption{     
  Continued.
}
\begin{tabular}{     l @{\hspace{0.60cm}}
                     l @{\hspace{0.80cm}}
                     r r r r r r r r r}
\hline
 PN\,G      &      &\multicolumn{2}{c}{302.2+02.5} &\multicolumn{2}{c}{305.1+01.4} &\multicolumn{2}{c}{307.3+05.0} &\multicolumn{2}{c}{308.2+07.7} \\
 name       &      &\multicolumn{2}{c}{Wray 16-120}&\multicolumn{2}{c}{ He 2- 90 } &\multicolumn{2}{c}{Wray 16-128}&\multicolumn{2}{c}{ MeWe 1- 3} \\
            &      &               &               &               &               &               &               &               &               \\
 C(H$\beta$)&      &               &   1.22~~~~    &               &   1.23~~~~    &               &   0.96~~~~    &               &   0.49~~~~    \\
            &      &               &               &               &               &               &               &               &               \\
 3727 & {[O\ II]}  &               &               &    12.8 ~     &    25.07 ~c   &               &               &               &               \\
 3869 & {[Ne III]} &    76.0 ;     &    10.89 ;c   &     2.5 ~     &     4.44 ~c   &    20.5 ~     &    35.26 ~c   &    55.9 ~     &    82.56 ~c   \\
 4068 & {[S II]}   &               &               &     1.4 ~     &     2.16 ~c   &               &               &               &               \\
\hline
\end{tabular}
\end{table*}

\setcounter{table}{0}
\begin{table*}
\caption{     
  Continued.
}
\begin{tabular}{     l @{\hspace{0.60cm}}
                     l @{\hspace{0.80cm}}
                     r r r r r r r r r}
\hline
 PN\,G      &      &\multicolumn{2}{c}{309.0+00.8} &\multicolumn{2}{c}{309.5-02.9} &\multicolumn{2}{c}{310.4+01.3} &\multicolumn{2}{c}{310.7-02.9} \\
 name       &      &\multicolumn{2}{c}{ He 2- 96 } &\multicolumn{2}{c}{ MaC 1- 2 } &\multicolumn{2}{c}{ Vo 4     } &\multicolumn{2}{c}{ He 2-103 } \\
            &      &               &               &               &               &               &               &               &               \\
 C(H$\beta$)&      &               &   1.78~~~~    &               &   1.27~~~~    &               &   3.25~~~~    &               &   1.05~~~~    \\
            &      &               &               &               &               &               &               &               &               \\
 3727 & {[O\ II]}  &    17.0 :     &    73.21 :c   &    91.04 :    &   122.57 :c   &               &               &   233.9 ~     &   433.55 ~c   \\
 3869 & {[Ne III]} &    20.3 :     &    71.46 :c   &    78.21 :    &   102.73 :c   &               &               &    68.1 :     &   116.88 :c   \\
 4068 & {[S II]}   &     2.2 ;     &     5.77 ;c   &               &               &               &               &               &               \\
\hline
\end{tabular}
\end{table*}

\setcounter{table}{0}
\begin{table*}
\caption{     
  Continued.
}
\begin{tabular}{     l @{\hspace{0.60cm}}
                     l @{\hspace{0.80cm}}
                     r r r r r r r r r}
\hline
 PN\,G      &      &\multicolumn{2}{c}{311.4+02.8} &\multicolumn{2}{c}{312.6-01.8} &\multicolumn{2}{c}{314.4+02.2} &\multicolumn{2}{c}{315.1-13.0} \\
 name       &      &\multicolumn{2}{c}{ He 2-102 } &\multicolumn{2}{c}{ He 2-107 } &\multicolumn{2}{c}{ PM 1- 81 } &\multicolumn{2}{c}{ He 2-131 } \\
            &      &               &               &               &               &               &               &               &               \\
 C(H$\beta$)&      &               &   1.03~~~~    &               &   1.31~~~~    &               &   4.69~~~~    &               &   0.09~~~~    \\
            &      &               &               &               &               &               &               &               &               \\
 3727 & {[O\ II]}  &     6.6 ~     &    11.20 ~c   &    50.7 ~     &   132.37 ~c   &               &               &    75.5 ~     &    93.26 ~c   \\
 3869 & {[Ne III]} &    50.9 ~     &    81.58 ~c   &               &               &               &               &               &               \\
 4068 & {[S II]}   &     2.3 ;     &     3.40 ;c   &               &               &               &               &     4.5 ~     &     5.17 ~c   \\
\hline
\end{tabular}
\end{table*} 

\setcounter{table}{0}
\begin{table*}
\caption{     
  Continued.
}
\begin{tabular}{     l @{\hspace{0.60cm}}
                     l @{\hspace{0.80cm}}
                     r r r r r r r r r}
\hline
 PN\,G      &      &\multicolumn{2}{c}{315.4+05.2} &\multicolumn{2}{c}{315.7+05.5} &\multicolumn{2}{c}{316.1+08.4} &\multicolumn{2}{c}{318.3-02.5} \\
 name       &      &\multicolumn{2}{c}{ He 2-109 } &\multicolumn{2}{c}{ LoTr 8   } &\multicolumn{2}{c}{ He 2-108 } &\multicolumn{2}{c}{ He 2-116 } \\
            &      &               &               &               &               &               &               &               &               \\
 C(H$\beta$)&      &               &   0.89~~~~    &               &   0.40~~~~    &               &   0.51~~~~    &               &   1.08~~~~    \\
            &      &               &               &               &               &               &               &               &               \\
 3727 & {[O\ II]}  &   269.5 ~     &   529.16 ~c   &               &               &    82.7 ~     &   151.31 ~c   &   162.8 ~     &   306.97 ~c   \\
 3869 & {[Ne III]} &    71.2 ~     &   128.02 ~c   &   284.3 ;     &   349.68 ;    &     3.0 :     &     5.04 :c   &    52.1 ~     &    90.77 ~c   \\
 4068 & {[S II]}   &         p     &          p ~  &               &               &     0.5 ;     &     0.74 ;c   &               &               \\
\hline
\end{tabular}
\end{table*}

\setcounter{table}{0}
\begin{table*}
\caption{     
  Continued.
}
\begin{tabular}{     l @{\hspace{0.60cm}}
                     l @{\hspace{0.80cm}}
                     r r r r r r r r r}
\hline
 PN\,G      &      &\multicolumn{2}{c}{319.2+06.8} &\multicolumn{2}{c}{320.9+02.0} &\multicolumn{2}{c}{321.3+02.8} &\multicolumn{2}{c}{321.8+01.9} \\
 name       &      &\multicolumn{2}{c}{ He 2-112 } &\multicolumn{2}{c}{ He 2-117 } &\multicolumn{2}{c}{ He 2-115 } &\multicolumn{2}{c}{ He 2-120 } \\
            &      &               &               &               &               &               &               &               &               \\
 C(H$\beta$)&      &               &   1.19~~~~    &               &   2.70~~~~    &               &   2.21~~~~    &               &   1.21~~~~    \\
            &      &               &               &               &               &               &               &               &               \\
 3727 & {[O\ II]}  &    51.1 ~     &   121.86 ~c   &     7.5 ~     &    32.60 ~c   &    16.5 ~     &    67.93 ~c   &   123.9 ~     &   252.22 ~    \\
 3869 & {[Ne III]} &    78.4 ~     &   165.14 ~c   &    14.7 ~     &    52.57 ~c   &     8.4 ~     &    29.01 ~c   &    41.8 ;     &    77.88 ;    \\
 4068 & {[S II]}   &     4.3 :     &     7.58 :c   &     2.2 ;     &     5.98 ;c   &               &               &               &               \\
\hline
\end{tabular}
\end{table*}

\setcounter{table}{0}
\begin{table*}
\caption{     
  Continued.
}
\begin{tabular}{     l @{\hspace{0.60cm}}
                     l @{\hspace{0.80cm}}
                     r r r r r r r r r}
\hline
 PN\,G      &      &\multicolumn{2}{c}{322.5-05.2} &\multicolumn{2}{c}{323.9+02.4} &\multicolumn{2}{c}{324.2+02.5} &\multicolumn{2}{c}{324.8-01.1} \\ 
 name       &      &\multicolumn{2}{c}{ NGC 5979 } &\multicolumn{2}{c}{ He 2-123 } &\multicolumn{2}{c}{ He 2-125 } &\multicolumn{2}{c}{ He 2-133 } \\ 
            &      &               &               &               &               &               &               &               &               \\
 C(H$\beta$)&      &               &   0.25~~~~    &               &   1.62~~~~    &               &   1.71~~~~    &               &   3.59~~~~    \\
            &      &               &               &               &               &               &               &               &               \\
 3727 & {[O\ II]}  &     7.7 ~     &     9.17 ~c   &    19.1 ~     &    57.86 ~c   &    17.4 :     &    67.97 :c   &               &               \\
 3869 & {[Ne III]} &    54.5 ~     &    63.74 ~c   &     3.7 ~     &     9.62 ~c   &               &               &    23.8 :     &   150.36 :    \\
 4068 & {[S II]}   &     1.9 ~     &     2.16 ~c   &               &               &               &               &               &               \\
\hline
\end{tabular}
\end{table*} 

\setcounter{table}{0}
\begin{table*}
\caption{     
  Continued.
}
\begin{tabular}{     l @{\hspace{0.60cm}}
                     l @{\hspace{0.80cm}}
                     r r r r r r r r r}
\hline
 PN\,G      &      &\multicolumn{2}{c}{327.1-01.8} &\multicolumn{2}{c}{327.8-01.6} &\multicolumn{2}{c}{330.9+04.3} &\multicolumn{2}{c}{336.9+08.3} \\
 name       &      &\multicolumn{2}{c}{ He 2-140 } &\multicolumn{2}{c}{ He 2-143 } &\multicolumn{2}{c}{Wray 16-189}&\multicolumn{2}{c}{ StWr 4-10} \\
            &      &               &               &               &               &               &               &               &               \\
 C(H$\beta$)&      &               &   1.94~~~~    &               &   0.74~~~~    &               &   1.15~~~~    &               &   0.95~~~~    \\
            &      &               &               &               &               &               &               &               &               \\
 3727 & {[O\ II]}  &    44.5 :     &   138.94 :c   &    97.4 ~     &   150.06 ~c   &               &               &               &               \\
 3869 & {[Ne III]} &               &               &   104.9 ~     &   153.14 ~c   &         p     &          p    &     34.0 ~    &    96.11 ~c   \\
 4068 & {[S II]}   &               &               &     3.8 :     &     5.14 :c   &               &               &               &               \\
\hline
\end{tabular}
\end{table*}

\setcounter{table}{0}
\begin{table*}
\caption{     
  Continued.
}
\begin{tabular}{     l @{\hspace{0.60cm}}
                     l @{\hspace{0.80cm}}
                     r r r r r r r r r}
\hline
 PN\,G      &      &\multicolumn{2}{c}{338.1-08.3}      & & & & & & \\
 name       &      &\multicolumn{2}{c}{ NGC 6326 }      & & & & & & \\
            &      &               &                    & & & & & & \\
 C(H$\beta$)&      &               &   0.11~~~~         & & & & & & \\
            &      &               &                    & & & & & & \\
 3727 & {[O\ II]}  &    40.5 ~     &    43.24 ~         & & & & & & \\
 3869 & {[Ne III]} &   102.2 ~     &   108.23 ~         & & & & & & \\
 4068 & {[S II]}   &     2.1 ~     &     2.20 ~         & & & & & & \\
\hline
\end{tabular}
\end{table*}

\clearpage

\setcounter{table}{1}
\begin{table}[t]
\caption{ New, reobserved, and mis-classified PNe with emission-line 
          central stars. Column 3 gives the width of C\,IV $\lambda$5805
          stellar emission feature if is broader than instrumental profile. 
          Columns 4 and 5 give stellar classification of this work and from
          Weidmann \& Gamen (2011).
}
\begin{tabular}{ l l @{\hspace{0.10cm}} r r r }
\hline
 & name          & FWHM   & \multicolumn{2}{c}{~~~~~classification} \\
 &               &        & t.w.~~~ & WG       \\
\hline
 & {\it new} WELs: &      &       &               \\
 & PB 5          &        & WEL~~ &     - ~~~~    \\
 & My 60         &        & WEL?  &     - ~~~~    \\
 & He 2-115      &        & WEL~~ &     - ~~~~    \\ \vspace{0.1cm}
 & {\it new} [WR]: &      &       &               \\
 & He 2- 67      & 43\AA  & [WC\,4]~~&     - ~~~~    \\
 & He 2-117      & 25\AA  & [WC\,5-6]~~&     - ~~~~    \\ \vspace{0.1cm}
 & He 2-125      &        & VL~~  &     - ~~~~    \\
 & {\it reobserved}: &  &       &               \\
 & Pe 1- 1       & 45\AA  & [WC\,5]  & [WO\,4]       \\
 & He 2- 63      &        & {\it undetect.} & WEL~~ \\
 & He 2- 86      & 22\AA  & [WC\,5-6]  & [WC\,4]       \\
 & He 2-107      &        & VL~~  & [WC10-11]     \\
 & He 2-131      &        & VL~~  & WEL~~         \\
 & He 2-108      &        & VL~~  & WEL~~         \\ \vspace{0.1cm}
 & He 2-123      &        & WEL~~ & WEL?          \\
 & {\it misclassified / mimics}:& &      &               \\
 & NGC 6326      &        & WEL?  &       \\
 & NGC 5979      &        & {\it not stellar}  & WEL~~         \\
 & NGC 3918      &        & {\it not stellar}  &     - ~~~~    \\
\hline
\end{tabular}\\
\end{table}

\setcounter{table}{2}
\begin{table*}
\caption{ Plasma parameters and abundances. The first row for each PN gives
 parameters computed from the nominal values of the observational data. The
 second and third row give the upper and lower limits, respectively, of
 these parameters.  Column (1) gives the PN\,G number; Column (2) usual
 name; Column (3) electron density deduced from \rSii, columns (4) and (5)
 electron temperatures from \rOiii\ and \rNii, respectively.  The value of
 \Te(N~{\sc ii}) is in parenthesis if \Te(O ~{\sc iii}) was chosen for all
 ions.  Columns (6) to (12) give the He/H, N/H, O/H, Ne/H, S/H, Ar/H, and
 Cl/H ratios, respectively.
}
\begin{tabular}{       @{\hspace{0.05cm}}
                     l @{\hspace{0.25cm}}
                     l @{\hspace{0.21cm}}
                     r @{\hspace{0.21cm}}
                     r @{\hspace{0.21cm}}
                     r @{\hspace{0.21cm}}
                     r @{\hspace{0.21cm}}
                     r @{\hspace{0.21cm}}
                     r @{\hspace{0.21cm}}
                     r @{\hspace{0.21cm}}
                     r @{\hspace{0.21cm}}
                     r @{\hspace{0.21cm}}
                     r } 
\hline
        PN G & Main Name & \Ne(S~{\sc ii})  &
                                      \Te(O~{\sc iii}) &
                                              \Te(N~{\sc ii})  &
                                                                He/H     &  N/H     &  O/H     &  Ne/H    &   S/H    &  Ar/H    &  Cl/H    \\
\hline
 268.4+02.4  & PB 5           & 1.82E+04  & (14413) &  13362  & 1.13E-01 & 1.34E-04 & 3.06E-04 & 6.48E-05 & 4.39E-06 & 1.77E-06 & 4.80E-07 \\
             &                & 4.43E+04  & (16435) &  13737  & 1.20E-01 & 1.80E-04 & 3.76E-04 & 7.84E-05 & 5.79E-06 & 2.03E-06 & 6.47E-07 \\
             &                & 1.12E+04  & ( 8523) &  12629  & 1.08E-01 & 1.19E-04 & 2.76E-04 & 5.67E-05 & 3.91E-06 & 1.64E-06 & 3.62E-07 \\
             &                &           &         &         &          &          &          &          &          &          &          \\
 269.7-03.6  & PB 3           & 2.50E+03  &  10829  &  11849  & 1.23E-01 & 2.21E-04 & 4.38E-04 & 1.02E-04 & 6.39E-06 & 2.51E-06 & 3.57E-07 \\
             &                & 3.08E+03  &  11190  &  12267  & 1.28E-01 & 2.52E-04 & 4.88E-04 & 1.19E-04 & 7.23E-06 & 2.77E-06 & 4.70E-07 \\
             &                & 2.11E+03  &  10446  &  11541  & 1.16E-01 & 1.89E-04 & 3.92E-04 & 9.02E-05 & 5.57E-06 & 2.24E-06 & 2.60E-07 \\
             &                &           &         &         &          &          &          &          &          &          &          \\
 279.6-03.1  & He 2- 36       & 5.50E+02  &  10793  &  14481  & 1.13E-01 & 6.38E-05 & 2.89E-04 & 7.77E-05 & 2.87E-06 & 1.25E-06 & 5.81E-07 \\
             &                & 6.42E+02  &  12013  &  14743  & 1.18E-01 & 9.13E-05 & 3.59E-04 & 9.24E-05 & 3.74E-06 & 1.44E-06 & 9.59E-07 \\
             &                & 4.33E+02  &   9822  &  13657  & 1.08E-01 & 5.31E-05 & 2.66E-04 & 6.56E-05 & 2.57E-06 & 1.17E-06 & 3.63E-07 \\
             &                &           &         &         &          &          &          &          &          &          &          \\
 283.8+02.2  & My 60          & 1.51E+03  &         &  13868  & 1.10E-01 & 5.11E-05 & 3.15E-04 & 6.02E-05 & 4.33E-06 & 1.68E-06 & 8.33E-06 \\
             &                & 2.45E+03  &         &  14412  & 1.16E-01 & 6.40E-05 & 3.80E-04 & 6.92E-05 & 5.10E-06 & 1.86E-06 & 1.04E-05 \\
             &                & 1.05E+03  &         &  13324  & 1.03E-01 & 4.17E-05 & 2.62E-04 & 5.24E-05 & 3.66E-06 & 1.48E-06 & 6.34E-06 \\
             &                &           &         &         &          &          &          &          &          &          &          \\
 283.8-04.2  & He 2-39        & 7.04E+02  &   9821  &  14289  & 1.09E-01 & 1.25E-04 & 4.77E-04 & 1.38E-04 & 2.85E-06 & 2.40E-06 &          \\
             &                & 7.99E+02  &  10751  &  14589  & 1.18E-01 & 1.53E-04 & 6.61E-04 & 2.01E-04 & 3.51E-06 & 3.12E-06 &          \\
             &                & 5.53E+02  &   8774  &  13546  & 9.90E-02 & 1.09E-04 & 4.19E-04 & 1.19E-04 & 2.42E-06 & 1.85E-06 &          \\
             &                &           &         &         &          &          &          &          &          &          &          \\
 285.4+01.5  & Pe 1- 1        & 6.08E+03  &  12859  &  10662  & 1.15E-01 & 1.13E-04 & 3.15E-04 & 7.50E-05 & 5.44E-06 & 2.60E-06 & 1.30E-06 \\
             &                & 8.96E+03  &  13611  &  10927  & 1.23E-01 & 1.44E-04 & 3.64E-04 & 8.28E-05 & 6.63E-06 & 2.99E-06 & 1.82E-06 \\
             &                & 4.56E+03  &  12264  &  10301  & 1.08E-01 & 1.04E-04 & 2.81E-04 & 6.65E-05 & 4.87E-06 & 2.38E-06 & 9.03E-07 \\
             &                &           &         &         &          &          &          &          &          &          &          \\
 286.0-06.5  & He 2- 41       & 6.74E+03  &  12917  &  11626  & 9.68E-02 & 4.53E-05 & 2.09E-04 & 4.53E-05 & 2.24E-06 & 9.05E-07 & 1.02E-06 \\
             &                & 1.30E+04  &  13591  &  11976  & 1.02E-01 & 5.12E-05 & 2.48E-04 & 5.37E-05 & 2.55E-06 & 1.03E-06 & 1.21E-06 \\
             &                & 3.76E+03  &  10618  &  11199  & 9.09E-02 & 3.73E-05 & 1.79E-04 & 4.15E-05 & 1.88E-06 & 8.02E-07 & 7.64E-07 \\
             &                &           &         &         &          &          &          &          &          &          &          \\
 289.8+07.7  & He 2- 63       & 9.86E+02  &         &  12459  & 9.87E-02 & 5.30E-05 & 2.66E-04 & 4.85E-05 & 2.61E-06 & 9.40E-07 &          \\
             &                & 1.93E+03  &         &  12885  & 1.03E-01 & 7.64E-05 & 3.11E-04 & 5.75E-05 & 3.62E-06 & 1.10E-06 &          \\
             &                & 4.38E+02  &         &  12078  & 9.16E-02 & 3.46E-05 & 2.31E-04 & 4.67E-05 & 1.70E-06 & 6.73E-07 &          \\
             &                &           &         &         &          &          &          &          &          &          &          \\
 291.4+19.2  & ESO 320-28     &           &         &  17898  & 9.38E-02 &          & 2.51E-05 & 2.17E-06 &          & 1.43E-06 &          \\
             &                &           &         &  23573  & 1.03E-01 &          & 4.65E-05 & 6.30E-06 &          & 2.84E-06 &          \\
             &                &           &         &  13839  & 8.75E-02 &          & 1.41E-05 & 1.17E-06 &          & 6.83E-07 &          \\
             &                &           &         &         &          &          &          &          &          &          &          \\
 292.8+01.1  & He 2- 67       & 3.07E+03  &  10228  &   9552  & 1.34E-01 & 4.86E-04 & 5.94E-04 & 1.85E-04 & 1.67E-05 & 4.71E-06 & 2.16E-06 \\
             &                & 3.80E+03  &  10637  &   9774  & 1.41E-01 & 5.60E-04 & 6.95E-04 & 2.26E-04 & 1.98E-05 & 5.23E-06 & 2.77E-06 \\
             &                & 2.50E+03  &   9825  &   9258  & 1.27E-01 & 3.90E-04 & 5.27E-04 & 1.76E-04 & 1.41E-05 & 4.25E-06 & 1.65E-06 \\
             &                &           &         &         &          &          &          &          &          &          &          \\
 293.1-00.0  & BMPJ1128-61    & 1.00E+05  &   9416  &  15416  & 1.10E-01 & 1.15E-04 & 3.33E-04 & 5.25E-05 & 4.75E-06 & 1.26E-06 &          \\
             &                & 1.00E+05  &  12660  &  21593  & 1.28E-01 & 2.68E-04 & 8.59E-04 & 1.58E-04 & 1.06E-05 & 2.27E-06 &          \\
             &                & 2.03E+04  &   7917  &  10081  & 9.74E-02 & 6.97E-05 & 1.49E-04 &          & 2.06E-06 & 7.29E-07 &          \\
             &                &           &         &         &          &          &          &          &          &          &          \\
 293.6+01.2  & He 2- 70       & 2.67E+02  &  11118  &  20203  & 2.71E-01 & 2.59E-04 & 1.78E-04 & 9.05E-05 & 4.32E-06 & 1.85E-06 &          \\
             &                & 3.64E+02  &  11571  &  24121  & 3.14E-01 & 2.99E-04 & 2.24E-04 & 1.37E-04 & 5.02E-06 & 2.16E-06 &          \\
             &                & 1.72E+02  &  10777  &  18398  & 2.32E-01 & 1.99E-04 & 1.45E-04 & 6.28E-05 & 3.14E-06 & 1.35E-06 &          \\
             &                &           &         &         &          &          &          &          &          &          &          \\
 294.6+04.7  & NGC 3918       & 4.37E+03  &  11286  &  12728  & 1.08E-01 & 1.64E-04 & 4.04E-04 & 7.24E-05 & 4.55E-06 & 2.48E-06 & 9.27E-07 \\
             &                & 6.07E+03  &  11744  &  13093  & 1.13E-01 & 1.95E-04 & 4.55E-04 & 8.59E-05 & 5.22E-06 & 2.97E-06 & 1.18E-06 \\
             &                & 3.52E+03  &  10861  &  12253  & 1.02E-01 & 1.33E-04 & 3.61E-04 & 6.60E-05 & 3.83E-06 & 2.15E-06 & 7.16E-07 \\
             &                &           &         &         &          &          &          &          &          &          &          \\
 294.9-04.3  & He 2- 68       & 1.49E+04  &   9666  &  10614  & 7.57E-02 & 4.57E-05 & 2.27E-04 & 1.49E-05 & 1.94E-06 & 8.00E-07 & 1.31E-07 \\
             &                & 6.42E+04  &  10402  &  11330  & 8.06E-02 & 7.18E-05 & 6.44E-04 & 4.49E-05 & 3.72E-06 & 1.02E-06 & 2.25E-07 \\
             &                & 8.86E+03  &   8135  &   9485  & 7.01E-02 & 3.79E-05 & 1.68E-04 & 1.09E-05 & 1.56E-06 & 6.75E-07 & 8.94E-08 \\
             &                &           &         &         &          &          &          &          &          &          &          \\
\hline
\end{tabular}
\end{table*}

\setcounter{table}{2}
\begin{table*}
\caption{ 
  Continued.
}
\begin{tabular}{       @{\hspace{0.05cm}}
                     l @{\hspace{0.25cm}}
                     l @{\hspace{0.21cm}}
                     r @{\hspace{0.21cm}}
                     r @{\hspace{0.21cm}}
                     r @{\hspace{0.21cm}}
                     r @{\hspace{0.21cm}}
                     r @{\hspace{0.21cm}}
                     r @{\hspace{0.21cm}}
                     r @{\hspace{0.21cm}}
                     r @{\hspace{0.21cm}}
                     r @{\hspace{0.21cm}}
                     r } 
\hline
        PN G & Main Name & \Ne(S~{\sc ii})  &
                                      \Te(O~{\sc iii}) &
                                              \Te(N~{\sc ii})  &
                                                         He/H     &  N/H     &  O/H     &  Ne/H    &   S/H    &  Ar/H    &  Cl/H    \\
\hline
 295.3-09.3  & He 2- 62       & 4.93E+03  &  16469  &  12945  & 1.03E-01 & 7.38E-05 & 1.41E-04 & 3.81E-05 & 3.18E-06 & 8.37E-07 & 1.00E-06 \\
             &                & 6.64E+03  &  17364  &  13389  & 1.11E-01 & 8.62E-05 & 1.64E-04 & 4.39E-05 & 3.75E-06 & 9.30E-07 & 1.30E-06 \\
             &                & 3.85E+03  &  15468  &  12481  & 9.64E-02 & 5.64E-05 & 1.22E-04 & 3.35E-05 & 2.57E-06 & 7.42E-07 & 6.44E-07 \\
             &                &           &         &         &          &          &          &          &          &          &          \\
 296.3-03.0  & He 2- 73       & 5.05E+03  &  12601  &  11580  & 1.22E-01 & 2.56E-04 & 4.15E-04 & 1.04E-04 & 8.50E-06 & 3.42E-06 & 1.40E-06 \\
             &                & 7.13E+03  &  13090  &  11942  & 1.28E-01 & 3.09E-04 & 4.74E-04 & 1.17E-04 & 9.78E-06 & 3.85E-06 & 1.50E-06 \\
             &                & 4.00E+03  &  11929  &  11241  & 1.16E-01 & 2.12E-04 & 3.59E-04 & 9.25E-05 & 7.16E-06 & 3.04E-06 & 1.30E-06 \\
             &                &           &         &         &          &          &          &          &          &          &          \\
 297.4+03.7  & He 2- 78       & 3.93E+03  &   7847  &         & 5.13E-02 & 7.43E-05 & 4.75E-04 &          & 5.63E-06 & 1.34E-06 &          \\
             &                & 5.17E+03  &   9025  &         & 5.44E-02 & 1.14E-04 & 1.23E-03 &          & 8.98E-06 & 2.05E-06 &          \\
             &                & 2.89E+03  &   6680  &         & 4.78E-02 & 5.39E-05 & 2.68E-04 &          & 4.26E-06 & 1.00E-06 &          \\
             &                &           &         &         &          &          &          &          &          &          &          \\
 299.0+18.4  & K 1- 23        & 1.07E+02  &  11854  &  11144  & 1.15E-01 & 1.18E-04 & 3.07E-04 & 8.77E-05 & 7.15E-06 & 1.74E-06 & 1.76E-06 \\
             &                & 1.82E+02  &  13232  &  11899  & 1.33E-01 & 1.67E-04 & 4.10E-04 & 1.23E-04 & 8.93E-06 & 2.15E-06 & 3.17E-06 \\
             &                & 6.35E+01  &  10822  &  10167  & 8.58E-02 & 8.21E-05 & 2.48E-04 & 7.03E-05 & 5.92E-06 & 1.45E-06 & 1.01E-06 \\
             &                &           &         &         &          &          &          &          &          &          &          \\
 299.5+02.4  & He 2- 82       & 1.56E+02  &   9922  &         & 1.27E-01 & 1.53E-04 & 5.45E-04 & 1.83E-04 & 9.25E-06 & 2.44E-06 &          \\
             &                & 2.24E+02  &  10627  &         & 1.35E-01 & 1.87E-04 & 7.66E-04 & 2.47E-04 & 1.33E-05 & 3.06E-06 &          \\
             &                & 1.13E+02  &   9225  &         & 1.18E-01 & 1.29E-04 & 3.85E-04 & 1.12E-04 & 6.06E-06 & 1.96E-06 &          \\
             &                &           &         &         &          &          &          &          &          &          &          \\
 300.2+00.6  & He 2- 83       & 5.41E+03  &   8019  &         & 9.98E-02 & 1.49E-04 & 2.72E-04 &          & 9.28E-06 & 3.13E-06 &          \\
             &                & 7.57E+03  &   8294  &         & 1.06E-01 & 1.76E-04 & 3.63E-04 &          & 1.24E-05 & 3.68E-06 &          \\
             &                & 4.18E+03  &   7595  &         & 9.34E-02 & 1.29E-04 & 2.38E-04 &          & 7.19E-06 & 2.79E-06 &          \\
             &                &           &         &         &          &          &          &          &          &          &          \\
 300.4-00.9  & He 2- 84       & 8.06E+02  &  11855  &  11504  & 1.46E-01 & 4.30E-04 & 3.51E-04 & 1.09E-04 & 1.27E-05 & 3.57E-06 &          \\
             &                & 9.53E+02  &  12431  &  12048  & 1.54E-01 & 5.22E-04 & 3.93E-04 & 1.30E-04 & 1.45E-05 & 3.94E-06 &          \\
             &                & 6.71E+02  &  11378  &  11169  & 1.38E-01 & 3.52E-04 & 2.94E-04 & 9.92E-05 & 1.06E-05 & 3.10E-06 &          \\
             &                &           &         &         &          &          &          &          &          &          &          \\
 300.5-01.1  & He 2- 85       & 3.01E+03  &  11663  &  11815  & 1.16E-01 & 2.00E-04 & 4.03E-04 & 8.86E-05 & 7.36E-06 & 3.28E-06 & 1.54E-06 \\
             &                & 3.80E+03  &  12094  &  12198  & 1.23E-01 & 2.42E-04 & 4.64E-04 & 9.50E-05 & 8.47E-06 & 3.63E-06 & 2.30E-06 \\
             &                & 2.51E+03  &  11204  &  11406  & 1.09E-01 & 1.62E-04 & 3.55E-04 & 7.56E-05 & 6.17E-06 & 2.94E-06 & 9.97E-07 \\
             &                &           &         &         &          &          &          &          &          &          &          \\
 300.7-02.0  & He 2- 86       & 9.96E+03  &  11073  &   8659  & 1.39E-01 & 4.25E-04 & 5.42E-04 & 1.82E-04 & 1.96E-05 & 6.54E-06 & 6.36E-06 \\
             &                & 1.45E+04  &  11741  &   9033  & 1.49E-01 & 5.88E-04 & 7.52E-04 & 2.58E-04 & 3.07E-05 & 8.27E-06 & 1.40E-05 \\
             &                & 6.77E+03  &   9808  &   8045  & 1.30E-01 & 3.35E-04 & 4.59E-04 & 1.34E-04 & 1.57E-05 & 5.77E-06 & 4.04E-06 \\
             &                &           &         &         &          &          &          &          &          &          &          \\
 300.8-03.4  & ESO 095-12     & 3.76E+02  &         &  12902  &          & 9.41E-05 & 2.76E-04 & 5.05E-05 & 1.35E-05 & 2.29E-06 &          \\
             &                & 4.78E+02  &         &  14770  &          & 1.40E-04 & 4.74E-04 & 9.93E-05 & 2.66E-05 & 3.46E-06 &          \\
             &                & 2.93E+02  &         &  10436  &          & 6.60E-05 & 1.77E-04 & 2.65E-05 & 8.22E-06 & 1.62E-06 &          \\
             &                &           &         &         &          &          &          &          &          &          &          \\
 305.1+01.4  & He 2- 90       & 1.87E+04  &         &  24953  & 5.25E-02 & 3.78E-06 & 8.88E-06 & 5.53E-07 & 6.93E-07 & 3.76E-07 & 3.96E-07 \\
             &                & 1.00E+05  &         &  25828  & 6.09E-02 & 7.98E-06 & 1.39E-05 & 7.40E-07 & 1.01E-06 & 4.74E-07 & 5.84E-07 \\
             &                & 7.21E+03  &         &  21122  & 5.15E-02 & 2.94E-06 & 8.26E-06 & 4.86E-07 & 6.32E-07 & 3.36E-07 & 3.33E-07 \\
             &                &           &         &         &          &          &          &          &          &          &          \\
 307.3+05.0  & Wray 16-128    & 3.00E+01  &         &  17216  & 1.06E-01 &          & 2.07E-04 & 3.84E-05 &          & 1.32E-06 &          \\
             &                & 6.48E+02  &         &  18192  & 1.13E-01 &          & 3.69E-04 & 5.98E-05 &          & 1.49E-06 &          \\
             &                & 3.00E+01  &         &  16303  & 9.83E-02 &          & 1.61E-04 & 3.08E-05 &          & 1.14E-06 &          \\
             &                &           &         &         &          &          &          &          &          &          &          \\
 308.2+07.7  & MeWe 1- 3      &           &         &  17803  & 9.84E-02 &          & 4.92E-05 & 1.39E-05 &          & 1.16E-06 &          \\
             &                &           &         &  18652  & 1.04E-01 &          & 5.77E-05 & 1.63E-05 &          & 1.33E-06 &          \\
             &                &           &         &  17050  & 9.27E-02 &          & 4.18E-05 & 1.19E-05 &          & 1.04E-06 &          \\
             &                &           &         &         &          &          &          &          &          &          &          \\
 309.0+00.8  & He 2- 96       & 7.60E+03  &  12410  &   8444  & 1.35E-01 & 1.73E-04 & 6.05E-04 & 1.78E-04 & 1.91E-05 & 4.63E-06 & 1.18E-05 \\
             &                & 1.22E+04  &  13139  &   8564  & 1.43E-01 & 2.22E-04 & 7.86E-04 & 2.40E-04 & 2.55E-05 & 5.28E-06 & 1.85E-05 \\
             &                & 5.93E+03  &  11439  &   8068  & 1.27E-01 & 1.41E-04 & 5.71E-04 & 1.43E-04 & 1.68E-05 & 4.33E-06 & 7.46E-06 \\
             &                &           &         &         &          &          &          &          &          &          &          \\
 309.5-02.9  & MaC 1- 2       & 1.63E+03  &  11936  &  11754  & 1.32E-01 & 3.55E-04 & 3.94E-04 & 7.51E-05 & 4.10E-06 & 2.63E-06 &          \\
             &                & 1.96E+03  &  13156  &  13256  & 1.39E-01 & 6.42E-04 & 7.09E-04 & 1.53E-04 & 8.03E-06 & 4.03E-06 &          \\
             &                & 1.38E+03  &  10827  &   9507  & 1.23E-01 & 2.17E-04 & 2.76E-04 & 4.27E-05 & 1.97E-06 & 1.96E-06 &          \\
             &                &           &         &         &          &          &          &          &          &          &          \\
\hline
\end{tabular}
\end{table*}

\setcounter{table}{2}
\begin{table*}
\caption{ 
  Continued.
}
\begin{tabular}{       @{\hspace{0.05cm}}
                     l @{\hspace{0.25cm}}
                     l @{\hspace{0.21cm}}
                     r @{\hspace{0.21cm}}
                     r @{\hspace{0.21cm}}
                     r @{\hspace{0.21cm}}
                     r @{\hspace{0.21cm}}
                     r @{\hspace{0.21cm}}
                     r @{\hspace{0.21cm}}
                     r @{\hspace{0.21cm}}
                     r @{\hspace{0.21cm}}
                     r @{\hspace{0.21cm}}
                     r } 
\hline
        PN G & Main Name & \Ne(S~{\sc ii})  &
                                      \Te(O~{\sc iii}) &
                                              \Te(N~{\sc ii})  &
                                                         He/H     &  N/H     &  O/H     &  Ne/H    &   S/H    &  Ar/H    &  Cl/H    \\
\hline
 310.7-02.9  & He 2-103       & 3.08E+02  &  10404  &  12355  & 1.27E-01 & 8.98E-05 & 2.76E-04 & 1.08E-04 & 1.06E-05 & 1.30E-06 &          \\
             &                & 3.60E+02  &  11348  &  13745  & 1.34E-01 & 1.88E-04 & 4.72E-04 & 1.66E-04 & 2.62E-05 & 2.16E-06 &          \\
             &                & 2.24E+02  &   9488  &   9873  & 1.17E-01 & 7.86E-05 & 2.02E-04 & 4.72E-05 & 7.40E-06 & 1.03E-06 &          \\
             &                &           &         &         &          &          &          &          &          &          &          \\
 311.4+02.8  & He 2-102       & 1.06E+03  &         &  10812  & 1.31E-01 & 2.10E-05 & 3.52E-04 & 8.33E-05 & 4.30E-06 & 2.02E-06 & 6.10E-06 \\
             &                & 1.29E+03  &         &  11135  & 1.37E-01 & 2.50E-05 & 4.08E-04 & 9.62E-05 & 5.01E-06 & 2.27E-06 & 7.52E-06 \\
             &                & 8.21E+02  &         &  10507  & 1.24E-01 & 1.90E-05 & 3.01E-04 & 7.47E-05 & 3.80E-06 & 1.82E-06 & 4.88E-06 \\
             &                &           &         &         &          &          &          &          &          &          &          \\
 312.6-01.8  & He 2-107       & 2.49E+03  &   7015  &         & 1.42E-01 & 1.51E-04 & 4.69E-04 &          & 8.80E-06 & 5.06E-06 &          \\
             &                & 3.09E+03  &   7371  &         & 1.51E-01 & 1.92E-04 & 7.18E-04 &          & 1.24E-05 & 6.40E-06 &          \\
             &                & 1.98E+03  &   6396  &         & 1.32E-01 & 1.26E-04 & 3.20E-04 &          & 6.37E-06 & 4.12E-06 &          \\
             &                &           &         &         &          &          &          &          &          &          &          \\
 315.1-13.0  & He 2-131       &           &   8918  &  20404  & 3.20E-02 & 3.22E-05 & 6.71E-05 &          & 1.64E-07 & 1.18E-07 &          \\
             &                &           &   9197  &  24008  & 3.49E-02 & 3.87E-05 & 8.08E-05 &          & 2.53E-07 & 1.55E-07 &          \\
             &                &           &   8729  &  17216  & 2.86E-02 & 2.57E-05 & 5.22E-05 &          & 9.84E-08 & 9.17E-08 &          \\
             &                &           &         &         &          &          &          &          &          &          &          \\
 315.4+05.2  & He 2-109       & 5.57E+02  &  11001  &  10816  & 1.32E-01 & 1.36E-04 & 4.80E-04 & 1.64E-04 & 2.85E-06 & 1.93E-06 &          \\
             &                & 6.53E+02  &  11237  &  11032  & 1.39E-01 & 1.58E-04 & 5.50E-04 & 1.86E-04 & 3.39E-06 & 2.16E-06 &          \\
             &                & 4.72E+02  &  10605  &  10506  & 1.24E-01 & 1.20E-04 & 4.42E-04 & 1.48E-04 & 2.27E-06 & 1.71E-06 &          \\
             &                &           &         &         &          &          &          &          &          &          &          \\
 316.1+08.4  & He 2-108       & 8.71E+02  &   8083  &  10396  & 1.26E-01 & 2.80E-05 & 1.91E-04 & 1.62E-05 & 1.67E-06 & 1.92E-06 & 4.16E-07 \\
             &                & 1.06E+03  &   8463  &  11712  & 1.33E-01 & 3.78E-05 & 2.80E-04 & 2.67E-05 & 3.21E-06 & 2.85E-06 & 1.03E-06 \\
             &                & 7.45E+02  &   7592  &   9057  & 1.17E-01 & 2.19E-05 & 1.39E-04 & 1.00E-05 & 1.06E-06 & 1.36E-06 & 1.89E-07 \\
             &                &           &         &         &          &          &          &          &          &          &          \\
 318.3-02.5  & He 2-116       & 1.75E+02  &  10299  &         & 1.44E-01 & 1.95E-04 & 3.52E-04 & 1.59E-04 & 1.64E-05 & 3.05E-06 &          \\
             &                & 2.37E+02  &  10641  &         & 1.54E-01 & 2.32E-04 & 4.27E-04 & 2.31E-04 & 1.86E-05 & 3.63E-06 &          \\
             &                & 1.17E+02  &   9872  &         & 1.35E-01 & 1.76E-04 & 3.17E-04 & 1.19E-04 & 1.54E-05 & 2.74E-06 &          \\
             &                &           &         &         &          &          &          &          &          &          &          \\
 319.2+06.8  & He 2-112       & 1.95E+03  &  12685  &  14422  & 1.40E-01 & 2.67E-04 & 2.93E-04 & 7.20E-05 & 5.29E-06 & 1.74E-06 & 5.26E-07 \\
             &                & 2.41E+03  &  13103  &  15135  & 1.47E-01 & 3.16E-04 & 3.24E-04 & 7.66E-05 & 5.98E-06 & 1.87E-06 & 6.16E-07 \\
             &                & 1.60E+03  &  12124  &  14126  & 1.31E-01 & 2.21E-04 & 2.54E-04 & 5.94E-05 & 4.40E-06 & 1.54E-06 & 4.05E-07 \\
             &                &           &         &         &          &          &          &          &          &          &          \\
 320.9+02.0  & He 2-117       & 6.84E+03  &  11101  &   8090  & 1.53E-01 & 5.92E-04 & 6.02E-04 & 1.58E-04 & 2.42E-05 & 6.85E-06 & 8.95E-06 \\
             &                & 1.03E+04  &  11727  &   8978  & 1.64E-01 & 9.02E-04 & 9.99E-04 & 2.70E-04 & 4.78E-05 & 9.91E-06 & 2.66E-05 \\
             &                & 4.74E+03  &  10264  &   7080  & 1.42E-01 & 3.69E-04 & 3.96E-04 & 8.86E-05 & 1.20E-05 & 4.96E-06 & 3.19E-06 \\
             &                &           &         &         &          &          &          &          &          &          &          \\
 321.3+02.8  & He 2-115       & 1.43E+04  &  11453  &   8926  & 1.17E-01 & 8.25E-05 & 3.71E-04 & 6.33E-05 & 7.15E-06 & 2.96E-06 & 4.43E-06 \\
             &                & 2.30E+04  &  12391  &   9618  & 1.24E-01 & 1.20E-04 & 8.64E-04 & 1.39E-04 & 1.23E-05 & 4.12E-06 & 1.01E-05 \\
             &                & 7.90E+03  &   7178  &   7803  & 1.08E-01 & 5.87E-05 & 2.63E-04 & 4.23E-05 & 4.44E-06 & 2.27E-06 & 1.68E-06 \\
             &                &           &         &         &          &          &          &          &          &          &          \\
 321.8+01.9  & He 2-120       & 2.91E+02  &   9709  &         & 1.47E-01 & 2.88E-04 & 4.54E-04 & 1.51E-04 & 8.43E-06 & 3.72E-06 &          \\
             &                & 3.70E+02  &  10064  &         & 1.55E-01 & 3.49E-04 & 5.27E-04 & 2.53E-04 & 1.09E-05 & 4.17E-06 &          \\
             &                & 2.20E+02  &   9419  &         & 1.39E-01 & 2.17E-04 & 3.81E-04 & 9.30E-05 & 5.92E-06 & 3.25E-06 &          \\
             &                &           &         &         &          &          &          &          &          &          &          \\
 322.5-05.2  & NGC 5979       & 5.09E+02  & (22456) &  14604  & 1.11E-01 & 6.70E-05 & 3.44E-04 & 5.75E-05 & 5.32E-06 & 2.27E-06 & 9.30E-06 \\
             &                & 8.49E+02  & (27308) &  15206  & 1.17E-01 & 7.62E-05 & 4.06E-04 & 6.59E-05 & 5.92E-06 & 2.48E-06 & 1.13E-05 \\
             &                & 2.40E+02  & ( 8121) &  14049  & 1.05E-01 & 5.62E-05 & 2.87E-04 & 5.13E-05 & 4.62E-06 & 2.03E-06 & 6.87E-06 \\
             &                &           &         &         &          &          &          &          &          &          &          \\
 323.9+02.4  & He 2-123       & 2.11E+03  &   7687  &         & 1.64E-01 & 2.44E-04 & 2.92E-04 & 5.82E-05 & 7.67E-06 & 3.19E-06 & 1.33E-06 \\
             &                & 2.51E+03  &   7859  &         & 1.74E-01 & 2.83E-04 & 3.31E-04 & 7.26E-05 & 8.75E-06 & 3.58E-06 & 1.58E-06 \\
             &                & 1.75E+03  &   7507  &         & 1.55E-01 & 2.22E-04 & 2.57E-04 & 4.16E-05 & 6.04E-06 & 2.79E-06 & 1.02E-06 \\
             &                &           &         &         &          &          &          &          &          &          &          \\
 324.2+02.5  & He 2-125       & 3.83E+03  &   6158  &         & 2.80E-02 & 2.38E-04 & 5.37E-04 &          & 1.34E-05 & 9.45E-07 &          \\
             &                & 5.83E+03  &   6492  &         & 3.03E-02 & 3.24E-04 & 9.27E-04 &          & 2.09E-05 & 1.20E-06 &          \\
             &                & 3.13E+03  &   5763  &         & 2.63E-02 & 1.74E-04 & 2.89E-04 &          & 8.83E-06 & 7.39E-07 &          \\
             &                &           &         &         &          &          &          &          &          &          &          \\
 324.8-01.1  & He 2-133       & 7.82E+03  &  14361  &  10072  & 1.40E-01 & 4.88E-04 & 4.28E-04 & 1.65E-04 & 1.25E-05 & 3.39E-06 & 4.67E-06 \\
             &                & 1.32E+04  &  15328  &  11171  & 1.49E-01 & 7.56E-04 & 6.82E-04 & 2.89E-04 & 2.19E-05 & 4.63E-06 & 1.07E-05 \\
             &                & 5.42E+03  &  13180  &   8719  & 1.29E-01 & 3.14E-04 & 2.88E-04 & 1.02E-04 & 7.20E-06 & 2.57E-06 & 1.61E-06 \\
             &                &           &         &         &          &          &          &          &          &          &          \\
\hline
\end{tabular}
\end{table*}

\setcounter{table}{2}
\begin{table*}
\caption{ 
  Continued.
}
\begin{tabular}{       @{\hspace{0.05cm}}
                     l @{\hspace{0.25cm}}
                     l @{\hspace{0.21cm}}
                     r @{\hspace{0.21cm}}
                     r @{\hspace{0.21cm}}
                     r @{\hspace{0.21cm}}
                     r @{\hspace{0.21cm}}
                     r @{\hspace{0.21cm}}
                     r @{\hspace{0.21cm}}
                     r @{\hspace{0.21cm}}
                     r @{\hspace{0.21cm}}
                     r @{\hspace{0.21cm}}
                     r } 
\hline
        PN G & Main Name & \Ne(S~{\sc ii})  &
                                      \Te(O~{\sc iii}) &
                                              \Te(N~{\sc ii})  &
                                                         He/H     &  N/H     &  O/H     &  Ne/H    &   S/H    &  Ar/H    &  Cl/H    \\
\hline
 327.1-01.8  & He 2-140       & 1.06E+04  &   8254  &         &          & 1.27E-04 & 2.86E-04 &          & 8.14E-06 & 2.29E-06 &          \\
             &                & 3.07E+04  &   8906  &         &          & 2.65E-04 & 1.08E-03 &          & 2.54E-05 & 3.92E-06 &          \\
             &                & 6.60E+03  &   7375  &         &          & 9.05E-05 & 1.39E-04 &          & 4.28E-06 & 1.74E-06 &          \\
             &                &           &         &         &          &          &          &          &          &          &          \\
 327.8-01.6  & He 2-143       & 6.37E+02  &  11164  &  12925  & 1.24E-01 & 3.53E-04 & 6.11E-04 & 1.34E-04 & 1.26E-05 & 3.22E-06 & 1.76E-06 \\
             &                & 7.70E+02  &  11580  &  13468  & 1.30E-01 & 4.10E-04 & 6.91E-04 & 1.52E-04 & 1.45E-05 & 3.52E-06 & 2.15E-06 \\
             &                & 5.43E+02  &  10822  &  12544  & 1.17E-01 & 2.83E-04 & 5.28E-04 & 1.18E-04 & 1.03E-05 & 2.87E-06 & 1.28E-06 \\
             &                &           &         &         &          &          &          &          &          &          &          \\
 336.9+08.3  & StWr 4-10      &           &         &  12825  & 1.00E-01 & 1.52E-05 & 1.37E-04 & 4.28E-05 & 2.17E-06 & 8.14E-07 &          \\
             &                &           &         &  13224  & 1.08E-01 & 1.94E-05 & 1.76E-04 & 4.96E-05 & 2.93E-06 & 1.04E-06 &          \\
             &                &           &         &  11893  & 9.39E-02 & 1.21E-05 & 1.21E-04 & 3.67E-05 & 1.96E-06 & 6.62E-07 &          \\
             &                &           &         &         &          &          &          &          &          &          &          \\
 338.1-08.3  & NGC 6326       & 8.09E+02  &  10845  &  11766  & 1.16E-01 & 9.76E-05 & 4.11E-04 & 9.24E-05 & 5.75E-06 & 1.94E-06 & 1.38E-06 \\
             &                & 9.61E+02  &  11721  &  12174  & 1.22E-01 & 1.24E-04 & 4.82E-04 & 1.07E-04 & 7.05E-06 & 2.25E-06 & 1.92E-06 \\
             &                & 6.84E+02  &   9864  &  11392  & 1.11E-01 & 6.99E-05 & 3.66E-04 & 8.18E-05 & 4.53E-06 & 1.73E-06 & 8.83E-07 \\
\hline
\end{tabular}
\end{table*}

\clearpage

\section{Supplementary figures}

\setcounter{figure}{0}

\begin{figure}[b]
\resizebox{\hsize}{!}{\includegraphics{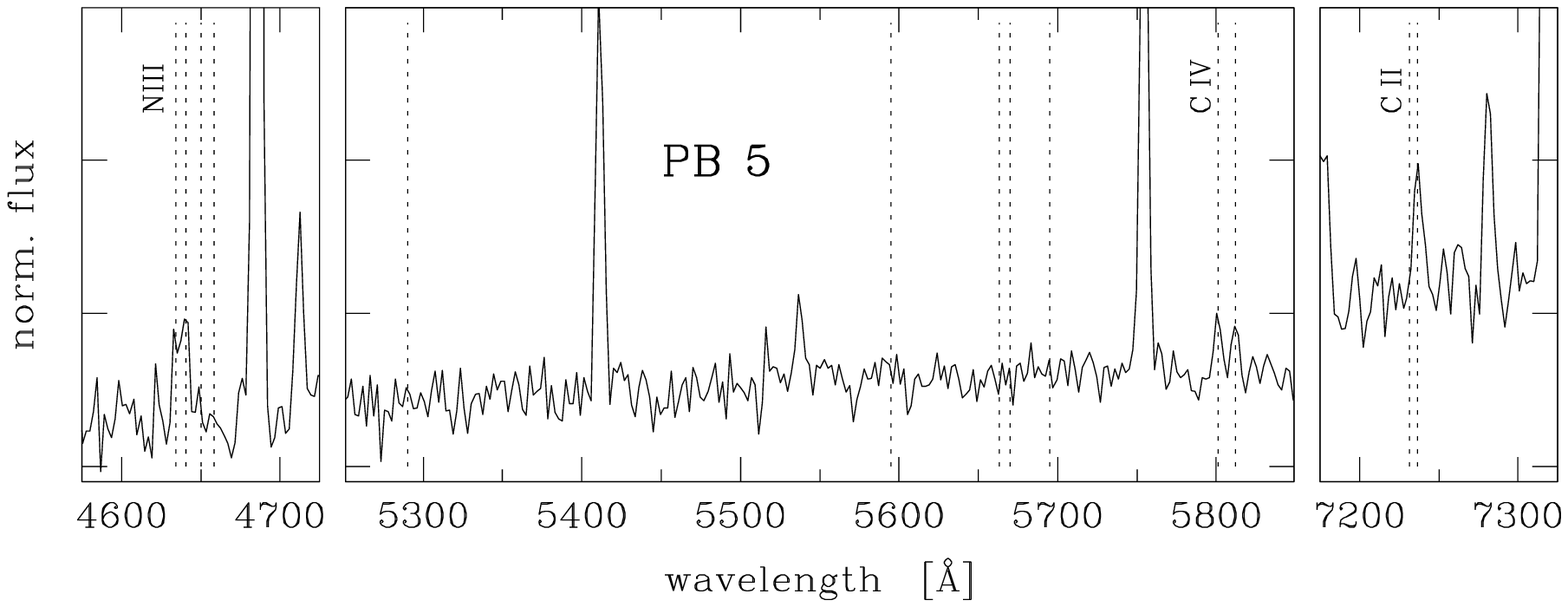}}
\resizebox{\hsize}{!}{\includegraphics{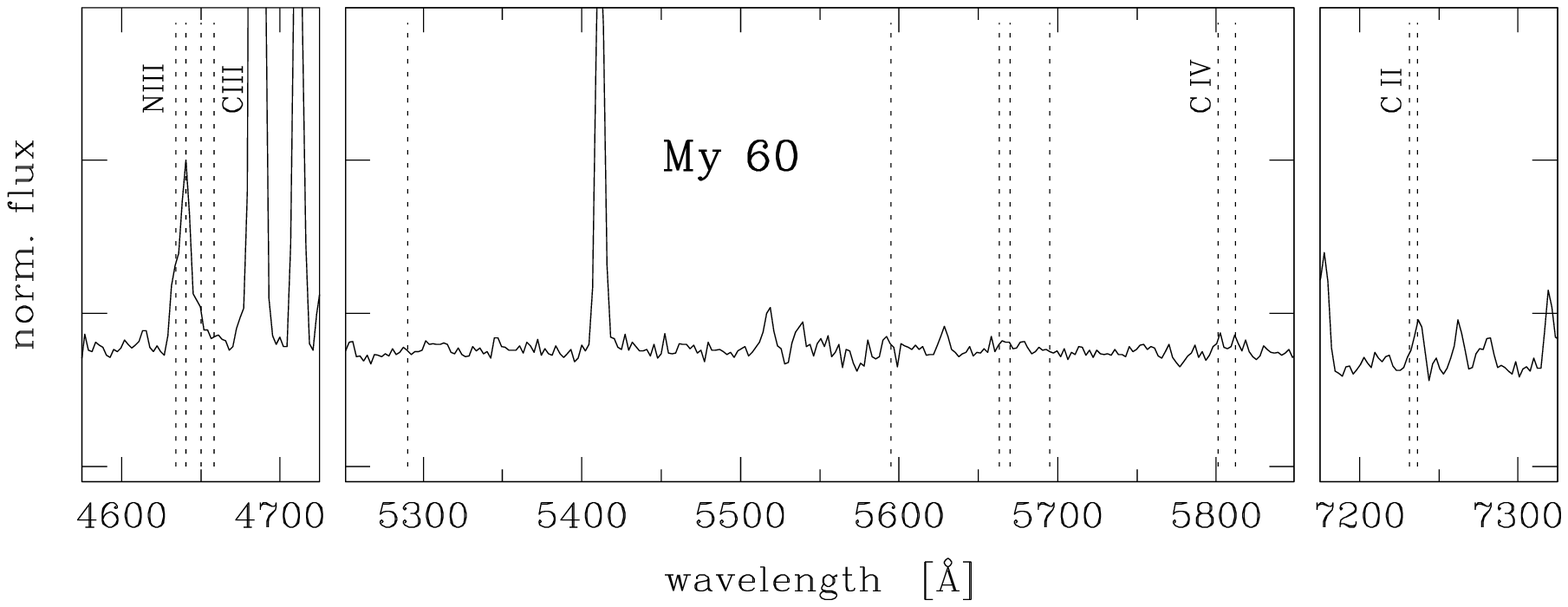}} 
\resizebox{\hsize}{!}{\includegraphics{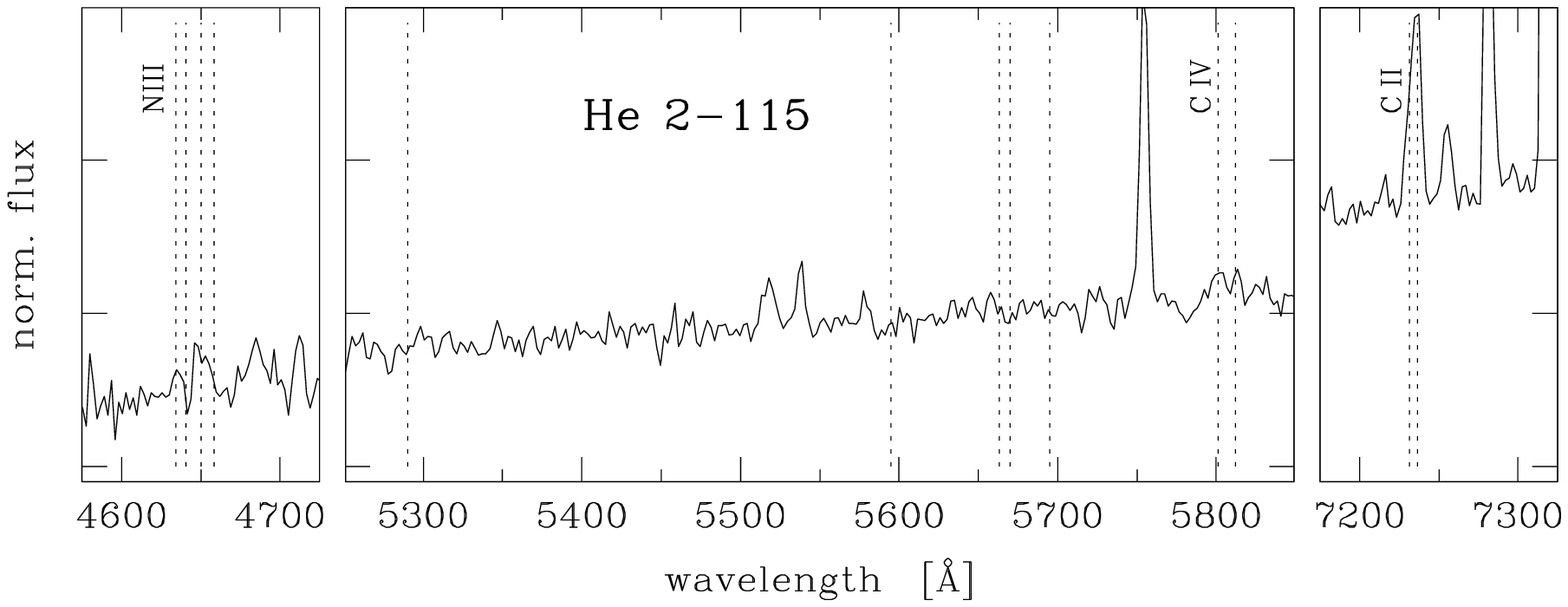}}  
\caption[]{
 Spectra of the new PNe with WELs stars. Dotted lines mark
 possible locations of stellar emission-lines identified with the ion name
 if the line was detected.
}
\label{spectra_wel}
\end{figure}

\begin{figure}[b]
\resizebox{\hsize}{!}{\includegraphics{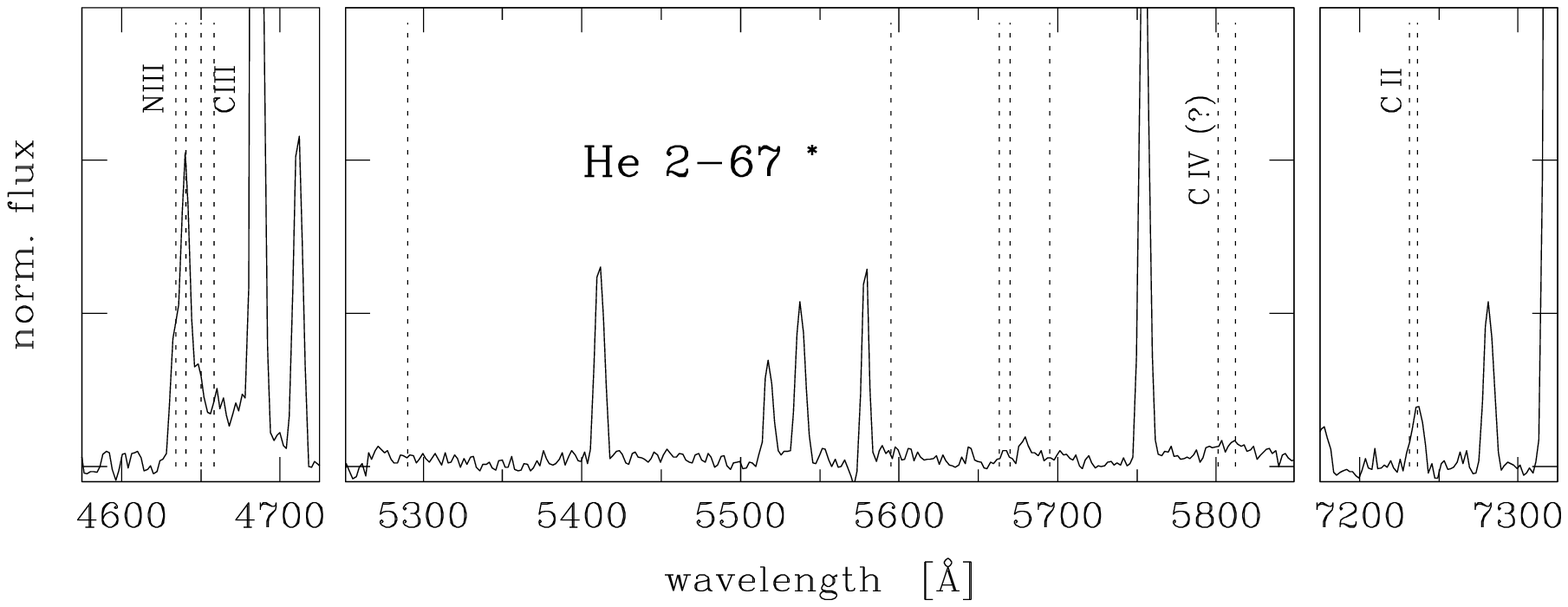}}
\resizebox{\hsize}{!}{\includegraphics{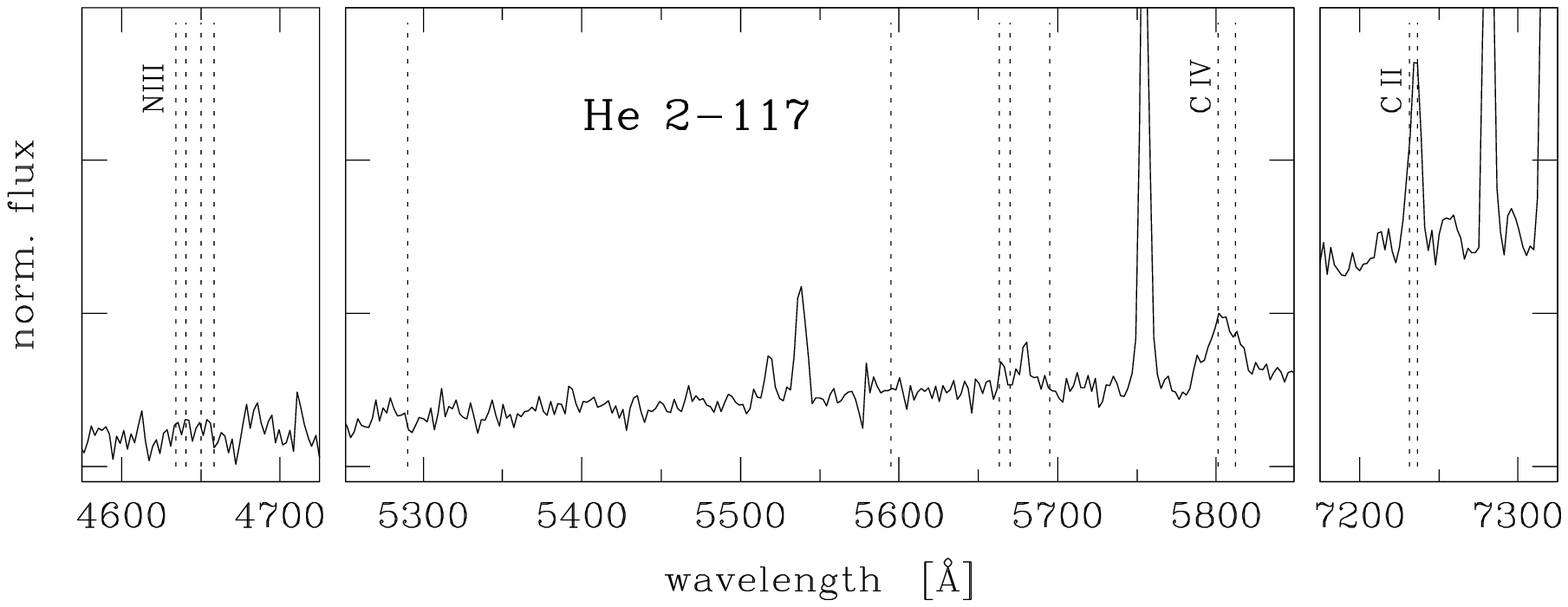}}
\resizebox{\hsize}{!}{\includegraphics{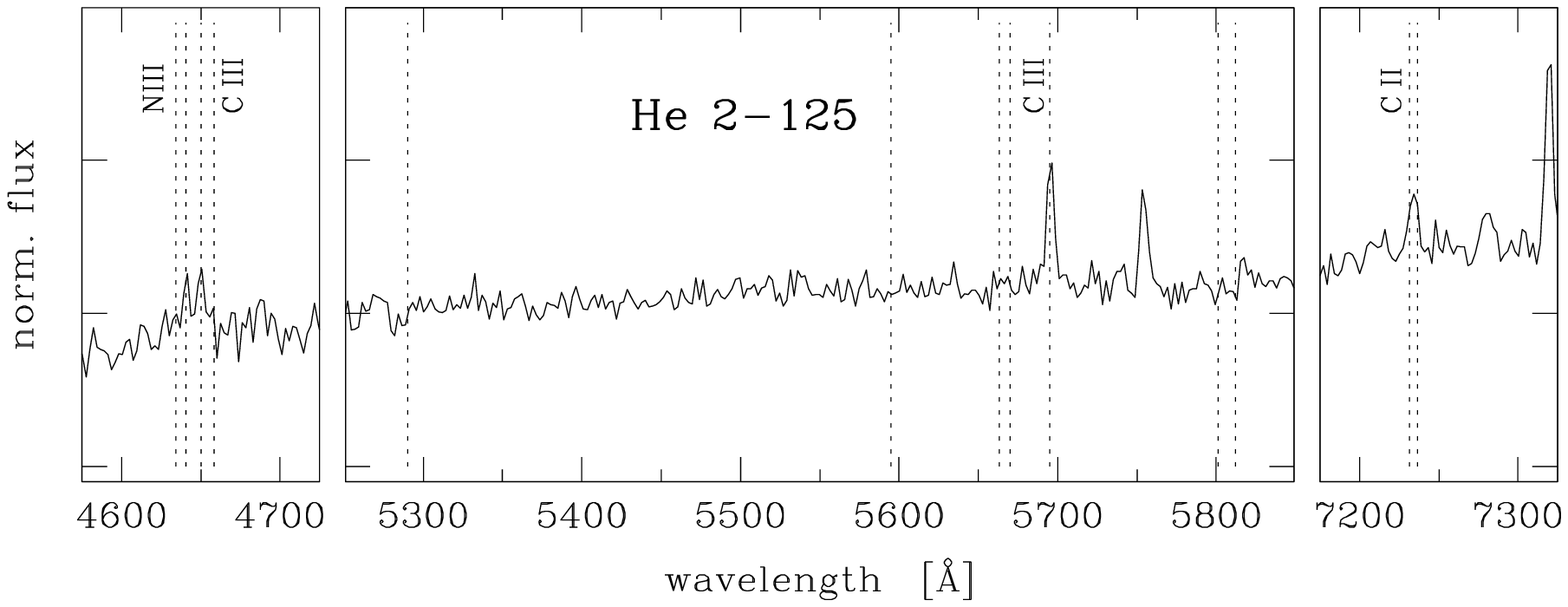}} 
\caption[]{
 Spectra of the possible new PNe with Wolf-Rayet type central stars. 
}
\label{spectra_WC}
\end{figure}

\begin{figure}
\resizebox{\hsize}{!}{\includegraphics{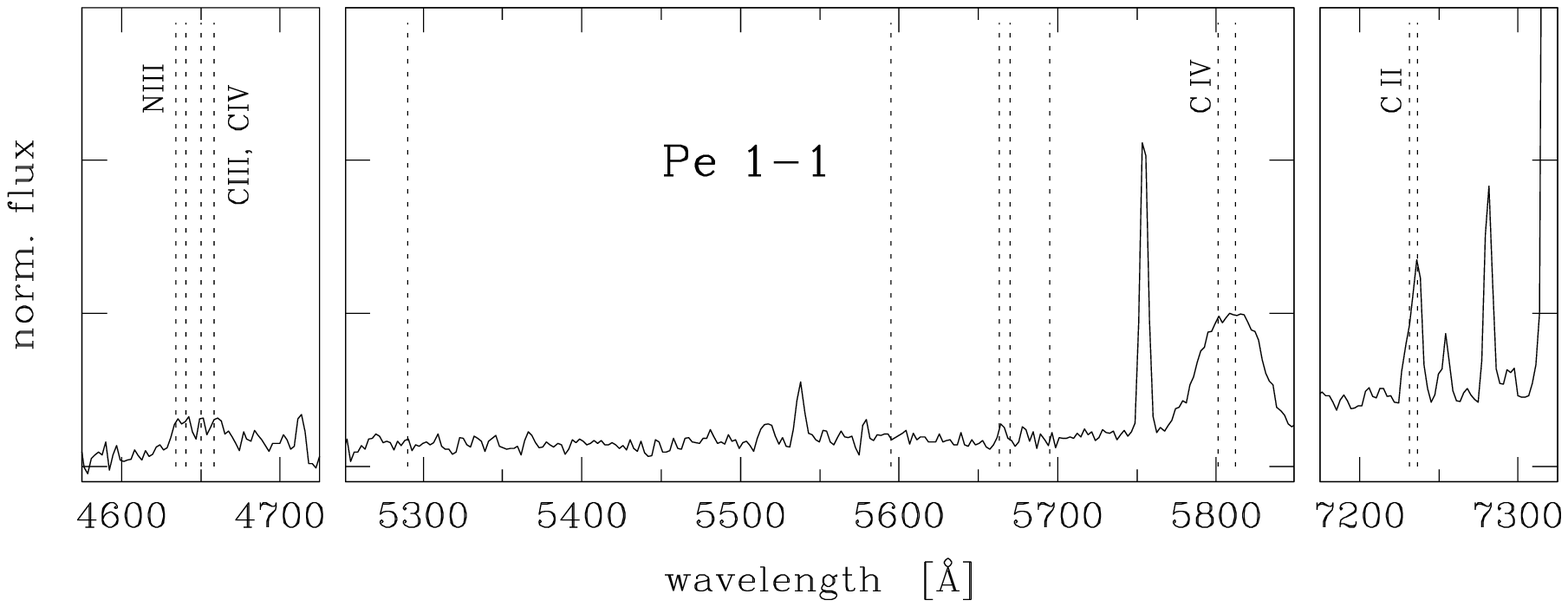}}
\resizebox{\hsize}{!}{\includegraphics{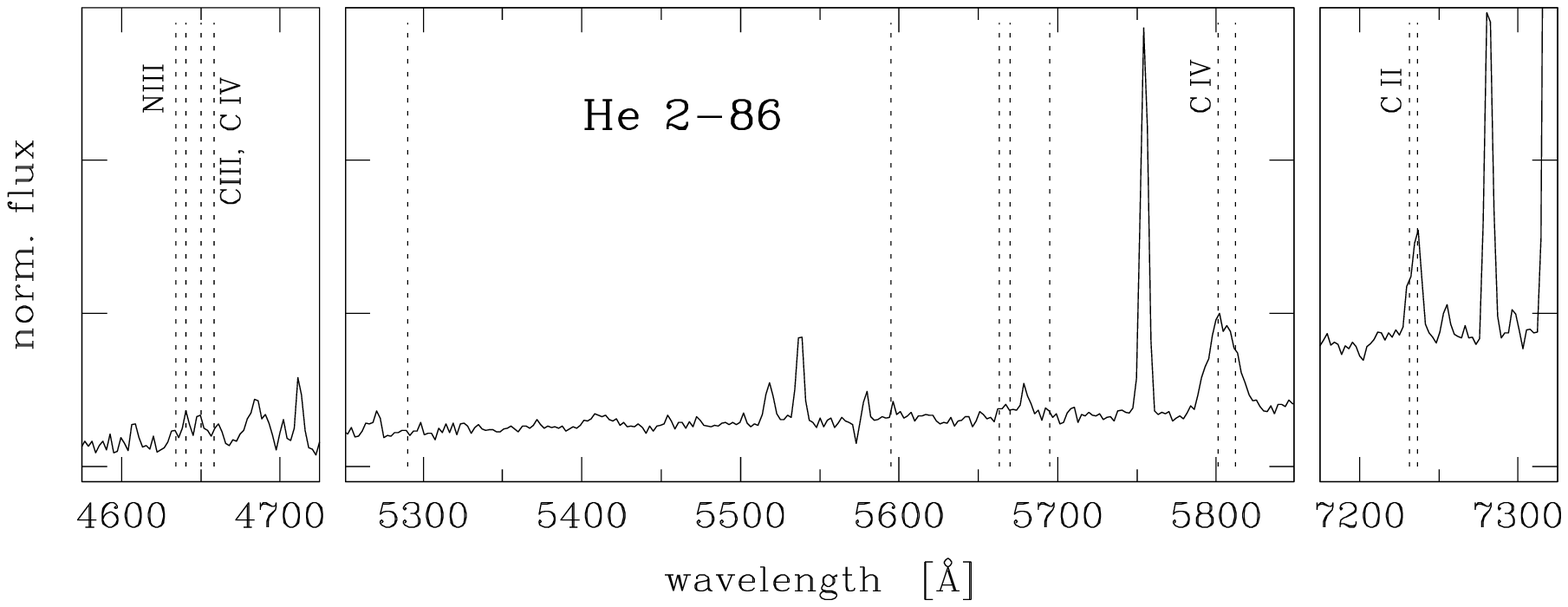}}
\resizebox{\hsize}{!}{\includegraphics{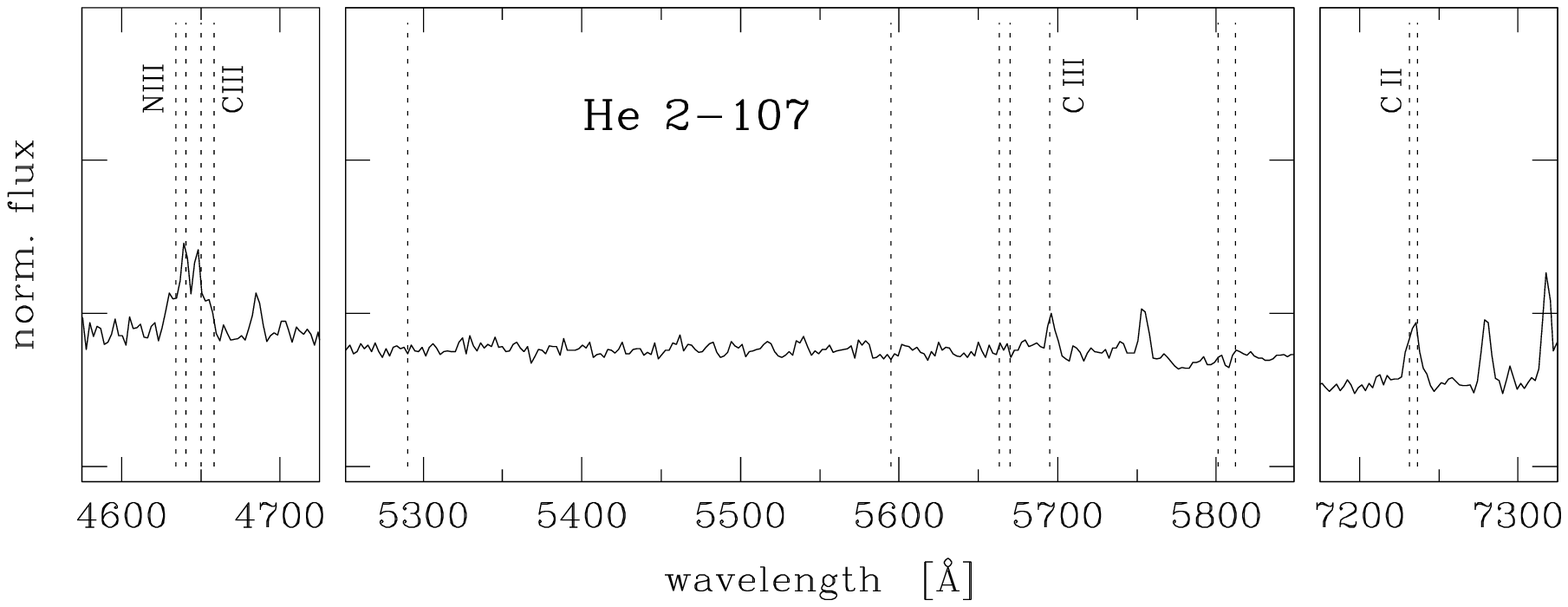}} 
\resizebox{\hsize}{!}{\includegraphics{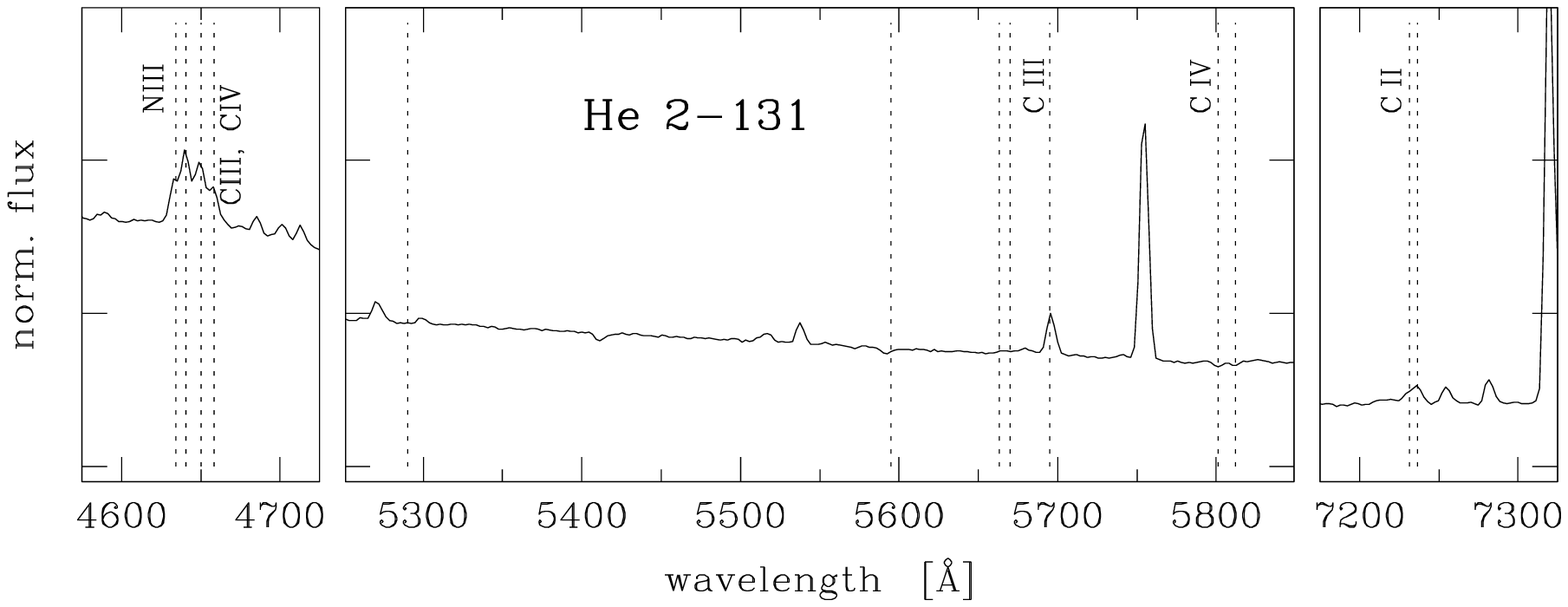}} 
\resizebox{\hsize}{!}{\includegraphics{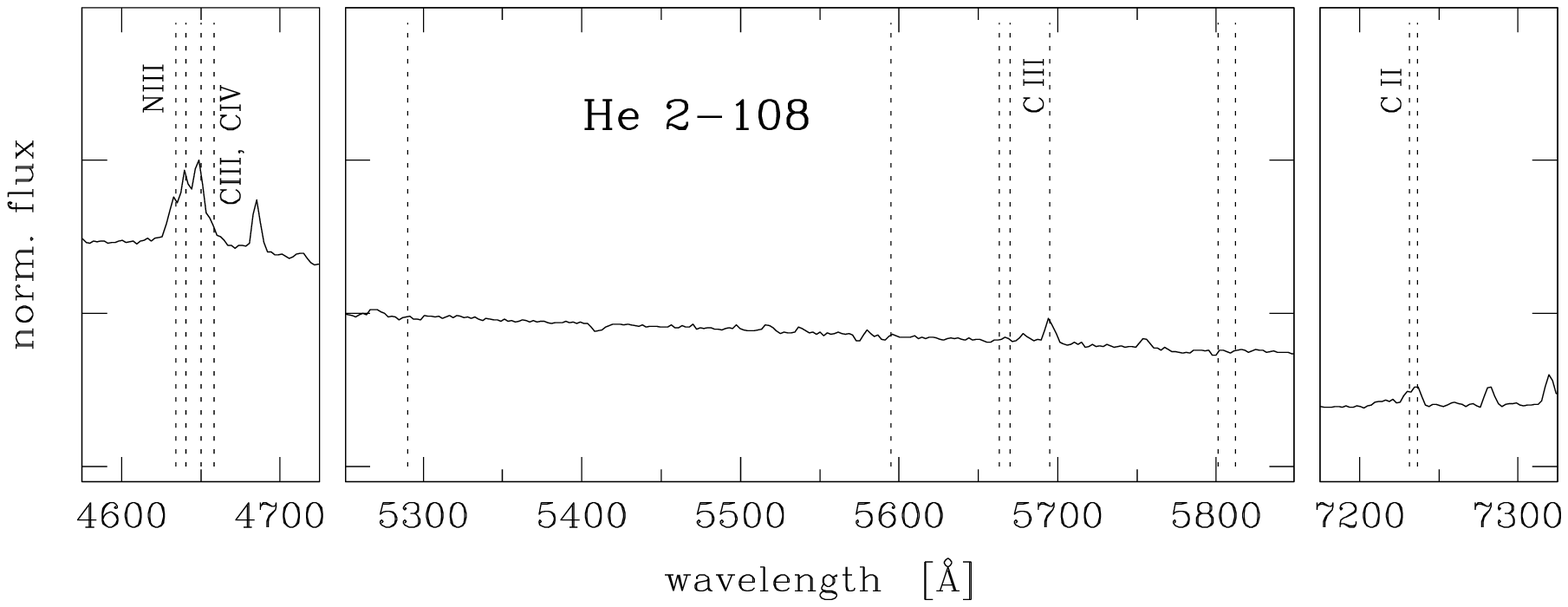}} 
\resizebox{\hsize}{!}{\includegraphics{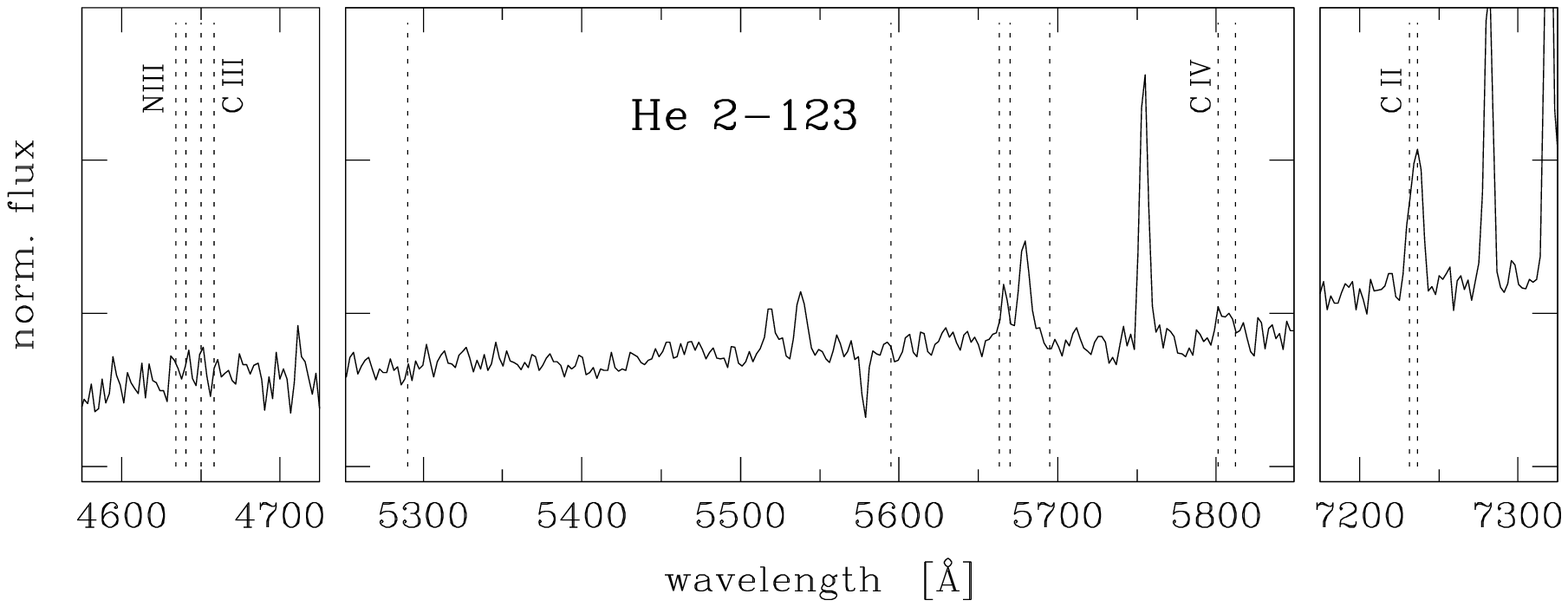}} 
\resizebox{\hsize}{!}{\includegraphics{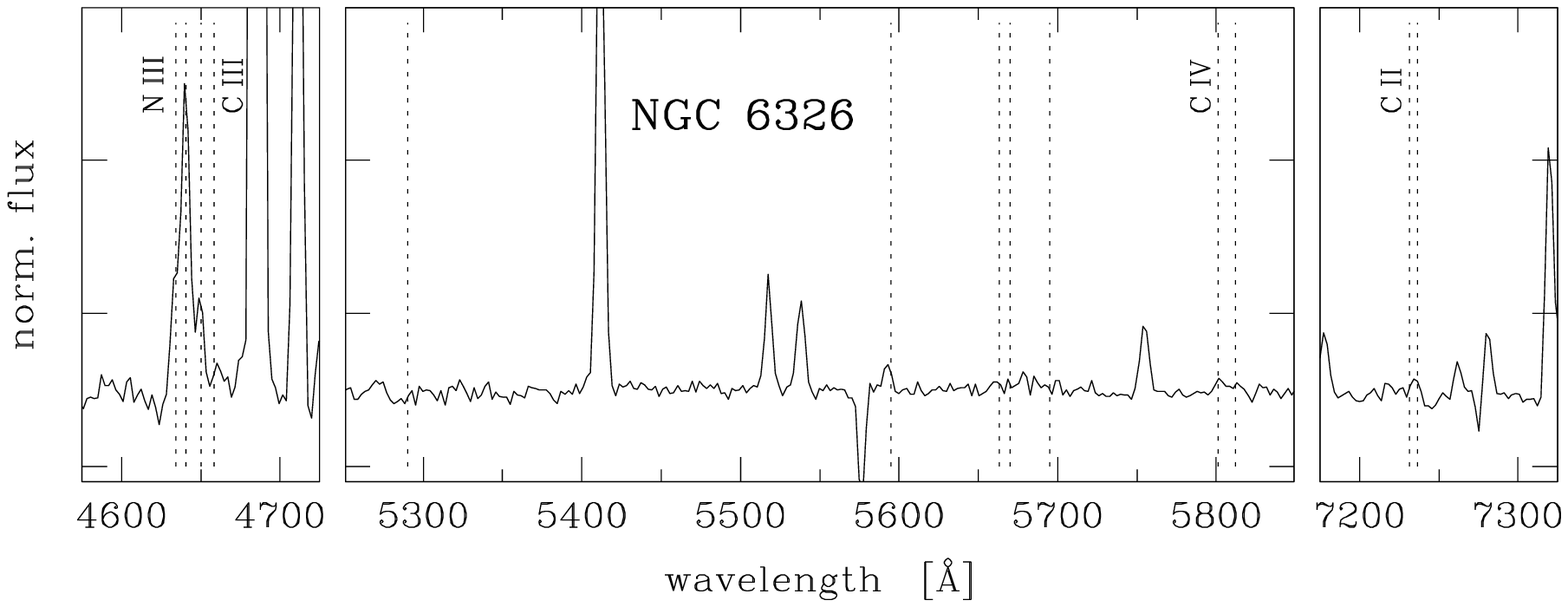}}  
\caption[]{
 Spectra of the reobserved PNe with known and misclassified 
 emission-line central stars. 
}
\label{spectra_reob}
\end{figure}

\begin{figure}
\resizebox{\hsize}{!}{\includegraphics{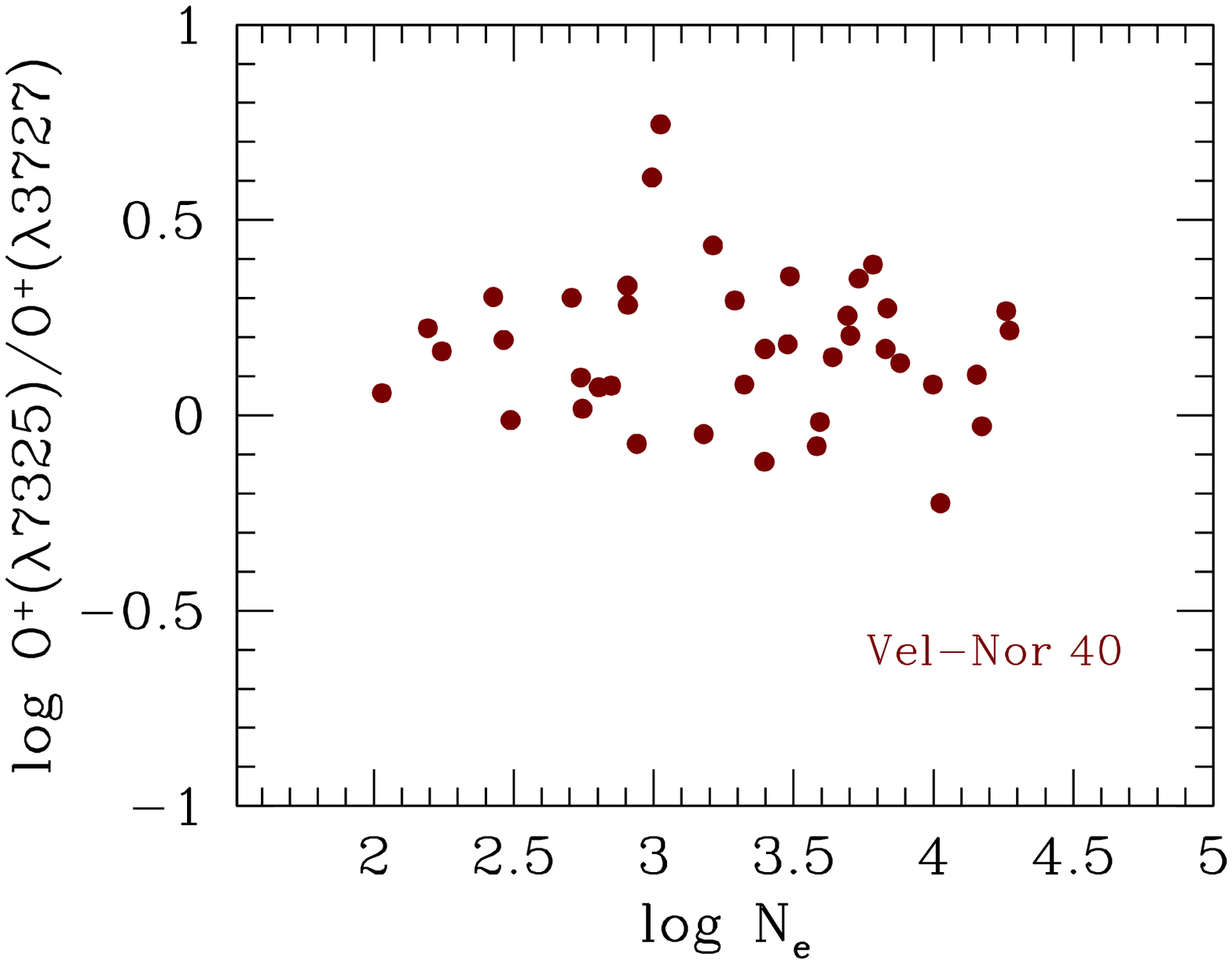}} 
\caption[]{
 Ratio of \Op\ ionic abundances derived from $\lambda$7325 and $\lambda$3727 
 lines as a function of electron density for observed PNe.
}
\label{opop}
\end{figure}

\begin{figure}
\resizebox{\hsize}{!}{\includegraphics{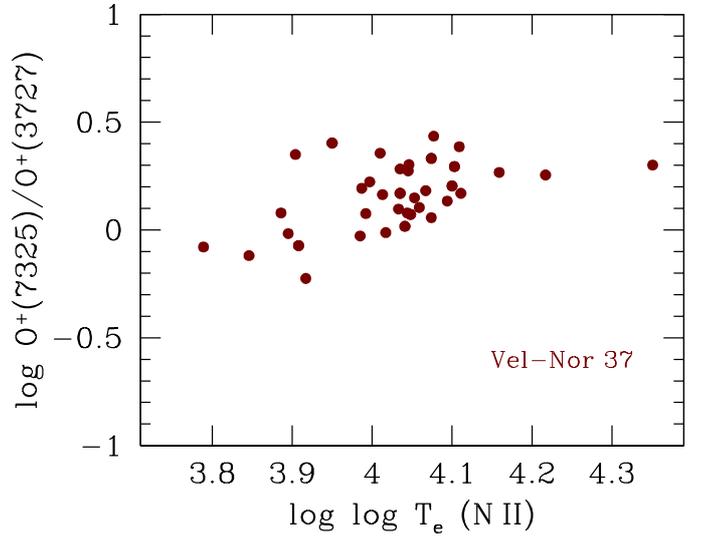}} 
\caption[]{
 Ratio of \Op\ ionic abundances derived from $\lambda$7325 and $\lambda$3727 
 lines as a function of electron temperature \Te(N~{\sc ii}) for observed PNe.
}
\label{opop2}
\end{figure}

\end{appendix}

\end{document}